\documentclass[12pt]{iopart}

\usepackage{iopams}
\usepackage{amssymb,bm,graphicx}
\usepackage{stmaryrd}
\usepackage[usenames,dvipsnames]{color}
\usepackage[normalem]{ulem}
\begin{document}

\title[Internal ring waves in a three-layer fluid over a linear shear current]{Internal ring waves in a three-layer fluid over a linear shear current}

\author{D Tseluiko$^1$, N S Alharthi$^{2,1}$, R Barros$^{1}$, K R Khusnutdinova$^{1}$\footnote{Author to whom any correspondence should be addressed.}}

\address{$^1$Department of Mathematical Sciences, Loughborough University, Loughborough LE11 3TU, UK}
\address{$^2$ Department of Mathematics, Faculty of Sciences and
Arts, King Abdulaziz University, Rabigh 25732, Saudi Arabia
%King Abdulaziz University, Department of Mathematics, Faculty of Sciences and Arts -- Rabigh Campus, Jeddah, Saudi Arabia}
}
\ead{K.Khusnutdinova@lboro.ac.uk  }
\vspace{10pt}
\begin{indented}
\item[] 30 June 2022
%\date{\today}
\end{indented}

\submitto{\NL}

\begin{abstract}
Oceanic internal waves often have curvilinear fronts and propagate over various currents. We present the first study of long weakly-nonlinear internal ring waves in a three-layer fluid in the presence of a background linear shear current. The leading order of this theory leads to the angular adjustment equation - a nonlinear first-order differential equation describing the dependence of the linear long-wave speed on the angle to the direction of the current. Ring waves correspond to singular solution (envelope of the general solution) of this equation, and they can exist only under certain conditions. The constructed solutions reveal qualitative differences in the shapes of the wavefronts of the two baroclinic modes: the wavefront of the faster mode is elongated in the direction of the current, while the wavefront of the slower mode is squeezed. Moreover, different regimes are identified according to the vorticity strength. When the vorticity is weak, part of the wavefront is able to propagate upstream. However, when the vorticity is strong enough, the whole wavefront propagates downstream. A richer behaviour can be observed for the slower mode. As the vorticity increases, singularities of the swallowtail-type may arise and, eventually, solutions with compact wavefronts crossing the downstream axis cease to exist. We show that the latter is related to the long-wave instability of the base flow. We obtain analytical expressions for the coefficients of the cKdV-type amplitude equation, and numerically model the evolution of the waves for both modes. The initial evolution is in agreement with the leading-order predictions for the deformations of the wavefronts. Then, as the wavefronts expand, strong dispersive effects in the upstream direction are revealed. Moreover, when nonlinearity is enhanced, fission of waves can occur in the upstream part of the ring waves.
\end{abstract}

\noindent{\it Keywords}: Internal ring waves, three-layer fluid, linear shear current 

%\maketitle

\section{Introduction}

Across the world's oceans, variations in seawater temperature and salinity stratify the water column. In many locations,  thin interfacial layers, known as pycnoclines, form whenever density varies sharply with depth. The stratification
%pycnocline 
supports internal waves which propagate horizontally over large distances. Such waves are frequently observed  in coastal oceans, and they are regularly generated in straits, river-sea interaction areas and around some localized topographic features. 
Internal waves have a strong effect on acoustic signalling and offshore structures and cables. Furthermore, they are responsible for much of the mixing vital to maintain the grand network of ocean currents that carries heat around the globe.
%as well as impacting submersibles, offshore structures and underwater pipelines. They also significantly contribute to the ocean mixing processes.
 Therefore, it is important to develop a good understanding of this geophysical phenomenon. 
 
The Korteweg--de Vries (KdV) model and its various extensions are an established paradigm for the description of long weakly- and moderately-nonlinear plane and nearly-plane internal waves commonly observed in the oceans around the world (see \cite{GOSS, HM, AOSL, GPTK, G, LGJ, KST, La} and references therein). However, on satellite images, internal waves generated in narrow straits \cite{KZ} or by river plums \cite{NM}, in particular, appear to have curvilinear wavefronts that resemble part of a ring wave and propagate over various currents, which motivates the present study. To describe ring waves and their relatives \cite{HKG}, we require the development of akin models in cylindrical geometry. Such models were developed and analysed in the case of no background  current \cite{M, J, L, WZ, WV, MS, RRS, G1,HFMS}, and extended to study the propagation of internal ring waves over a parallel shear current~\cite{KZ, KZ1, K, HKG}, generalizing the work by Johnson on surface ring waves in a homogeneous fluid over a parallel shear current \cite{J1, Jbook}. 
%Another recent development concerns the higher-order extensions of the KdV- and cKdV-type models (see \cite{KST, HFMS} and references therein). A recent review of the strongly-nonlinear long-wave models for water waves can be found in \cite{La}. 

%
%
%for the long surface ring waves propagating in a homogeneous fluid over a parallel shear current being developed in 
% \cite{J1, Jbook}, and to the case of the surface and internal waves propagating over a parallel shear current  in stratified fluids in   \cite{KZ, KZ1, K, HKG}. Another recent development concerns the higher-order extensions of the KdV- and cKdV-type models (see \cite{KST, HFMS} and references therein). A recent review of the lstrongly-nonlinear long-wave models for water waves can be found in \cite{La}. 

Although a general theoretical framework has been developed in \cite{KZ} for arbitrary density stratification and background shear current, using a linear modal decomposition in the far-field set of Euler equations with the boundary conditions appropriate for oceanographic applications, 
%particular features of 
solutions can only be constructed and analysed
%appreciated 
when specific physical configurations (i.e. stratification and current) are fixed. Indeed, we need to solve a spectral problem defining three-dimensional modal functions and an angular adjustment equation (a nonlinear first-order ODE defining the linear long wave speed at any angle to the current), as well as a nonlinear amplitude equation of cylindrical KdV (cKdV) type (generally, 2+1-dimensional) with coefficients dependent on solutions of the modal equations. 
%examined in detail. 

The simplest stratified configuration amenable to analytical studies is a two-layer system (see \cite{KZ, KZ1, K, HKG}). For example, in \cite{KZ, KZ1}, ring waves in a two-layer flow with piecewise-constant current, bounded above by a free surface, were considered. For this system, there are two modes of propagation and for the fast (barotropic) mode describing surface waves, it was found that wavefronts were elongated along the current. This feature is aligned with the findings of Johnson for a homogeneous fluid \cite{J1, Jbook}. In contrast, it was shown that wavefronts for the slow (baroclinic) mode describing interfacial waves displayed a counter-intuitive squeezing along the current. The nonlinear evolution for the round dam-break problem was also considered and the formation of 2D dispersive shock waves and oscillatory wave trains was revealed for axisymmetric ring waves  in the absence of a current \cite{KZ1}.      

Linear theory predicts infinitely many baroclinic modes of propagation in continuously stratified oceans. We note that the two-layer system can only describe the first baroclinic mode (mode-1). To model more realistic situations and be able to describe higher baroclinic modes, while taking advantage of the simplicity of a layered model, more layers need to be considered. In this work, we investigate internal ring waves of the second baroclinic mode (mode-2), by adopting a three-layer system. Moreover, vorticity effects on such waves are explored and numerically modelled for the first time by including a background linear shear current.  
% bounded by two rigid walls. 

%composed by two homogeneous immiscible fluids, relevant for oceanic conditions when a sharp density transition is present.  
%
% 
% , such as the elongation of the wave fronts along the current for surface ring waves
% 
% Previous studies based on  this weakly-nonlinear theory were devoted only to two-layer flows \cite{KZ, KZ1, K, HKG}, and there were no attempts of systematic numerical modelling of the propagation of these two-dimensional internal waves in the presence of any shear flow and for any stratification. This new combined analytical and numerical research is the first attempt to study the  interfacial ring waves in a more realistic and complicated setting of a  three-layer fluid over a parallel shear current,  supporting not one but two interfacial ring modes. The three-layer stratification is frequently used to model internal waves in oceanic contexts. 
This paper is organised as follows. In Section~2, we 
%formulate 
briefly overview the theoretical framework
%the mathematical problem 
(following closely \cite{KZ}). In Section~3, the modal equations and the angular adjustment equation (which can be regarded as a 2D long-wave dispersion relation in the form of a nonlinear first-order ODE) is obtained and its singular solution (envelope of the general solution) is constructed, which are essential building blocks of the theory. 
%In particular, we derive the analytical expressions for the coefficients of the cKdV-type amplitude equation which depend on the respective solutions of the modal and angular adjustment equations. 
Section~4 is devoted to describing interesting three-dimensional effects of the shear flow on the wavefronts and vertical structure of two modes of interfacial ring waves. %which were revealed using the constructed analytical solution of the modal and angular adjustment equations.
Different regimes are unveiled according to the vorticity strength and it is shown that mode-2 ring waves can cease to exist provided the vorticity is strong enough. Moreover, swallowtail-type singularity of the wavefront may be formed before that.
In Section~5 we relate these findings with the transition to a long-wave instability. In Section~6 we develop and compare two different approaches to the construction of the singular solution of the angular adjustment equation. We then consider an initial-value problem where ring waves are generated by a localised source, and numerically solve the nonlinear cKdV-type amplitude equation for non-axisymmetric ring waves on a linear shear current in Section~7. 
%The initial evolution of the waves is described by an analytical solution of the 2D linear wave equation assuming that there is no current. We then consider the case where part of the ring wave enters the region where it propagates over a linear shear current, and we numerically solve the amplitude equation in order to describe the behaviour of the relevant part of the  ring wave. 
%Numerical results agree with the analytical predictions concerning the deformation of the wavefronts for both modes of the ring waves. The knowledge of the analytical coefficients of the amplitude equation allows us to explain very strong dispersive effects observed in the upstream direction, as well as other features of the numerical solutions. 
Concluding remarks are given in Section~8.

\section{Problem formulation}
Consider the Euler equations for a density stratified fluid:
\begin{eqnarray*}
& \rho (u_t + u u_x + v u_y + w u_z) + p_x = 0, \\
%\label{1}
 & \rho (v_t + u v_x + v v_y + w v_z) + p_y = 0, \\
 %\label{2} \\
& \rho (w_t + u w_x + v w_y + w w_z) + p_z + \rho g = 0, \quad \\
%\label{3} 
& \rho_t + u \rho_x + v \rho_y + w \rho_z = 0, \quad \\
 %\label{4} \\
&u_x + v_y + w_z = 0, 
%\label{5} 
\end{eqnarray*}
with the free surface and rigid bottom boundary conditions typical for the oceanic applications:
\begin{eqnarray*}
&w = h_t + u h_x + v h_y, \quad p = p_a \quad \mbox{at} \quad z = h(x,y,t), \\
&w = 0 \quad \mbox{at} \quad z = 0. 
\end{eqnarray*}
Here, $u,v,w$ are the velocity components in the $x,y,z$ directions, respectively, $p$ is the pressure, $\rho$ is the density, $g$ is the gravitational acceleration, $z = h(x,y,t)$ is the free-surface height (with $z = 0$ at the bottom), and $p_a$ is the constant atmospheric pressure at the surface. It is assumed that in the basic state $u_0 = u_0(z), ~ v_0 = w_0 = 0, ~p_{0z} = - \rho_0 g, ~h = h_0$,  where $h_0$ is the unperturbed depth of the fluid. Also, $u_0(z)$ is a horizontal  shear flow in the $x$-direction, and  $\rho_0 = \rho_0(z)$ is a stable background density stratification (i.e. $\rho_0'(z)<0$).	
We introduce the vertical particle displacement $\zeta$ defined by the equation
%\begin{equation*}
$$\zeta_t + u \zeta_x + v \zeta_y + w \zeta_z = w,$$ 
%\label{9}
%\end{equation*}
satisfiying the surface boundary condition 
%\begin{equation*}
$\zeta = h - h_0 \  \mbox{at} \  z = h(x,y,t). $
%\label{10}
%\end{equation*}

%We aim to derive an amplitude equation for the amplitudes of the long surface and internal waves. %Thus, 
The following non-dimensionalisation is adopted:
%\begin{eqnarray*}
$x \to \lambda x, \ y \to \lambda y, \ z \to h_0 z, \ t \to \frac{\lambda}{c^*}t, \
u \to c^* u, \ v \to c^* v, \quad w \to \frac{h_0 c^*}{\lambda} w, \ 
 (\rho_0, \rho) \to \rho^*( \rho_0, \rho), \ h \to h_0 + a \eta, \
p \to p_a + \int_{z}^{h_0} \rho^* \rho_0(s) g ~\mathrm{d} s + \rho^* g h_0 p,$
%\end{eqnarray*}
where $\lambda$ is the wave length, $a$ is the wave amplitude, $c^*= \sqrt{g h_0}$ is the  long-wave speed of surface waves, 
%($\sqrt{g h_0}$ or $h^* N^*$, respectively, where $N^*$ is a typical value of the buoyancy frequency, %and $h^*$ is a typical depth of the stratified layer), 
$\rho^*$ is the dimensional reference density of the fluid, while $\rho_0(z)$ is the non-dimensional function describing stratification in the basic state, and $\eta = \eta(x,y,t)$ is the non-dimensional free-surface perturbation. 
Non-dimensionalisation leads to the appearance of two small parameters in the problem, the amplitude parameter $\varepsilon = a/h_0$ and the wavelength parameter $\delta = h_0/\lambda$.  The maximum balance condition $\delta^2 = \varepsilon$ is imposed here.
% although this is not the necessary condition. Indeed, 
%Variables can be scaled further to replace $\delta^2$ with $\varepsilon$ in the equations  \citep{Johnson_book}. 

% Thus, it is natural to non-dimensionalise the general problem formulation, including both surface and internal waves,  using the parameters of the faster surface waves, and measure the speeds of the internal waves as fractions of the surface wave speed, etc. However, if one is primarily interested in the study of internal waves, it is more natural to use the typical speed of the internal waves. 

A cylindrical coordinate system moving at a constant speed $c$ is considered,  and the same notations $u$ and $v$ are used for the projections of the velocity vector
%of the deviations of the speed 
on the new coordinate axes, with the scalings using the amplitude parameter $\varepsilon$,
 \begin{eqnarray*}
&x \to ct + r \cos \theta, ~~ y \to r \sin \theta, ~~ z \to z, ~~ t \to t, \\
&u \to u_0(z) + \varepsilon (u \cos \theta - v \sin \theta), ~~
v \to \varepsilon (u \sin \theta + v \cos \theta), ~~w \to \varepsilon w,\\
& p \to  \varepsilon p, ~~ \rho \to \rho_0 + \varepsilon \rho.
 \end{eqnarray*}
% we can arrive the non-dimensional problem formulation.
 
 A weakly-nonlinear solution of the problem can be constructed \cite{KZ} in the form of an asymptotic multiple-scale expansion 
 $
 \zeta = \zeta_1 + \varepsilon \zeta_2 + \dots,
 $
 and similar expansions for other variables. Here,
 %\begin{equation*}
$$ \zeta_1 = A(\xi, R, \theta) \phi(z, \theta), $$
%\label{11}
 %\end{equation*}
 with 
 %the following set of fast and slow variables:
 %\begin{eqnarray}
$ \xi = r k(\theta) - s t, \ R = \varepsilon r k(\theta)$,
 %\end{eqnarray}
and  $s$ defined as the wave speed in the absence of a shear flow. When a shear flow is present, the function $k(\theta)$ describes  the angular adjustment of the linear long wave speed in the direction of the polar angle $\theta$ to the current -- the speed is $s/k(\theta)$,  and $k(\theta)$ is to be found as part of the solution of the problem.  To leading order, this defines the distortion of the wavefront $ k(\theta) r - s t=\rm{constant}$ in a particular direction. 
 %In this description, when a shear flow is present, the wave speed in the direction $\theta$ is not equal to $s$, but to $s/k(\theta)$.  
% The choice of the fast and slow variables is similar to that in the derivation of the cKdV-type %equation for the surface waves \citep{Johnson90}, with 
 The formal range of the asymptotic validity of the model is defined by the conditions $\xi\sim R\sim O(1)$.  
 %To leading order, the wavefront at any fixed moment of time $t$ is described by the equation
%$$rk(\theta)=\text{constant},$$
%\begin{equation}
%$rk(\theta)=\mbox{constant},$
%\end{equation}
% and we consider  outward propagating ring waves, requiring that 
% the function 
% $k = k(\theta) > 0$.
 % is strictly positive.  
 
 To leading order, 
 %assuming that perturbations of the basic state are caused only by the propagating wave, 
 one obtains:
\begin{eqnarray*}
&& u_1 = - A \phi u_{0z} \cos \theta - \frac{k F}{k^2 + k'^{2}} A\phi_z,  \quad \\
&&v_1 = A \phi u_{0z} \sin \theta -  \frac{k' F}{k^2 + k'^{2}} A\phi_z,  \quad
w_1 =  A_{\xi} F \phi,  \\
&& p_1 = \frac{\rho_0}{k^2 + k'^{2}} A F^2 \phi_z,  \quad
\rho_1 = - \rho_{0z} A \phi,  
\end{eqnarray*}
and also 
$$\eta_1 = A \phi \quad \mbox{at} \quad z = 1,$$  
where the function $\phi = \phi(z, \theta)$ satisfies the set of modal equations:
\begin{eqnarray}
&&\left (\frac{\rho_0 F^2}{k^2 + k^{'2}} \phi_z\right )_z -  \rho_{0z} \phi = 0 \quad \mbox{for} \quad  0 < z < 1, \nonumber \\
&&\frac{F^2}{k^2 + k^{'2}} \phi_z -  \phi = 0 \quad \mbox{at} \quad z=1, \label{m} \\
&&\phi = 0 \quad \mbox{at} \quad z=0 \nonumber.
\end{eqnarray}
Here, $F = F(z, \theta) = -s + (u_0(z)  - c) (k \cos \theta - k' \sin \theta)$, and $c$ is set to be equal to the speed of the shear flow at the bottom, i.e.~$c =u_0(0)$. 
% We will fix the speed of the moving coordinate frame $c$ to be equal to the speed of the shear flow at the bottom, $c =u_0(0)$. 
 %Then, $F = -s \ne 0$ at $z = 0$, and the condition $F \phi = 0$ at $z=0$ implies (\ref{m3}), simplifying the mathematical formulation. 
 %Of course, the physics does not depend on the choice of $c$ (see a discussion in \cite{Johnson90}), but our derivation shows that the mathematical formulation simplifies if we choose $c = u_0(0)$. 
% The values of the wave speed $s$ in the absence of the shear flow, and the pair of functions $\phi(z, \theta)$ and $k(\theta)$,  for a given shear flow,  constitute solution of the modal equations (\ref{m}).  
 %Unlike the surface wave problem considered by \cite{Johnson90}, the exact form of equations for the wave speed $s$, in the absence of a shear flow, and the function $k(\theta)$, for a given shear flow, depend on stratification. 
 
Considering the higher-order corrections, the following cKdV-type amplitude equation for the ring waves was derived in \cite{KZ}:
\begin{equation}
\mu_1 A_R + \mu_2 A A_{\xi} + \mu_3 A_{\xi \xi \xi} + \mu_4 \frac{A}{R} + \mu_5 \frac{A_\theta}{R} = 0.\label{cKdV}
\end{equation}
The coefficients are given in terms of the solutions of the modal equations (\ref{m}) by the formulae:
\begin{eqnarray}
&& \mu_1 = 2 s \int_0^1 \rho_0 F \phi_z^2 ~ \mathrm{d}z, \quad 
%\label{c1}
\mu_2 = - 3 \int_0^1 \rho_0 F^2 \phi_z^3 ~ \mathrm{d}z, \quad \nonumber \\
%\quad\qquad\qquad\qquad\qquad\qquad\qquad
%\label{c2}\\
&&\mu_3 = - (k^2 + k'^2) \int_0^1  \rho_0 F^2 \phi^2 ~ \mathrm{d}z, 
%\qquad\qquad\qquad\qquad\qquad  
%\label{c123}\\
\nonumber \\
&&\mu_4 = - \int_0^1 \bigg(\frac{\rho_0 \phi_z^2 k (k+k'')}{(k^2+k'^2)^2} \big [ (k^2-3k'^2) F^2 
%\qquad \qquad\qquad \quad 
\nonumber \\
&& -4 k' (k^2 + k'^2) W_0 F \sin \theta     - W_0^2(k^2 + k'^2)^2  \sin^2 \theta \big ] 
% \qquad 
 \nonumber\\
 &&+  \frac{2 \rho_0 k}{k^2 + k'^2} F \phi_z \phi_{z\theta} [k' F + (k^2 + k'^2) W_0 \sin \theta ] \bigg) ~ \mathrm{d}z,
% \quad \quad 
% \label{c4}\\
\nonumber \\
&&  \mu_5 = - \frac{2k}{k^2 + k'^2} \int_0^1 \rho_0 F \phi_z^2 [k' F + W_0 (k^2+k'^2) \sin \theta ]~\mathrm{d}z, 
  \qquad 
   \label{mu}
\end{eqnarray}
where $W_0 = u_0-c$.

%\newpage
\section{Modal and angular adjustment equations for a three-layer system}\label{sec:angular_eq}

\begin{figure}[h]
\centering
\includegraphics[width = 0.4\textwidth] {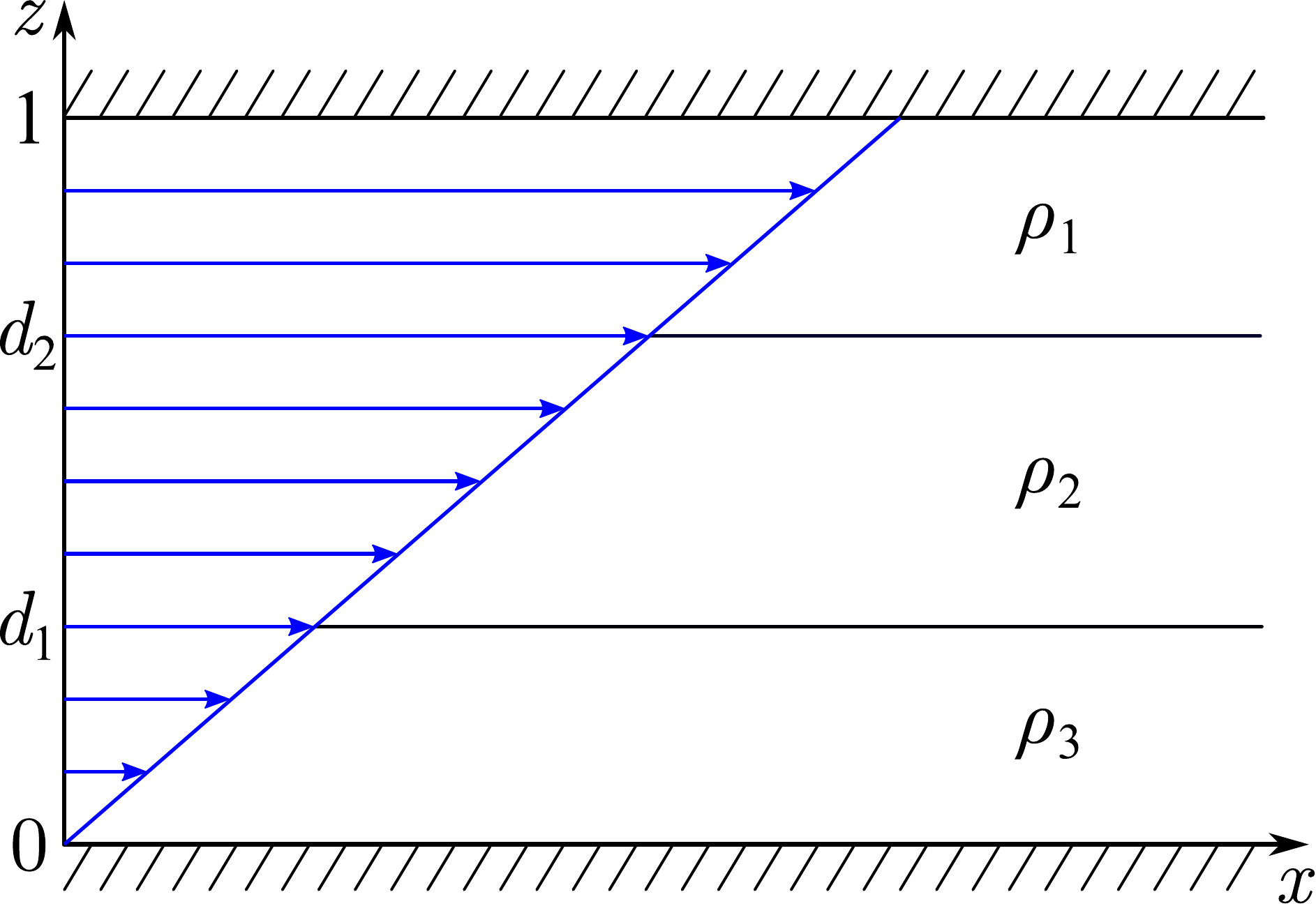}
\caption{Three-layer fluid with a linear shear current in the rigid-lid approximation.}
\label{fig:Pic20}
\end{figure}

In this section, we consider internal ring waves in a three-layer fluid over a linear current $u_0 = \gamma z$, shown in Figure~\ref{fig:Pic20}. Linear ring waves in a homogeneous fluid with that current have been extensively studied in \cite{ET}. Without loss of generality we assume that the vorticity $\gamma$ is nonnegative. Here, the density of the fluid is given by
%\vspace{0.3cm}
%\begin{eqnarray}
$\rho_{0}(z) = \rho_{3}H(z) + (\rho_{2}-\rho_{3})H(z-d_{1}) + (\rho_{1}-\rho_{2})H(z-d_{2})$, 
%\nonumber
%&& u_{0} = U_{3}H(z) + (U_{1}-U_{3})H(z-d_{1}) + (U_{1}-U_{2})H(z-d_{2}). \label{50} 
%\end{eqnarray}
%\vspace{0.3cm}
%\noindent 
where $d_{1}$ is the thickness of the lower layer,  $d_{2}- d_1$ is  the thickness of the middle layer and $1-d_{2}$ is the thickness of the top layer, and $H(z)$ is the Heaviside function. We assume the fluid to be stably stratified, so that $\rho_3>\rho_2>\rho_1$. 

%\vspace*{-1.5cm}
%\vspace{0.6cm}
%\noindent
We note that, for simplicity, we adopt the rigid-lid approximation. This approximation was shown to be in a very good agreement with the exact solution for the internal ring waves in the two-layer configuration \cite{HKG},  when the density contrast is small.
As a consequence, in (\ref{m}) the top boundary condition must be adapted, so that 
the modal equations take the form
%\vspace{0.3cm}
\begin{eqnarray}
&&\Big (\frac {\rho_{0} F^2}{k^2 + k'^2} \phi_{z}\Big)_{z} -\rho_{0z} \phi = 0 \quad  \mbox{for} \quad  0 < z < 1,  \nonumber \\
&&\phi = 0 \quad  \mbox{at}\quad   z = 1,   \label{modal}\\ 
&&\phi = 0 \quad \mbox{at}\quad   z = 0,  \nonumber       
\end{eqnarray}
%\newpage
%\vspace{0.6cm}
%\noindent  
with 
$$F = F(z,\theta) = -s + \gamma z\ (k \, \cos \theta - k' \, \sin \theta),$$ 
where $k = k(\theta)$ is an unknown function which should be found as part of the solution of this spectral problem (i.e. a spectral function). For the sake of brevity, in what follows, we do not explicitly indicate the dependence of the function $F$ on $\theta$. Then, the solution of the modal equations (\ref{modal}) in the respective layers is given by %linear functions of $z$ ( and parameter $\theta$) as:
%\begin{equation}
%\begin{aligned}
%\vspace{0.3cm}
\begin{eqnarray}
\hspace{-2.4cm}\phi_{1} = \frac {C_1(k^2+k'^2) (z-1)}{\rho_{1} F(1) F(z)} , \, \,
\phi_{2} =  \frac {C_3(k^2+k'^2)(z-d_{2})}{\rho_{2} F(d_{2}) F(z)} + C_4, \label{phi_all} \,\,
\phi_{3} = -\frac {C_2(k^2+k'^2) z}{\rho_{3} s F(z)},  
\end{eqnarray}
%\end{aligned}
%\end{equation}
%\vspace{0.4cm}
%\noindent 
%where $\phi_{1}$ is the modal function in the upper layer, $\phi_{2}$ is the modal function in the middle layer and $\phi_{3}$ is the modal function in the lower layer; 
where $C_1,\ldots, C_4$ are $z$-independent parameters (they are generally dependent on $\theta$).

\noindent Requiring the continuity of $\phi$, i.e.
$\phi_{1}(d_{2}) = \phi_{2}(d_{2})$ and 
$\phi_{2}(d_{1}) = \phi_{3}(d_{1})$, we obtain
%It is possible to eliminate one of them by using the continuity of $\phi$ at the interface between the two fluids ($z=d_{1}$ and $z=d_{2}$), we have
\begin{eqnarray}
\hspace{-1.5cm} C_3 =\frac{\rho_2}{d_1 - d_2} \left [ \frac{(1-d_2) F(d_1)}{\rho_1 F(1)}C_1 + \frac{d_1 F(d_2)}{\rho_3 F(0)}C_2 \right ] , \quad
C_4 = \frac {(k^2+k'^2) (d_2-1)}{\rho_{1} F(1) F(d_2)}C_1. \label{CD}
\end{eqnarray}

%The solution to the modal equations takes the form
%\vspace{0.3cm}
%\begin{eqnarray}\label{e-eqnarray}
%&&\phi_{1} = \frac {A(k^2+k'^2) (z-1)}{\rho_{1} F(1) F(z)} , \nonumber\\
%&&\phi_{2} =  \frac {(k^2+k'^2)[A(d_{2}-1)F(d_{1})-Bd_{1}F(d_{2})] + \rho_{2}F(z)[Bd_{1}-A(d_{2}-1)]}{\rho_{2}F(z)[F(d_{1})-F(d_{2})]},  \nonumber \\
%&&\phi_{3} = \frac {{B}(k^2+k'^2) z}{\rho_{3} s F(z)}. \label{78} 
%\end{eqnarray}
%\noindent Thus, the jump condition at two the interface:
%\noindent 
The function $\phi$ is continuous, but its derivative is discontinuous because of the discontinuities of $\rho_0(z)$. The modal equations (\ref{modal}) imply two jump conditions:
\begin{eqnarray*}
&&\lim_{\epsilon \rightarrow 0} \int_{d_{i}-\epsilon}^{d_{i}+\epsilon} \Big(\frac {\rho_{0} F^2}{k^2 + k'^2} \phi_{z} \Big)_{z} dz  =  \lim_{\epsilon \rightarrow 0} \int_{d_{i}-\epsilon}^{d_{i}+\epsilon} \rho_{0z} \phi \; dz \quad  i = {1,2}, \label{53} 
\end{eqnarray*}
where
%\begin{eqnarray}
$\rho_{0z} = \rho_{3}\delta(z) + (\rho_{2}-\rho_{3}) \delta(z-d_{1})+(\rho_{1}-\rho_{2}) \delta(z-d_{2})$,
%\label{55} 
%\end{eqnarray}
and $\delta(z)$ is the Dirac delta function.  Thus, we obtain the equations
%\begin{eqnarray}\label{e-eqnarray}
%&&\lim_{\epsilon \rightarrow 0} \bigg [\frac {\rho_{0} F^2}{k^2 + k'^2} \phi_{z} \bigg] _{d_{1}-\epsilon}^{d_{1}+\epsilon} - \rho_{3} \lim_{\epsilon \rightarrow 0}\int_{d_{1}-%\epsilon}^{d_{1}+\epsilon}\delta(z) \phi(z) dz - (\rho_{2}-\rho_{3}) \lim_{\epsilon \rightarrow 0}\int_{d_{1}-\epsilon}^{d_{1}+\epsilon} \delta(z-d_{1}) \phi(z) dz = 0 %\nonumber \\
%&&\lim_{\epsilon \rightarrow 0} \bigg [\frac {\rho_{0} F^2}{k^2 + k'^2} \phi_{z} \bigg] _{d_{1}-\epsilon}^{d_{1}+\epsilon}  - (\rho_{2}-\rho_{3}) \lim_{\epsilon \rightarrow %0}\int_{d_{1}-\epsilon}^{d_{1}+\epsilon} \delta(z-d_{1}) \phi(z) dz = 0 \nonumber \\
%&&\lim_{\epsilon \rightarrow 0} \bigg [\frac {\rho_{0} F^2}{k^2 + k'^2} \phi_{z} \bigg] _{d_{1}-\epsilon}^{d_{1}+\epsilon}  
%= (\rho_{2}-\rho_{3})  \lim_{\epsilon \rightarrow 0}\int_{d_{1}-\epsilon}^{d_{1}+\epsilon} \delta(z-d_{1}) \phi(z) dz \nonumber \\
%&&\lim_{\epsilon \rightarrow 0} \bigg [\frac {\rho_{0} F^2}{k^2 + k'^2} \phi_{z} \bigg] _{d_{1}-\epsilon}^{d_{1}+\epsilon} = (\rho_{2}-\rho_{3}) \phi(d_{1}) \nonumber
%\end{eqnarray}
%\vspace{0.3cm}
\begin{eqnarray*}
&&\frac{\rho_{2} F^2(d_{1})}{k^2 + k'^2} \phi_{2z}(d_{1}) - \frac{\rho_{3} F^2(d_{1})}{k^2 + k'^2} \phi_{3z}(d_{1}) =(\rho_{2}-\rho_{3}) \phi_{2}(d_{1}),\label{112}
\end{eqnarray*}
%\vspace{0.3cm}
%and 
%\vspace{0.3cm}
\begin{eqnarray*}
&&\frac{\rho_{1} F^2(d_{2})}{k^2 + k'^2} \phi_{1z}(d_{2}) - \frac{\rho_{2} F^2(d_{2})}{k^2 + k'^2} \phi_{2z}(d_{2}) = (\rho_{1}-\rho_{2}) \phi_{1}(d_{2}), \label{113}
\end{eqnarray*}
%\vspace{0.3cm}
%\noindent  
which yield the following system:
\vspace{0.3cm}
\begin{eqnarray}
&& C_3 - C_2 = (\rho_3 - \rho_2) \left [\frac{(d_2 - d_1) (k^2 + k'^2) C_3}{\rho_2 F(d_1) F(d_2)} - C_4\right ], \nonumber \\
&& C_1 - C_3 = (\rho_2 - \rho_1) \left [ \frac {(1-d_2) (k^2+k'^2)  }{\rho_1 F(d_2) F(1)}C_1\right]. 
\label{system}
\end{eqnarray}

%\vspace{0.3cm}
%%\newpage
%\noindent  Here, we can obtain $C$ and $B$ as a function of $A$:
%\vspace{0.3cm}
%\begin{eqnarray}
%&& B  = -\bigg [\frac{\rho_{3} (d_{2}-d_{1})F(0)}{\rho_{2} d_{1} F(d_{2})} + \frac{\rho_{3}(\rho_{2}-\rho_{1})(d_{2}-d_{1})(d_{2}-1) F(0)}{\rho_{1} \rho_{2} d_{1} F(1) F^2(d_{2})}(k^2+k'^2) \nonumber\\
%&&\hspace{1cm} + \frac{\rho_{3} (1- d_{2})F(0) F(d_{1})}{\rho_{1} d_{1} F(1) F(d_{2})}\bigg] A, \label{paraB}\\
%&& C  = \bigg [1- \frac{(\rho_2-\rho_1)(1-d_2)}{\rho_1F(1) F(d_2)}(k^2+k'^2) \bigg] A.  \label{paraC}
%\end{eqnarray}

\vspace{0.3cm}
\noindent  Substituting formulae (\ref{CD}) for $C_3$ and $C_4$ into (\ref{system}), we obtain a linear system for the parameters $C_1$ and $C_2$:
\begin{eqnarray}
&& \hspace{-2cm}\bigg [\frac{\rho_{2}(1-d_{2})F(d_{1})}{(d_{2}-d_{1}) \rho_{1} F(1)}\bigg] C_1 +  \bigg [1 + \frac{\rho_{2}d_{1}F(d_{2})}{(d_{2}-d_{1})\rho_{3}F(0)} - \frac{(\rho_{3}-\rho_{2})d_{1}(k^2+k'^2)}{\rho_{3}F(0)F(d_{1})} \bigg] C_2 = 0,\nonumber\\
&& \hspace{-2cm}\bigg [1+\frac{\rho_{2}(1 - d_{2})F(d_{1})}{(d_{2}-d_{1})\rho_{1}F(1)} - \frac{(\rho_{2}-\rho_{1})(k^2+k'^2)(1-d_{2})}{\rho_{1}F(1)F(d_{2})}\bigg] C_1 +\bigg [\frac{\rho_{2}d_{1}F(d_{2})}{(d_{2}-d_{1}) \rho_{3} F(0)}\bigg] C_2 = 0.\nonumber
\end{eqnarray}

\vspace{0.3cm}
\noindent To have a non-trivial solution we require that the determinant of the matrix of the coefficients of this system is equal to zero, which results in the nonlinear first-order  {\it angular adjustment equation} \cite{HKG} (2D long-wave dispersion relation) which constitutes 
%a further generalisation of the Burns' \cite{B} and 
an analogue of the generalised Burns'  condition for the surface ring waves in a homogeneous fluid \cite{J1,B}, obtained for a three-layer stratification and a linear shear current:
%\vspace{0.3cm}
%\begin{eqnarray}
%&& \bigg[\frac{\rho_{2}^2 d_{1}(1- d_{2})F(d_{1})F(d_{2})}{\rho_{1}  \rho_{3}(d_{2}-d_{1})^2 F(0)F(1)}\bigg]- \bigg [1 + \frac{\rho_{2} d_{1}F(d_{2})}{(d_{2}-d_{1})\rho_{3}F(0)} - \frac{(\rho_{3}-\rho_{2}) d_{1}(k^2+k'^2)}{\rho_{3} F(0)F(d_{1})} \bigg] \nonumber\\ 
%&&\hspace{0.1 cm} \times \bigg[ 1+ \frac{\rho_{2} (1 - d_{2}) F(d_{1})}{(d_{2}-d_{1})\rho_{1}F(1)} - \frac{(\rho_{2}-\rho_{1})(1-d_{2})(k^2+k'^2)}{\rho_{1}F(1)F(d_{2})} \bigg] = 0 .\label{new_eq}
%\end{eqnarray}
%\vspace{0.3cm}
%\noindent This condition provide us the required dispersion relation:
%\vspace{0.3cm}
%\begin{eqnarray}
%&&\bigg[\frac{(\rho_{1}-\rho_{2})(\rho_{2} -\rho_{3})d_{1}(1-d_{2})}{\rho_{1} \rho_{3}F(0,\theta) F(1,\theta)F(d_{1},\theta)F(d_{2},\theta)} \bigg] (k^2 +k'^2)^2  + \bigg[ \frac{(\rho_{2}-\rho_{3}) d_{1}}{\rho_{3}F(0, \theta)F(d_{1}, \theta)} \nonumber\\
%&&\hspace{0.1 cm} - \frac{\rho_{2}(\rho_{2} -\rho_{3})d_{1}(1-d_{2})}{\rho_{1}  \rho_{3}(d_{1}-d_{2})F(0,\theta) F(1,\theta)}+\frac{\rho_{2}(\rho_{1} -\rho_{2})d_{1}(d_{2}-1)}{\rho_{1}   \rho_{3}(d_{1}-d_{2})F(0,\theta) F(1,\theta)}  \nonumber\\
%&&\hspace{0.1 cm} - \frac{(\rho_{1}-\rho_{2})(d_{2}-1)}{\rho_{1}F(1,\theta)F(d_{2},\theta)} \bigg]  (k^2 +k'^2) + \bigg [ 1- \frac{\rho_{2}d_{1}F(d_{2},\theta)}{\rho_{3}(d_{1}-d_{2})F(0,\theta)} - \frac{\rho_{2}(1-d_{2})F(d_{1},\theta)}{\rho_{1}(d_{1}-d_{2})F(1,\theta)} \bigg] = 0.  \nonumber
%\end{eqnarray}
%\vspace{0.3cm}
%\noindent We can re-write the above equation as:
%\vspace{0.3cm}
\begin{equation}
A (k^2+k'^2)^2 + B(\theta,k,k') (k^2+k'^2) + C(\theta,k,k') = 0, \label{dr}
\end{equation}
where
%\vspace{0.3cm}
\begin{eqnarray}
&&\hspace{-1.7cm}A = (\rho_{2}-\rho_{1})(\rho_{3}-\rho_{2})d_{1}(1-d_{2})(d_{2}-d_{1}), \nonumber\\[0.2cm]
&&\hspace{-1.7cm}B(\theta,k,k') = - \big[\rho_{1}(\rho_{3}-\rho_{2})d_{1}(d_{2}-d_{1})F(d_{2}) F(1) \nonumber\\
&&\hspace{-1.2cm}+\rho_{2}(\rho_{3}-\rho_{1})d_{1}(1-d_{2})F(d_{1}) F(d_{2})  +\rho_{3}(\rho_{2}-\rho_{1})(1-d_{2})(d_{2}-d_{1})F(0)F(d_{1})\big], \nonumber\\[0.2cm]
&&\hspace{-1.7cm}C(\theta,k,k') = \rho_{1}\rho_{2}d_{1} F(d_{1}) F^2(d_{2})F(1)+\rho_{2}\rho_{3}(1-d_{2}) F(0) F^2(d_{1})F(d_{2})\nonumber\\
&&\hspace{-1.2cm}+\rho_{1}\rho_{3}(d_{2}-d_{1}) F(0) F(d_{1})F(d_{2})F(1).  \nonumber
\end{eqnarray}
We recall that $F(z)$ is a brief notation for   $F(z, \theta) = -s + \gamma z\ (k \, \cos \theta - k' \, \sin \theta)$. Thus, the angular adjustment equation (\ref{dr}) is a highly non-trivial nonlinear first-order differential equation for the function $k(\theta)$. Unlike the generalised Burns' condition, the form of the equation depends on the choice of stratification and current, and is obtained as part of solution of the spectral problem (modal equations). We note that a useful discussion of the linear internal waves in a general setting can be found in \cite{BVV}. 

This angular adjustment equation extends the results in \cite{HKG} for a two-layer fluid. To see this, consider the limiting cases when $\rho_2\rightarrow \rho_1$, or $d_2\rightarrow 1$. It can be shown that, in the former case, the angular adjustment equation reduces to 
\begin{eqnarray}
&&k^2+k'^2 =  \frac{\rho_{1}d_{1} F(d_{1}) F(1) + \rho_{3} (1-d_{2}) F(0) F(d_{1})}{(\rho_{3}-\rho_{1}) d_{1} (1-d_{1})}, \nonumber
\end{eqnarray}
and in the latter, 
\begin{eqnarray*}
&&k^2+k'^2 = \frac{\rho_{2}d_{1} F(d_{1}) F(1) + \rho_{3} (1-d_{1}) F(0) F(d_{1})}{(\rho_{3}-\rho_{2}) d_{1} (1-d_{1})}, 
\end{eqnarray*}
which both agree with the results for a two-layer fluid in \cite{HKG}. 
Other reductions to the two-layer case can be treated similarly.

%\vspace{0.3cm}
%\noindent Similarly, we can consider the limits $d_{2} \to d_{1}$, $d_{1} \to 0$, $\rho_{2} \to \rho_{3}$, which also reduce to the result for two layers. 

%\vspace{0.3cm}
%\noindent 
%The dispersion relation is obtained in the form:
%\vspace{0.3cm}
%\begin{eqnarray}
%\begin{split}
%&& (k^2 +k'^2)^2  + \bigg[ \frac{\rho_{1}F(1) F(d_{2})}{(\rho_{1}-\rho_{2})(1-d_{2})} + \frac{\rho_{2}F(d_{1}) F(d_{2})}{(\rho_{1}-\rho_{2}) (d_{2}-d_{1})} - \frac{\rho_{3}F(0) F(d_{1})}{(\rho_{2}-\rho_{3}) d_{1}} \\%\nonumber\\
%&&\hspace{0.001 cm} + \frac{\rho_{2}F(d_{1}) F(d_{2})}{(\rho_{2}-\rho_{3}) (d_{2}-d_{1})} \bigg]  (k^2 +k'^2) +\bigg[ - \frac{\rho_{1}\rho_{3}sF(1)F(d_{1}) F(d_{2})}{(\rho_{2} -\rho_{3})(\rho_{1}-\rho_{2})d_{1}(1-d_{2})} \\ %\nonumber\\
%&&\hspace{0.001 cm} -\frac{\rho_{1}\rho_{2}F(1) F^2(d_{2})F(d_{1})}{(\rho_{2}-\rho_{3})(\rho_{1}-\rho_{2})(1-d_{2})(d_{1}-d_{2})}+\frac{\rho_{2}\rho_{3}sF^2(d_{1}) F(d_{2})}{(\rho_{2}-\rho_{3})(\rho_{1}-\rho_{2})d_{1}(d_{1}-d_{2})} \bigg] = 0,   \label{new_eq1}
%\end{split}
%\end{eqnarray}
%
%\vspace{0.3cm}
%%\newpage
%\noindent with $F(z) = -s +\gamma z(k \cos \theta + k' \sin \theta)$. This nonlinear first order differential equation \eqref{new_eq1} for the function  $k(\theta)$ is further generalisation of both Burns and generalized Burns conditions \cite{J1}.  

%\noindent  
Let us first consider the case of no current, when $\gamma = 0$. Then, $k(\theta)\equiv 1, \, \theta \in[-\pi,\pi]$ (which physically corresponds to a concentric ring wave) is clearly a solution to  equation~(\ref{dr}), provided the wave speed $s$ satisfies the following algebraic equation: 

%of two concentric interfacial waves:

%\vspace{0.3cm}
\begin{equation}\label{lw_no_shear}
a_{4} s^4  + a_{2} s^2 + a_{0} = 0,
\end{equation}
where 
%\vspace{0.3cm}
%\noindent
\begin{eqnarray*}
&&\hspace{-2cm}a_{4} =  \rho_1 \rho_2 d_1 + \rho_1 \rho_3 (d_2-d_1) + \rho_2 \rho_3 (1-d_2),  \nonumber\\
&&\hspace{-2cm}a_{2} = \rho_1(\rho_2-\rho_3) d_1 (d_2-d_1) + \rho_2 (\rho_1-\rho_3) d_1 (1-d_2)+ \rho_3 (\rho_1-\rho_2) (d_2-d_1)(1-d_2), \nonumber\\
&&\hspace{-2cm}a_{0} = (\rho_{1}-\rho_{2})(\rho_{2}-\rho_{3})d_{1}(d_2-d_1)(1-d_{2}).  \nonumber
\end{eqnarray*}
From (\ref{lw_no_shear}) we obtain
%\vspace{0.3cm}
\begin{eqnarray}
s^2= \frac{-a_2 \pm \sqrt{a_2^2-4a_0 a_4}}{2a_4}, \label{speed}
\end{eqnarray}
%
%\begin{eqnarray}
%s^2= \frac{-b_{1} \pm \sqrt{b^2_{1}-4a_{1}c_{1}}}{2a_{1}} \label{speed}
%\end{eqnarray}
and both roots are real, as these coincide with the values ${c_0^\pm}^2$ corresponding to the linear long-wave speeds for a three-layer fluid at rest (see Section \ref{sec:lw_instability} and Appendix in \cite{BCM}). The upper sign corresponds to the faster (first baroclinic) mode, or mode-1, and the lower sign corresponds to the slower (second baroclinic) mode, or mode-2, of internal waves.  We note that for each mode, the constant solution $k(\theta)\equiv 1$ is in fact a {\it singular solution} of (\ref{dr}), which can be found as the envelope of the one-parameter family of solutions 
$$
k(\theta)=a \cos \theta + \sqrt{1-a^2} \sin \theta, \quad \theta\in[-\pi,\pi].
$$

Assuming that the current is sufficiently weak guarantees the existence of a part of the wavefront that is able to propagate in the upstream direction, and therefore $\theta \in [-\pi,\pi]$. This corresponds to the so-called {\it elliptic regime} as defined in \cite{K}, with the singular solution having just once branch. However, 
%other regimes are also possible, they will be considered later in the text. 
%So far, the ring waves presented are within the elliptic regime, where $\theta$ is within $[-\pi, \pi]$, and so part of wavefront propagates upstream. However, 
when the vorticity is stronger, it is possible that the whole wavefront propagates downstream, with $\theta$ taking values only in some subdomain of $(-\frac{\pi}{2}, \frac{\pi}{2})$, and the singular solution containing two branches. This regime is called {\it hyperbolic}, and the transition between the two occurs when the wavefront has a fixed point at the origin, which is referred to as the {\it parabolic regime} \cite{K}.

\begin{figure}
\centering
\includegraphics[width = 0.7\textwidth] {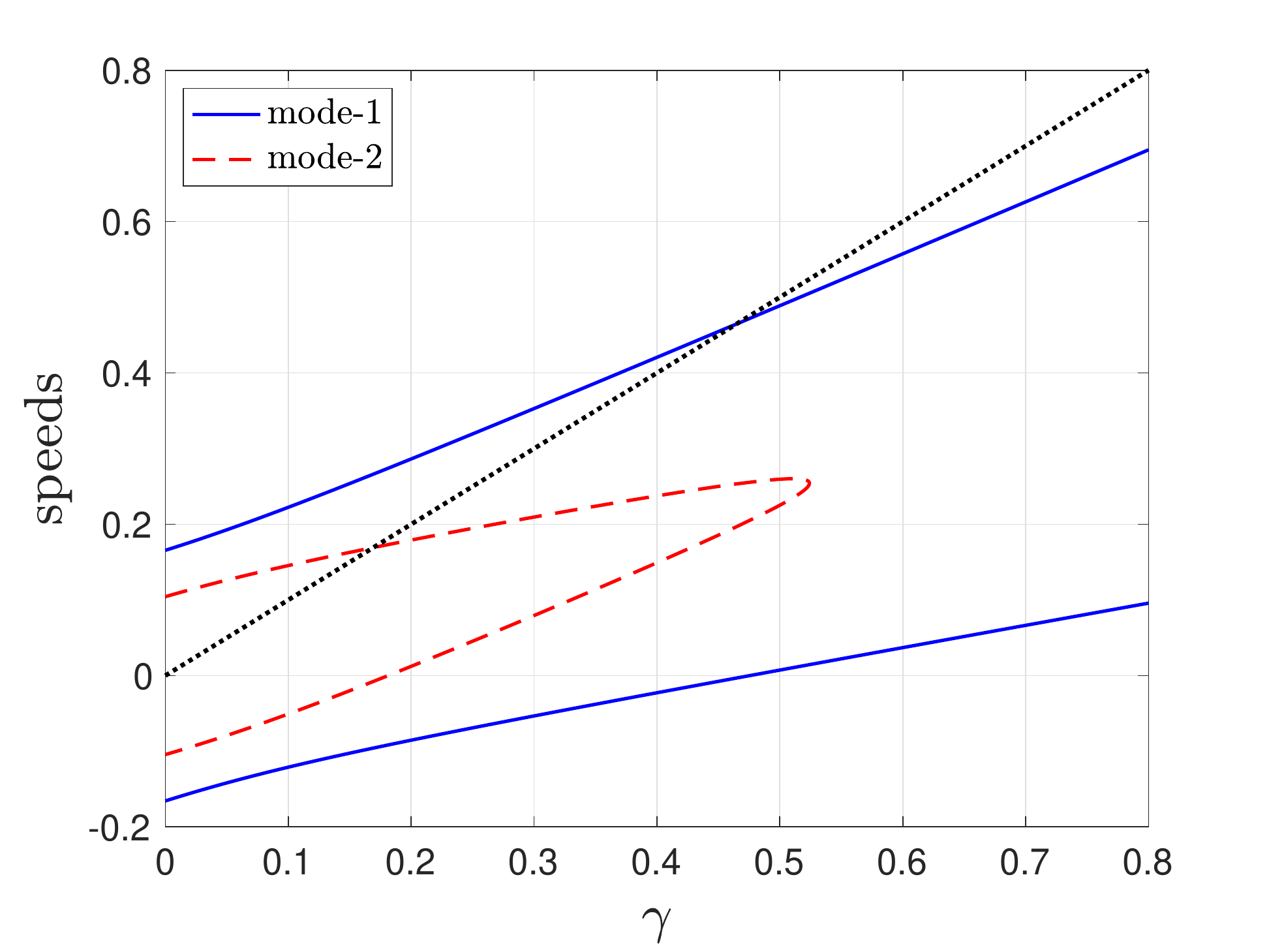}
\caption{Wave speeds as functions of $\gamma$ for mode-1 (blue solid lines) and mode-2 (red dashed line) ring waves along the flow direction when $\rho_1=1$, $\rho_2=1.1$, $\rho_3=1.2$ and $d_{1} = 0.3$, $d_{2} = 0.7$. The black dotted line shows the speed of the background current at the top surface, i.e. $u_0(1)$. %The speeds match the vorticity strength $\gamma$ at the dotted line. 
The intersection between this straight line and the curves for mode-1 and mode-2 speeds correspond to the emergence of {\it critical surfaces} in the fluid, as defined in section~\ref{sec:wave_fronts}.}
\label{fig:speeds}
\end{figure}

In the elliptic regime, the sought solutions of (\ref{dr}) are required to satisfy the conditions $k'(0)=k'(\pi)=0$ due to the symmetry of the problem. At $\theta=0,\pi$, equation (\ref{dr}) reduces to an algebraic equation for $k$:
\begin{eqnarray}
&&\hspace{-2cm}\Big\{ (\gamma d_1-c)^2 \left[ \rho_3(d_2-d_1) + \rho_2 d_1\right] - (\rho_3-\rho_2)d_1 (d_2-d_1) \left[ \gamma (\gamma d_1 -c)+1 \right] \Big\}  \nonumber \\
&&\hspace{-2cm}\times \Big\{ (\gamma d_2-c)^2 \left[ \rho_1 (d_2-d_1) + \rho_2 (1-d_2)\right]+(\rho_1-\rho_2)(d_2-d_1)(1-d_2)\left[ \gamma (\gamma d_2 -c)+1 \right] \Big\} \nonumber \\
&&\hspace{-2cm}=\rho_2^2 d_1 (1-d_2) (\gamma d_1-c)^2 (\gamma d_2 -c)^2. \label{lin_lw_speeds}
\end{eqnarray}
%\begin{eqnarray}
%\hspace{-3cm} (\rho_2-\rho_1)(\rho_3-\rho_2)d_1 (1-d_2)(d_2-d_1) - \Big\{ \rho_1(\rho_3-\rho_2)d_1 (d_2-d_1) (-c+ \gamma d_2) (-c+ \gamma)\nonumber\\
%\hspace{-3cm}+\rho_2 (\rho_3-\rho_1)d_1 (1-d_2) (-c+\gamma d_1)(-c+ \gamma d_2)+\rho_3(\rho_2-\rho_1) (1-d_2)(d_2-d_1)(-c) (-c+ \gamma d_1) \Big\}\nonumber\\
%\hspace{-2cm}+\rho_1 \rho_2 d_1 (-c + \gamma d_1) (-c+ \gamma d_2)^2 (-c+ \gamma) + \rho_2 \rho_3 (1-d_2) (-c) (-c+ \gamma d_1)^2 (-c+ \gamma d_2)\nonumber\\
%\hspace{-1cm}+ \rho_1 \rho_3 (d_2-d_1) (-c) (-c+ \gamma d_1) (-c + \gamma d_2)(-c+ \gamma)=0,\nonumber
%\end{eqnarray}
with $c=s/k(0)$ or $c=-s/k(\pi)$, according to $\theta=0$ or $\pi$.  In Section 5 we will show that this equation for $c$ is precisely the one found for the linear long-wave speeds of the so-called Taylor's configuration \cite{T}. For fixed densities and thicknesses of the layers, a diagram in the $(c,\gamma)$-plane can be obtained, as in Figure~\ref{fig:speeds}. Here, we can observe four branches of solutions for small values of $\gamma$, which can be interpreted as mode-1 and mode-2 solutions. As the value of $\gamma$ increases, the speeds of the mode-2 wave fronts along the current 
%in the upstream and downstream directions 
coincide at $\gamma=\gamma^-$ and  {mode-2} solutions with compact wavefronts crossing the $x$-axis cease to exist for $\gamma>\gamma^-$.  For the parameter values of Figure~\ref{fig:speeds} this critical value is found to be $\gamma^-\approx 0.524063$, beyond which only mode-1 speeds can be found up to the value $\gamma^+ \approx 0.829932$, at which the mode-2 solutions with the wavefront crossing the $x$-axis reappear (outside the range of values in Figure~\ref{fig:speeds}).

%$$
%s^2= \frac{-b_{2} \pm \sqrt{(b^2_{2}-4a_{2}c_{2})}}{2a_{2}}.
%$$

\begin{figure}
\centering
\includegraphics[width = 0.49\textwidth] {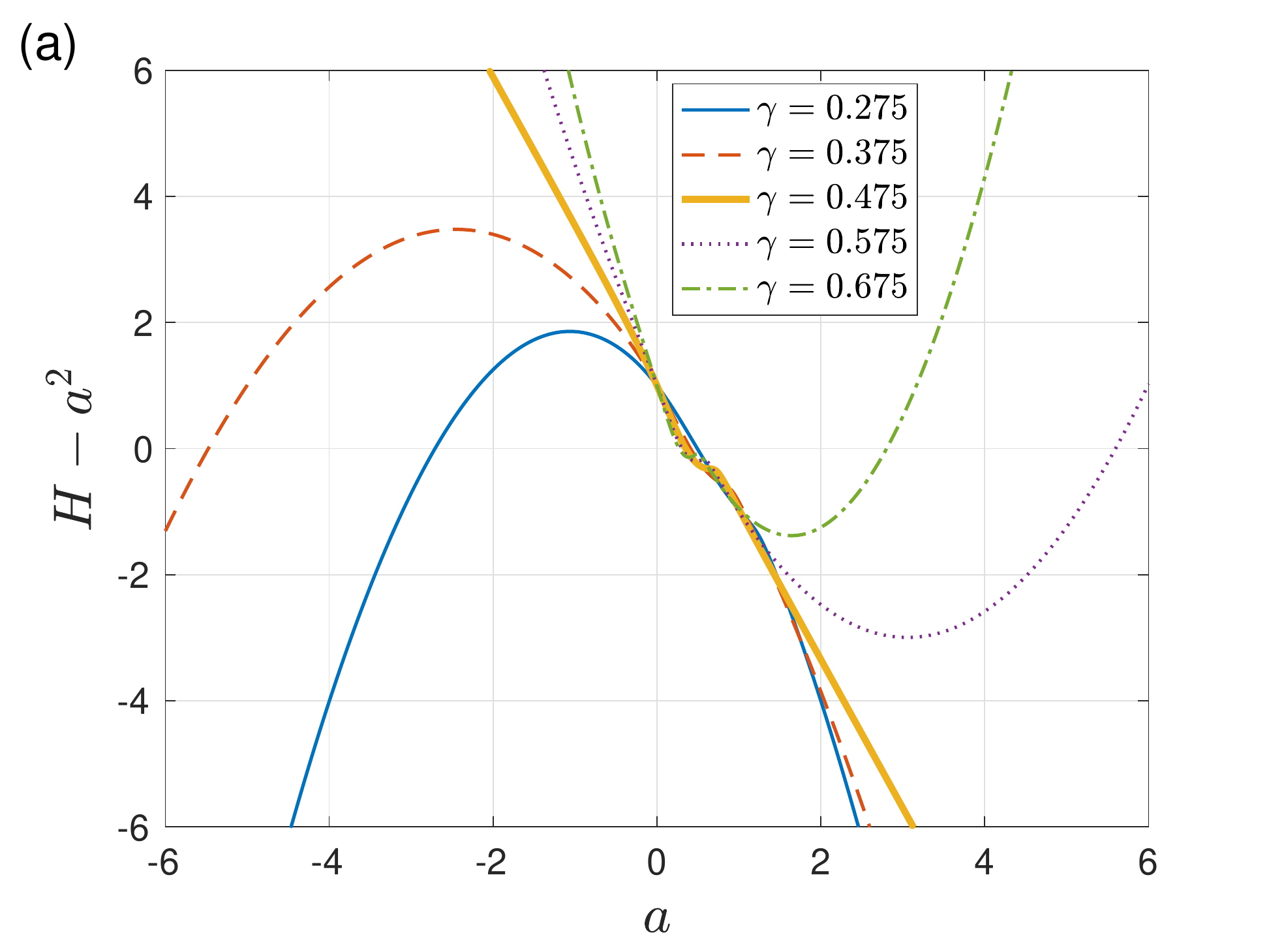}
\includegraphics[width = 0.49\textwidth] {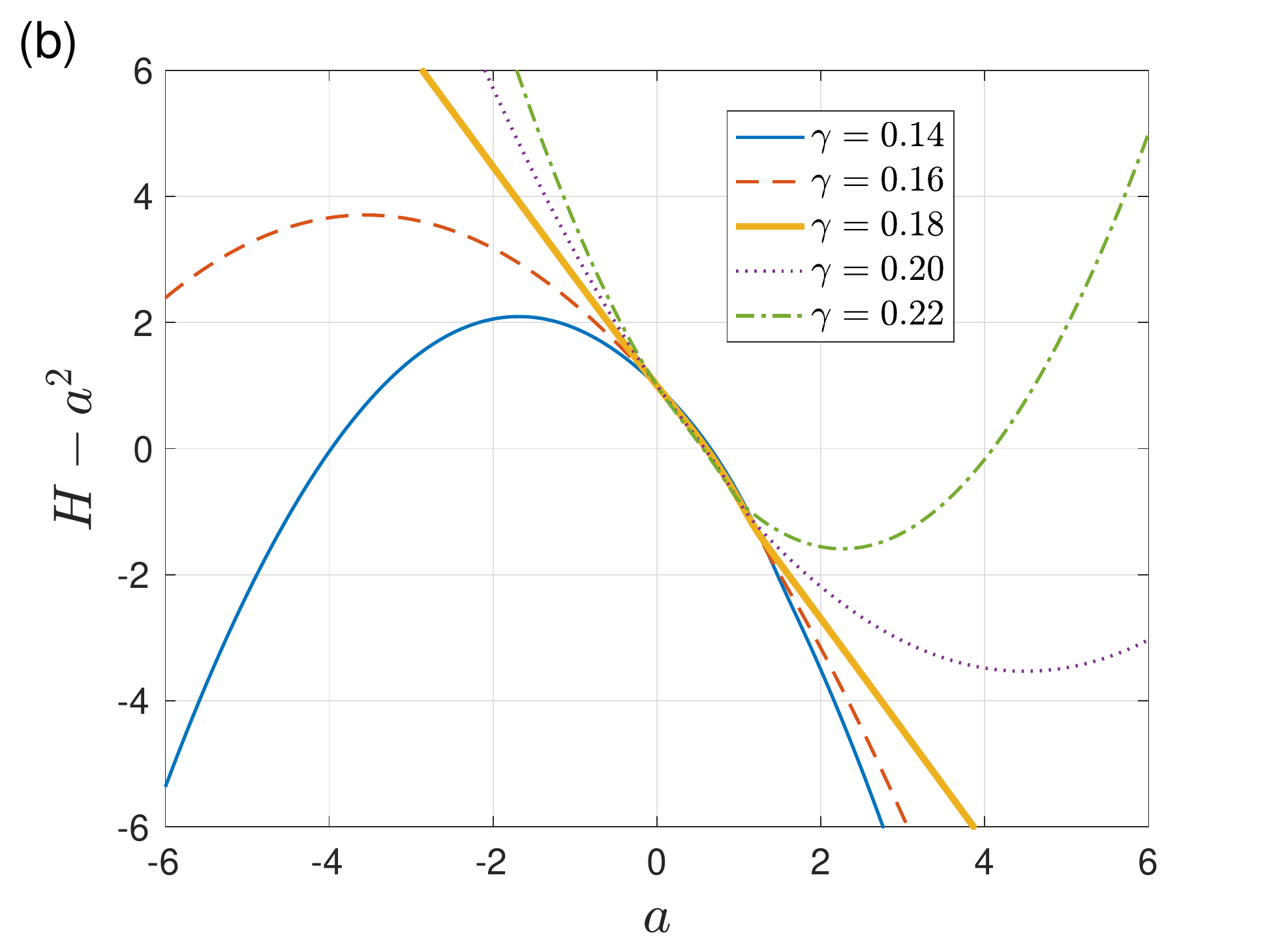}
\caption{Plots of $H-a^2$ for (a) mode-1 and (b) mode-2 as functions of $a$ for various values of $\gamma$, as is indicated in the legends, showing the transitions from the elliptic to the hyperbolic regime for each of the modes when $\rho_1=1$, $\rho_2=1.1$, $\rho_3=1.2$ and $d_{1} = 0.3$, $d_{2} = 0.7$.}
\label{fig:Q_a2}
\end{figure}

%\noindent 
The {\it general solution} of equation (\ref{dr}) can be found in the form $k(\theta) = a \cos\theta+b(a)\sin\theta$, similar to  \cite{J} (see \cite{HKG} for the physical interpretation of the general solution as the solution defining plane waves tangent to the ring wave, which explains why the general solution has this form), allowing us then to find the singular solution (envelope of the general solution) defining the ring waves in the three-layer fluid over a linear shear current  in the form
 \begin{equation}
\left\{
\begin{array}{l}
k(\theta)= a \cos\theta+b(a)\sin\theta,\\
b'(a)=-1/\tan\theta,\\
\displaystyle a^2+ b^2(a)= \frac{-B(a) \pm \sqrt{\Delta(a)}}{2A},
\end{array}
 \right.
 \label{ss}
 \end{equation}
 where $\Delta (a)  = B^2-4AC$ and the ``$+/-$" sign is chosen for mode-2/mode-1, respectively. 
  %CHECK: when is $\Delta \ge 0$? 
In what follows, the singular solution $k = k(\theta)$ is given in the parametric form: $k = k(a), \theta = \theta(a)$, where $a$ is a parameter.
%
%%\noindent 
%Here, we consider an {\it elliptic regime} \cite{K}, where a part of the wavefront is able to propagate upstream. Then, $k(\theta)$ is defined for $\theta \in [-\pi, \pi]$. In the absence of any current, $\hat a^2 + \hat b^2(\hat a) = 1$, and the wave is concentric. By continuity, the locus of parameters $\hat a$ and $\hat b$ will be close to a circle for a sufficiently weak current, which motivates the terminology.
We require $k(\theta)$ to be positive everywhere in order to describe the outward propagating ring wave.   Let us denote
$a^2+ b^2 = ({-B(a) \pm \sqrt{\Delta(a)}})/{2A}\equiv H$, where $\ b = b(a).$
Then the condition $b^2 =H- a^2\ge0$ determines  the domain of $a$. The behaviour of $H- a^2$ for {mode-1} (panel (a)) and for \mbox{mode-2} (panel (b)) is given in Figure~\ref{fig:Q_a2} for a particular set of parameters for which the elliptic regime (smaller values of $\gamma$) yields $a\in[a_{\mathrm{min}}, a_{\mathrm{max}}]$, while the hyperbolic regime (larger values of $\gamma$) yields $a \in (-\infty, a_{\mathrm{min}}] \cup [a_{\mathrm{max}}, +\infty)$ for some $a_{\mathrm{min}}$ and $a_{\mathrm{max}}$, which depend on $\gamma$. For these parameters, the transition between the two regimes occurs for mode-1 at $\gamma=\gamma_{p1}\approx0.4754$ and for mode-2 at $\gamma=\gamma_{p2}\approx0.1812$. %$\gamma=0.181192$. 
At the transition, the curve $H-a^2$ intersects the horizontal axis only once at $a=a^*$, and we have the parabolic regime with $a \in (-\infty, a^*]$.  
%  where we choose the interval $[\hat a_{\mathrm{min}},\hat a_{\mathrm{max}}]$ containing $\hat a=0$, since $\hat a$ should take both positive and negative %values (in particular, $k(\theta)$ should be positive both at $\theta = 0$ and $\theta = \pi$).
We observe that the values of $\gamma_{p1}$, $\gamma_{p2}$ can be found in explicit form by setting $c=0$ in equation (\ref{lin_lw_speeds}), which yields a biquadratic equation for $\gamma$.

Differentiating $b^2 = H- a^2$ with respect to $a$ and using (\ref{ss}), we obtain 
\begin{equation}\label{eq_theta}
\tan\theta=-({2 b})/({H_{a}-2 a}).
\end{equation}
Therefore, $k(a) = k(\theta (a))$ can be written in the form
%\begin{equation*}
%\hspace{-2cm}k(a)=a\cos\theta+ b\sin\theta=\cos\theta(a+b\tan\theta)=\mbox{sign}\left (-2 b \frac{\cos\theta}{\tan\theta}\right )\frac{aH_{a}-2H}{\sqrt{(H_{a}-2 a)^2+4 b^2}}.
%\end{equation*}
\begin{equation*}
\hspace{-1cm}k(a)=a\cos\theta+ b\sin\theta=\cos\theta(a+b\tan\theta)=\cos\theta \left( \frac{aH_a-2H}{H_a-2a}\right).
\end{equation*}
Notice that in (\ref{eq_theta}) both positive and negative signs can be assigned for $b$. Depending on the range of values of $a$ prescribed for each of the regimes mentioned above, appropriate signs for $b$ must be taken and appropriate integer multiples of $\pi$ must be added to the principal values of the $\arctan$ when computing $\theta$ from (\ref{eq_theta}) in order to guarantee that $\theta(a)$ is a continuous function. By symmetry about the $x$-axis, we are required only to construct the solution $k(\theta)$ within $[0,\pi]$ (which corresponds to the upper half of the wave front). Details can be found in Appendix B.

%for $\theta\in(\pi,2\pi)$.
%Thus, we obtained the required singular solution analytically,  in the parametric form (\ref{k})-(\ref{t2}). 
%We fix $\rho_1=1, \rho_2=1.1, \rho_3 = 1.2$ and consider two cases: $d_1=0.4, d_2 = 0.6$ (thin middle layer) and $d_1 = 0.3, d_2 = 0.7$ (thick middle layer) as well as several values of the strength of the current.

\begin{figure}%[!p!t]
\centering
\includegraphics[width=0.4\textwidth]{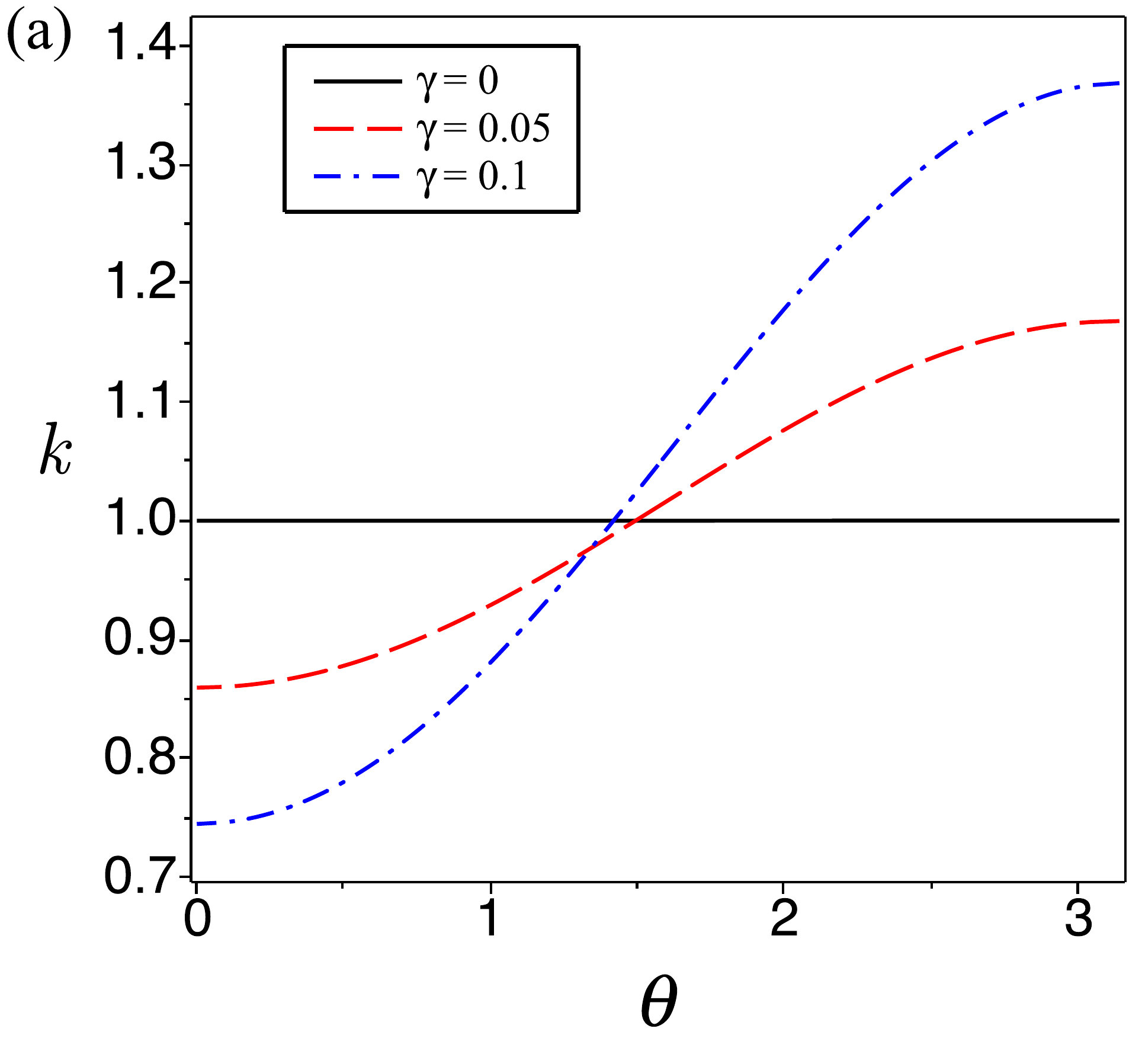}
%\caption{Left figure}
%\label{fig:left}
\hspace{0.5cm}
%\hfill
\includegraphics[width=0.4\textwidth]{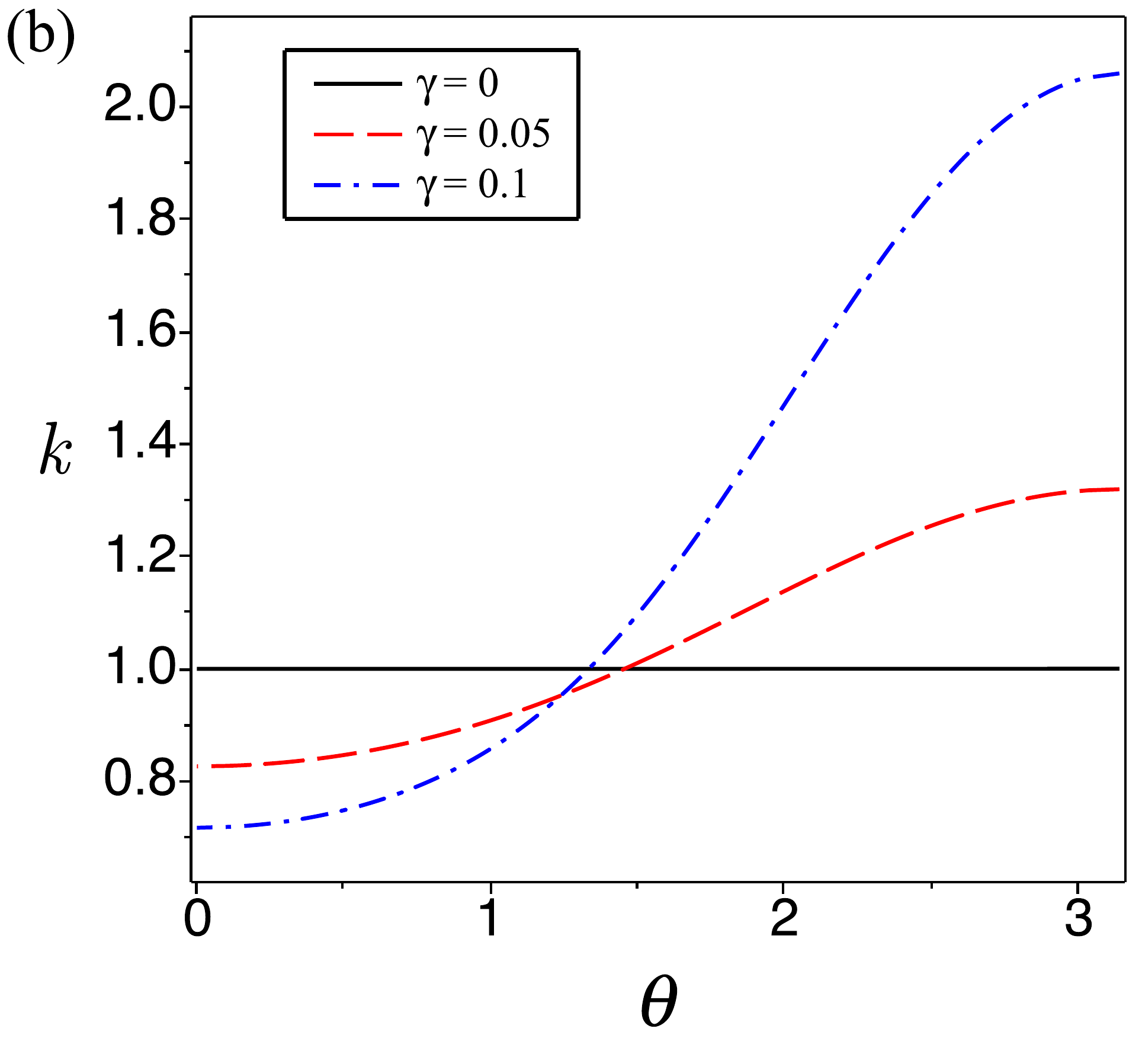}
%\caption{Right figure}
%\label{fig:right}
\caption{%\footnotesize 
The function $k(\theta)$ of (a) the mode-1 internal ring waves and (b) the mode-2 internal ring waves for 
 $\rho_1=1, \rho_2=1.1, \rho_3 = 1.2$ and $d_{1} = 0.3, d_{2} = 0.7$. Different (small) values of $\gamma$ are considered here: $\gamma=0, 0.05, 0.1$.}
%in terms of piecewise constant and piecewise linear model, respectively.}
\label{fig:combined_6_9}
\end{figure}

%%%%%%%%%%%%%%%%%%%%%%%%%%%%%%%%%%%%%%%%%%%
%%%%%%%%%%%%%%%%%%%%%%%%%%%%%%%%%%%%%%%%%%%
%%%%%%%%%%%%%%%%%%%%%%%%%%%%%%%%%%%%%%%%%%%

%\newpage
\section{Wavefronts and vertical structure}\label{sec:wave_fronts}

In this section, we use the constructed singular solutions for the function $k (\theta)$ in order to visualise the shape of the wavefronts of interfacial ring waves in a three-layer fluid with a linear current. We recall that the wavefronts are described to leading order by $k(\theta) r- s t =\mathrm{constant}$. 

\subsection{Elliptic regime}

\begin{figure}
\centering
\includegraphics[height=5.5cm]{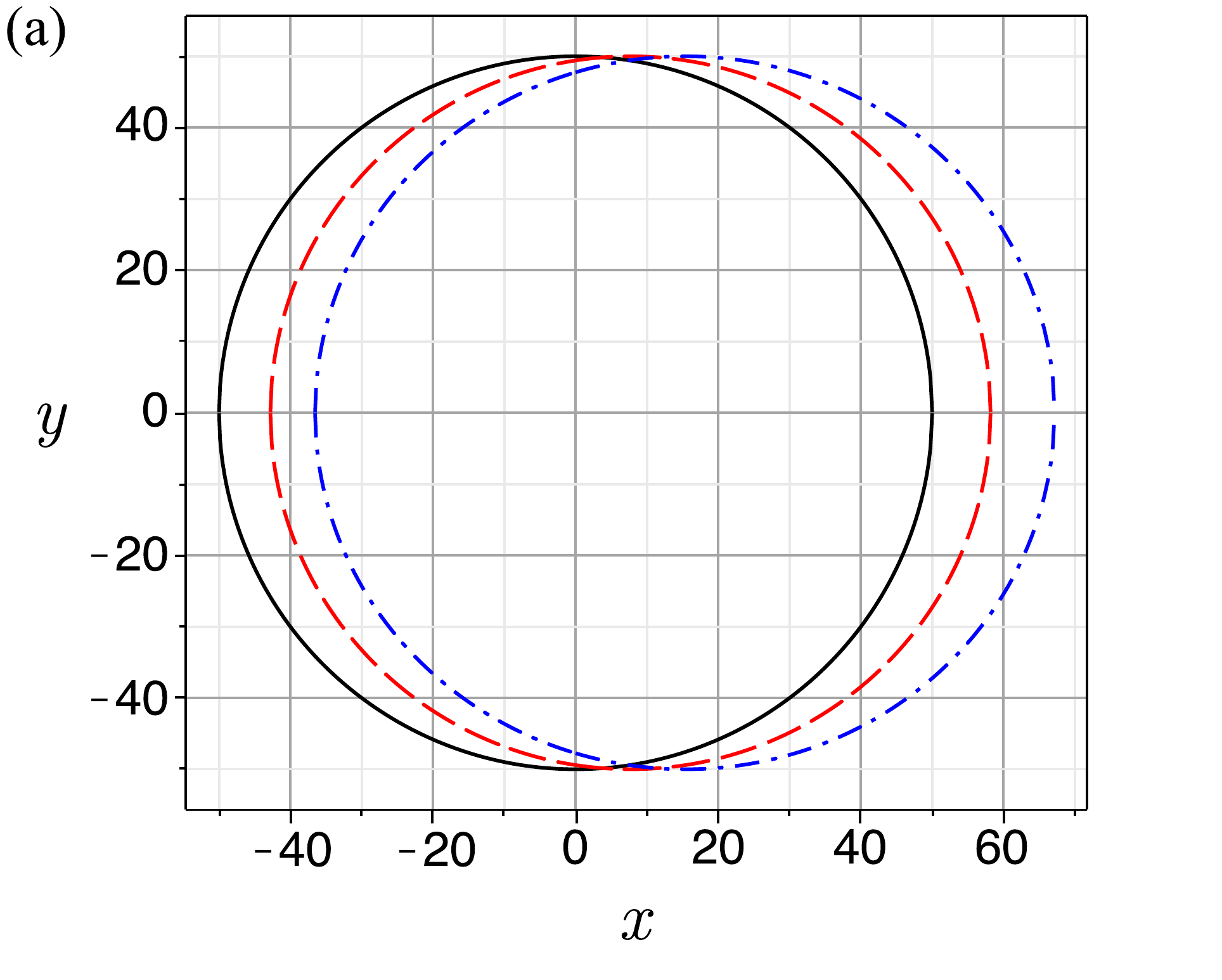}
\qquad
\includegraphics[height=5.5cm]{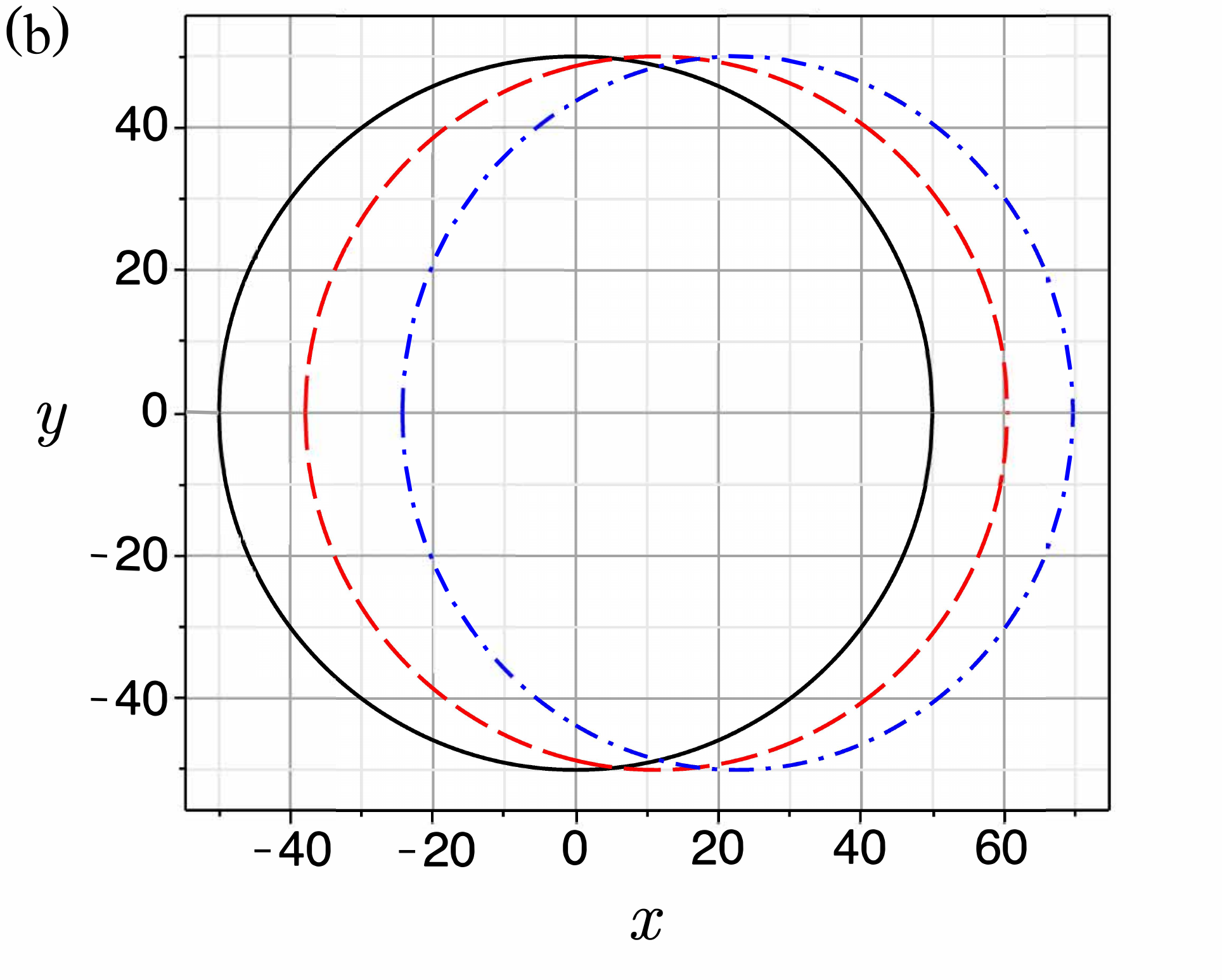}
\caption{%\footnotesize 
Wavefronts of (a) mode-1 and (b) mode-2 ring waves for $\rho_1=1, \rho_2=1.1, \rho_3 = 1.2$ and  $d_{1} = 0.3$, $d_{2} = 0.7$ described by $k(\theta)r = 50$ when 
%\eqref{depth1} and \eqref{depth2} with difference of flow values: 
$\gamma =0$ (solid black), $\gamma =0.05$ (dashed red) and $\gamma =0.1$ (dash-dotted blue), respectively.}
\label{fig:combined_6_10}
\vspace{0.5cm}
\includegraphics[height=5.5cm]{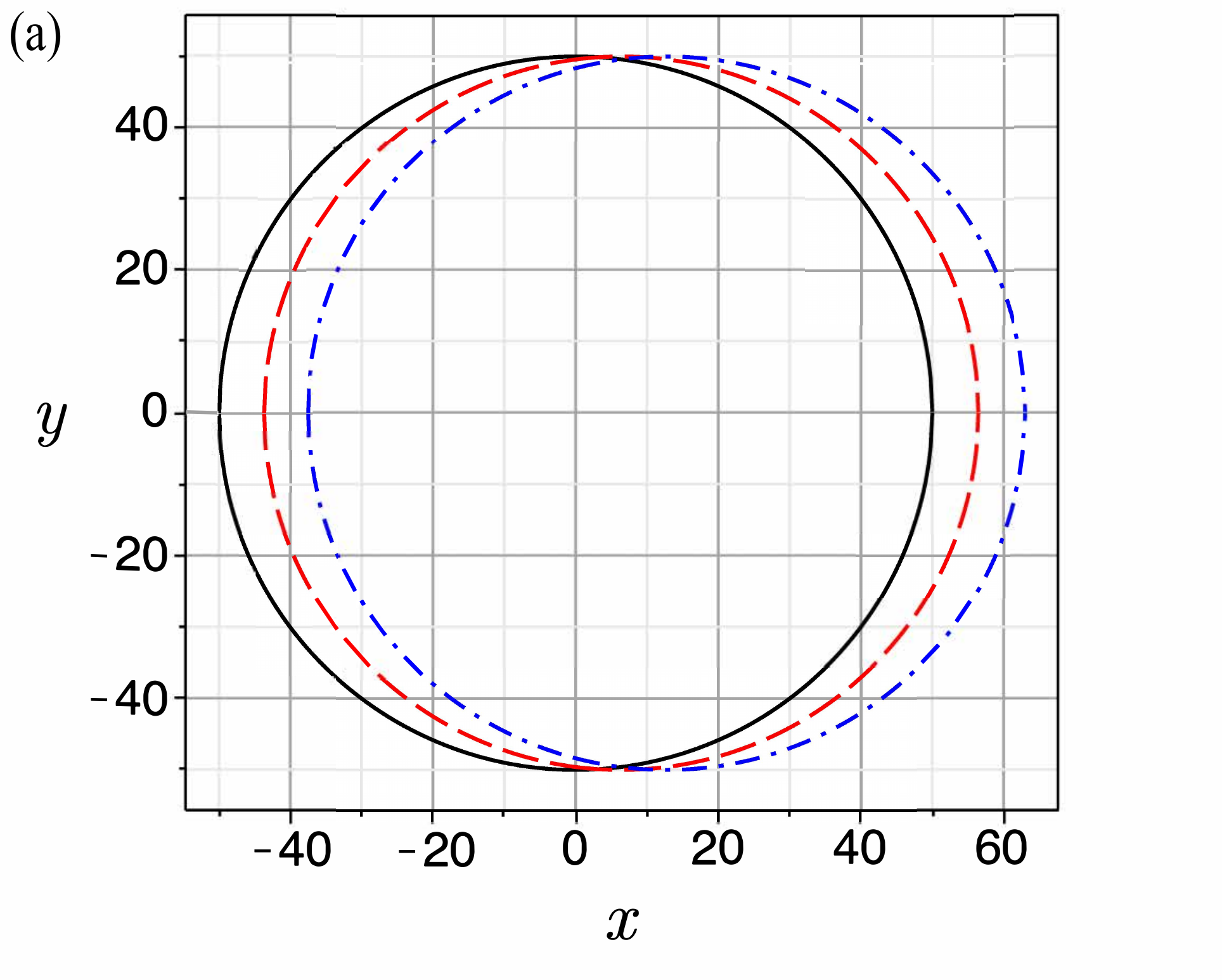}
\qquad
\includegraphics[height=5.5cm]{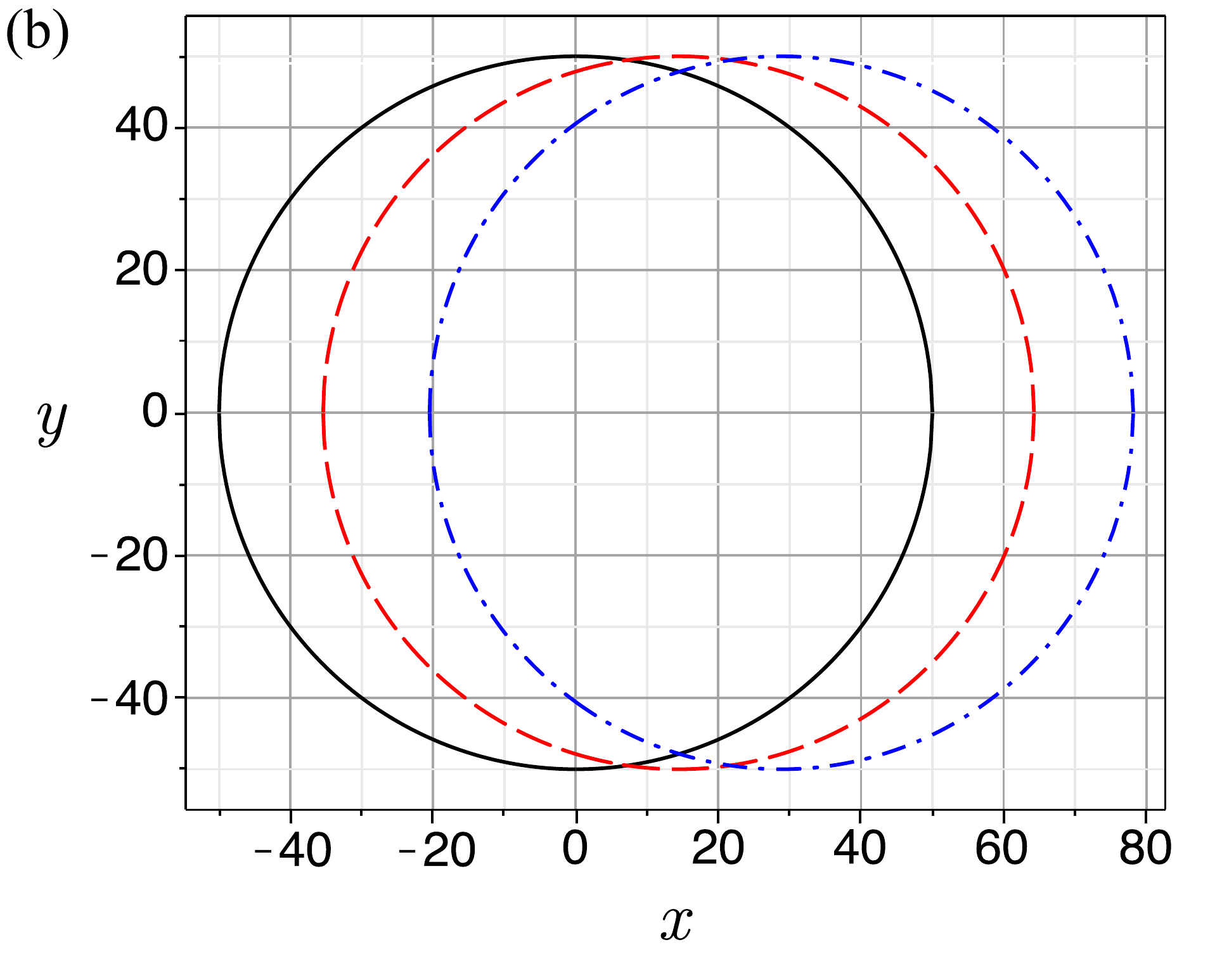}
\caption{%\footnotesize 
Wavefronts of (a) mode-1 and (b) mode-2 ring waves for  $\rho_1=1, \rho_2=1.1, \rho_3 = 1.2$ and $d_{1} = 0.4$, $d_{2} = 0.6$ described by $k(\theta)r = 50$ when 
%\eqref{depth1} and \eqref{depth2} with difference of flow values: 
$\gamma =0$ (solid black), $\gamma =0.05$ (dashed red) and $\gamma =0.1$ (dash-dotted blue), respectively.}
\label{fig:combined_6_11}
\end{figure}

The functions $k(\theta)$ for both internal ring modes are shown in Figure~\ref{fig:combined_6_9} for weak vorticity when $\rho_1=1, \rho_2=1.1, \rho_3 = 1.2$ and $d_1 = 0.3, d_2 = 0.7$. As $k(\theta)$ is even and $2\pi$-periodic, it is sufficient to show the range $0\leqslant \theta \leqslant \pi$. 
%since the symmetric profile is generated for $-\pi\leqslant \theta \leqslant 0$. 
We note that these same values for the fluid densities will be used throughout the text. The corresponding wavefronts of the two interfacial modes are shown in  
Figure~\ref{fig:combined_6_10}. For this set of parameters, the speeds of the first and second modes are $s \approx 0.1656$ and $s \approx 0.1043$, respectively. We also show the effect of having a thinner intermediate layer in Figure \ref{fig:combined_6_11}, where we set $d_1=0.4$, $d_2=0.6$, for which the speeds are $s \approx 0.1909$ and $s \approx 0.0853$.
We see that the shear flow has qualitatively different effect on the first and second interfacial ring modes: mode-1 ring waves are elongated in the direction of the shear flow. On the other hand, mode-2 ring waves are squeezed in that direction. 
%There is a slight difference in the shape of wavefronts between \eqref{depth1} and \eqref{depth2}. 
We also note the deformation of the wavefronts is enhanced in the case of a thicker intermediate layer.
% stronger for $d_1 = 0.3, d_2 = 0.7$ than for $d_1= 0.4, d_2=0.6$, and the wavefront of the second mode appears to be mainly convected in the second case.
%{\color{red} For example, let $\rho_1 = 1, \rho_2 = 1.1$ and $\rho_2 = 1.2$. For $d_1 = 0.3$ and $d_2 = 0.7$, we obtain the following speeds of the first and second modes: $s \approx 0.1656$ and $s \approx 0.1043$; while for $d_1 = 0.4$ and $d_2 = 0.6$ the respective speeds are $s \approx 0.1909$ and $s \approx 0.0853$.}

%%%%%%%%%%%%%%%%%%%%%%%%%%%%%%%%%%%%%%%%%%%%%%%%%%%%%%%%
%\newpage
%\newpage
\begin{figure}%[!p!t]  %hbt! %h
      \centering
      \includegraphics[width=0.3\linewidth]{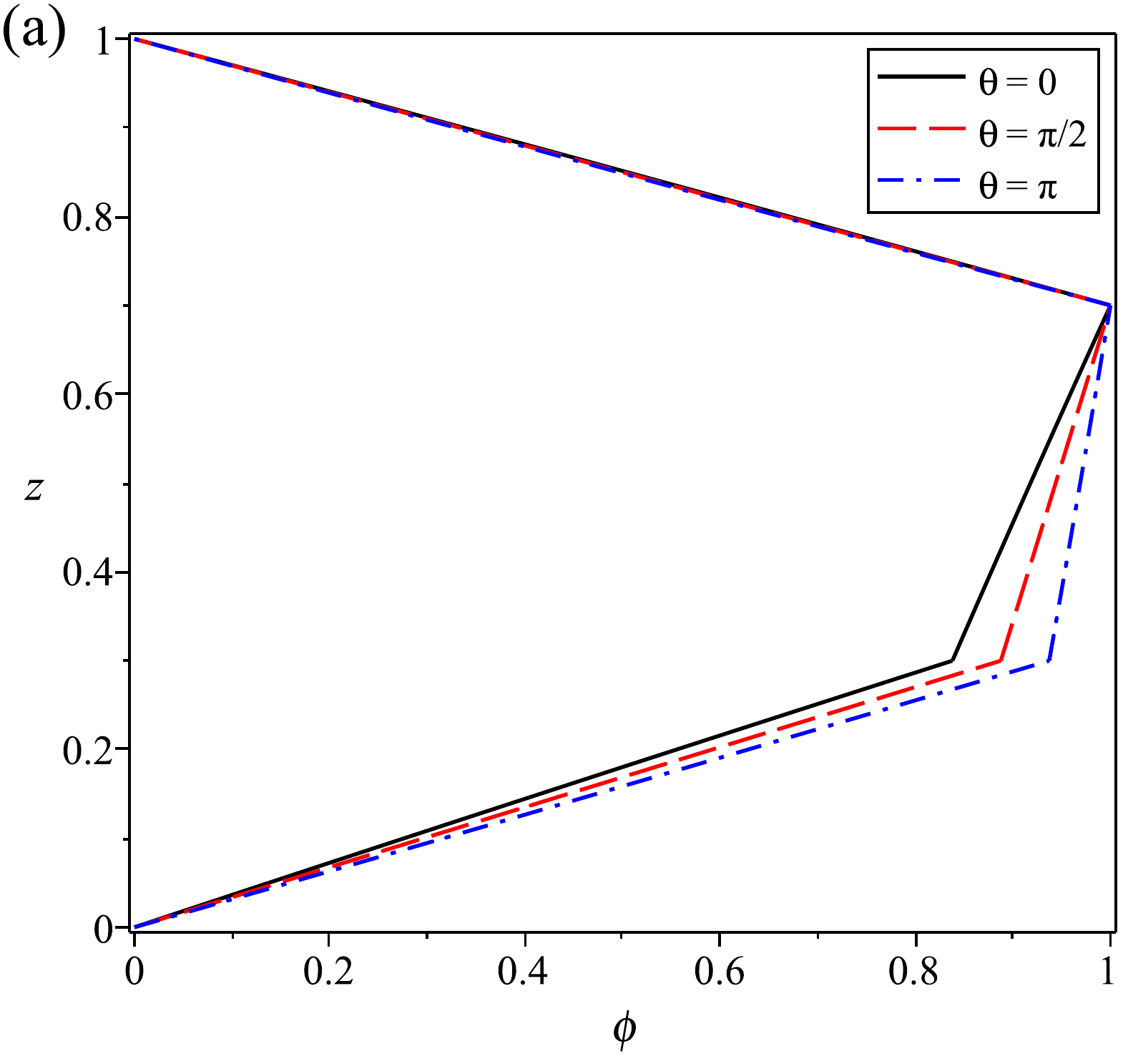} 
      \includegraphics[width=0.3\linewidth]{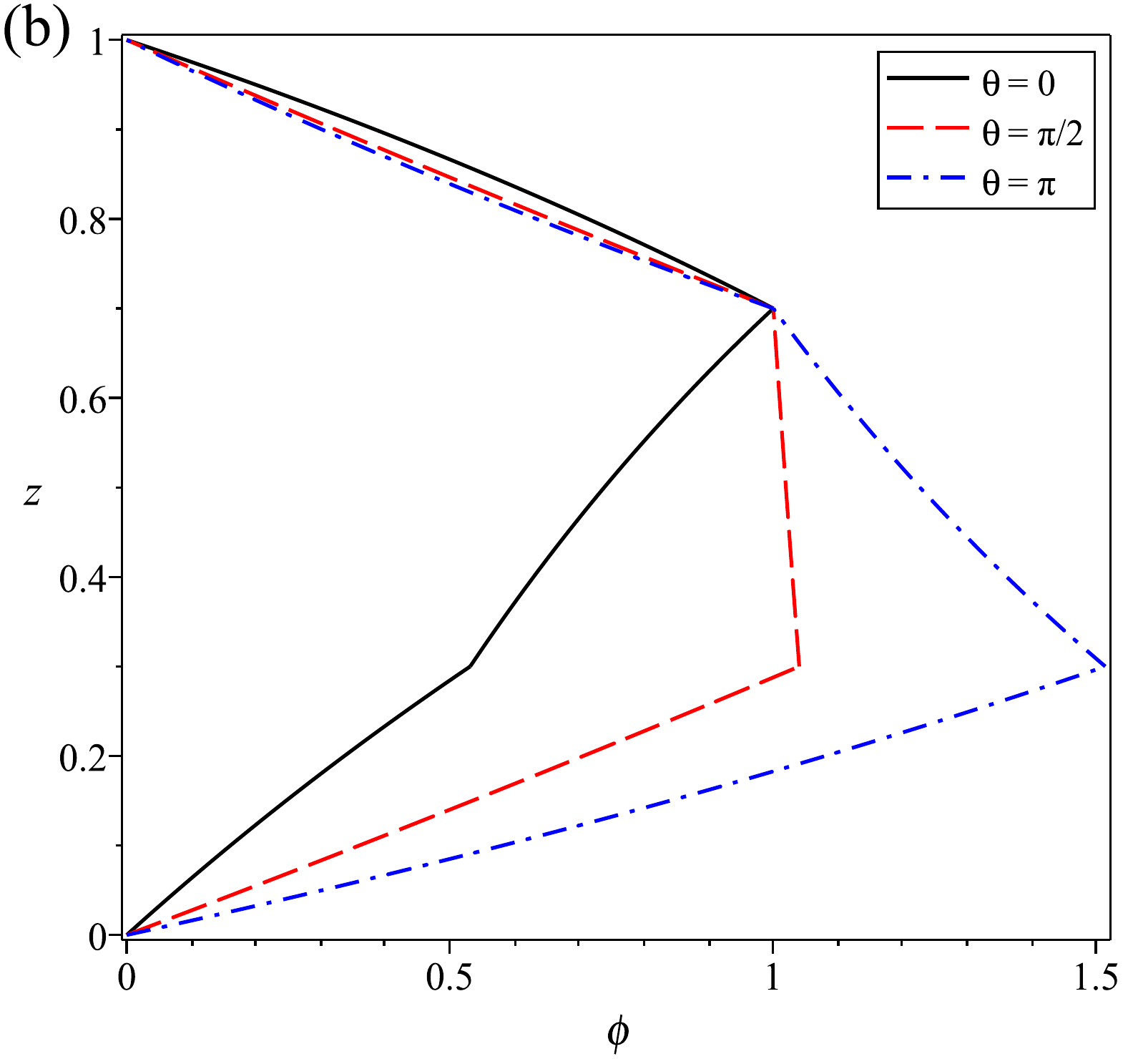} 
      \includegraphics[width=0.3\linewidth]{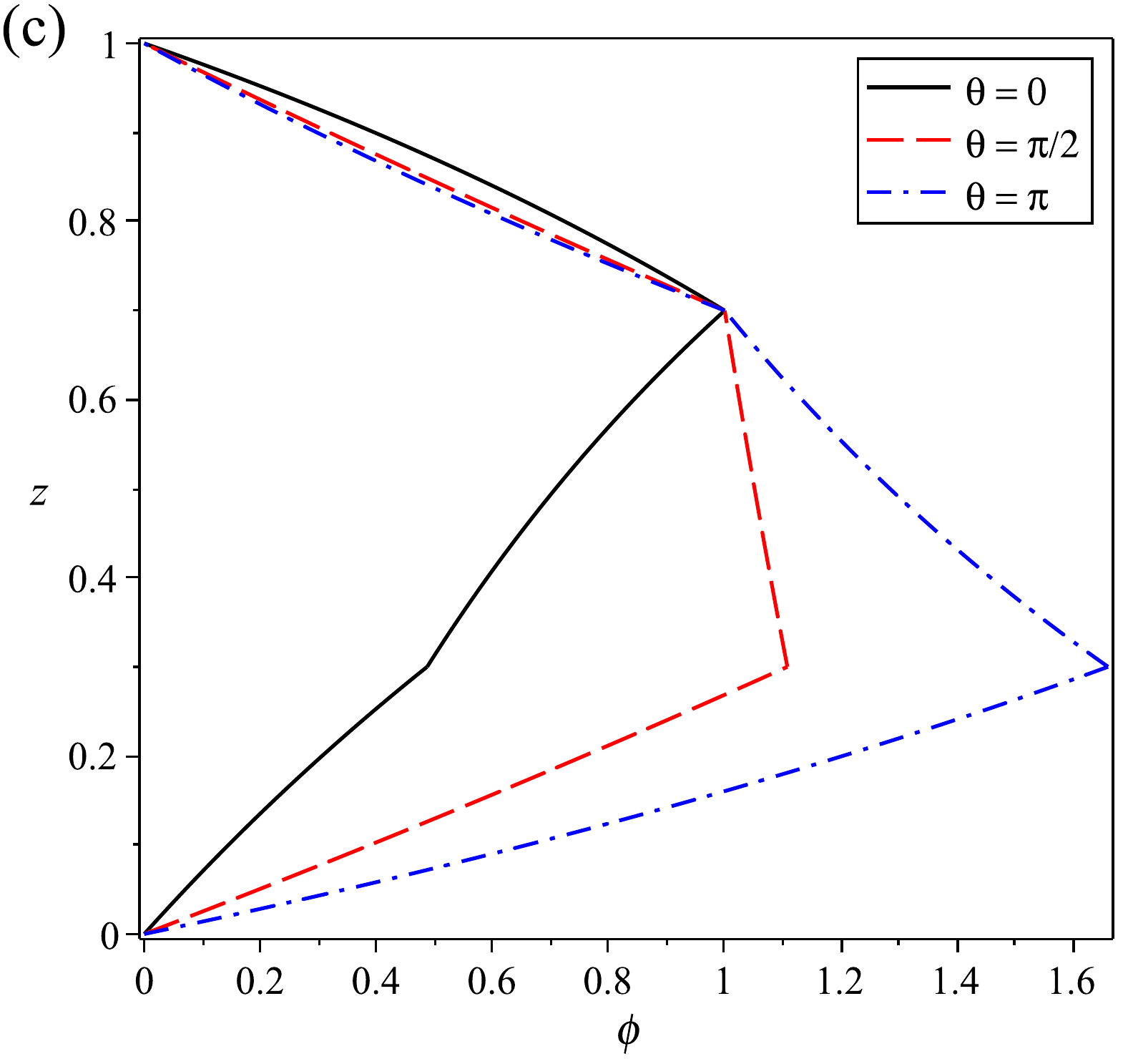} \\
      \includegraphics[width=0.3\linewidth]{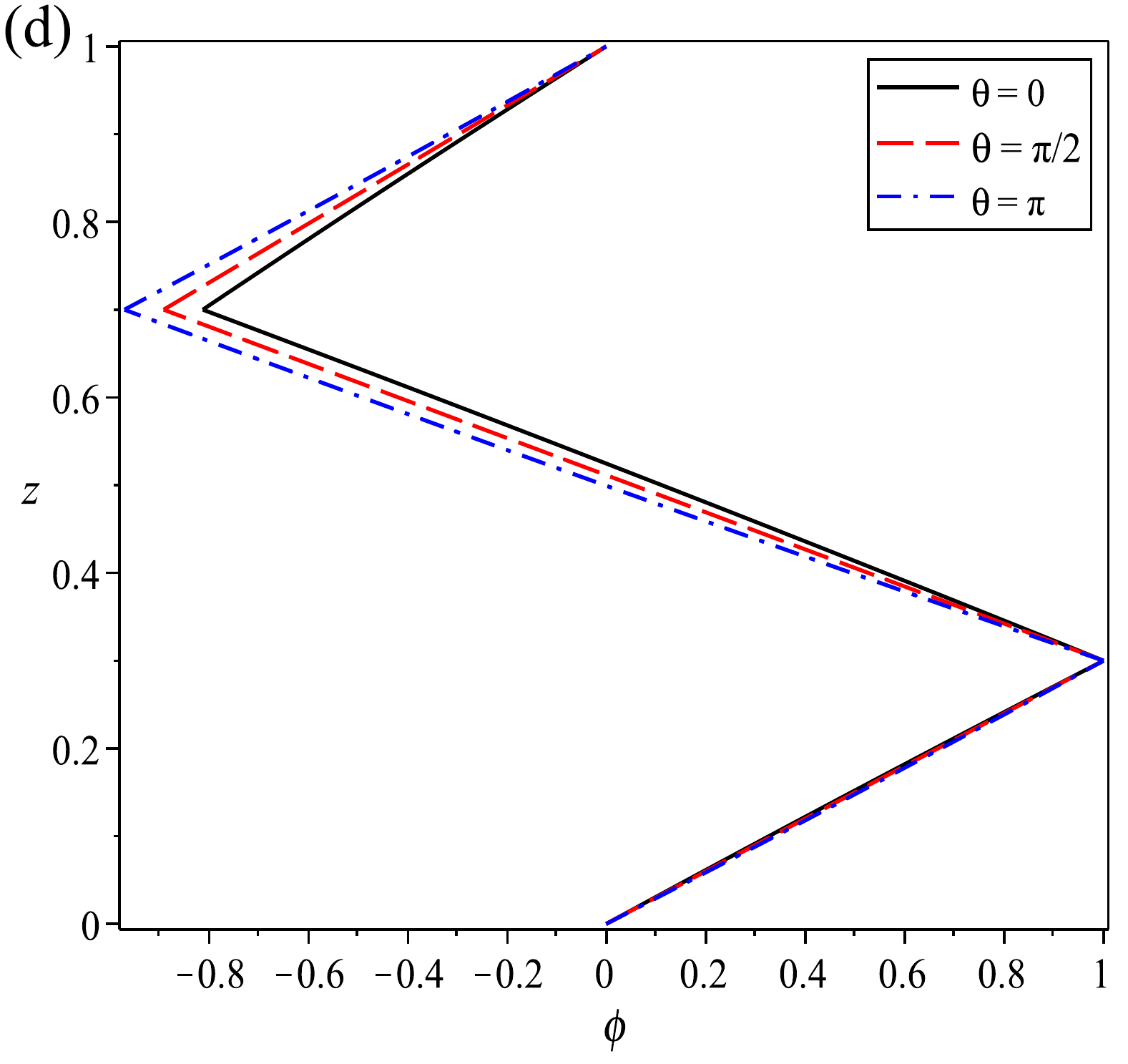} 
      \includegraphics[width=0.3\linewidth]{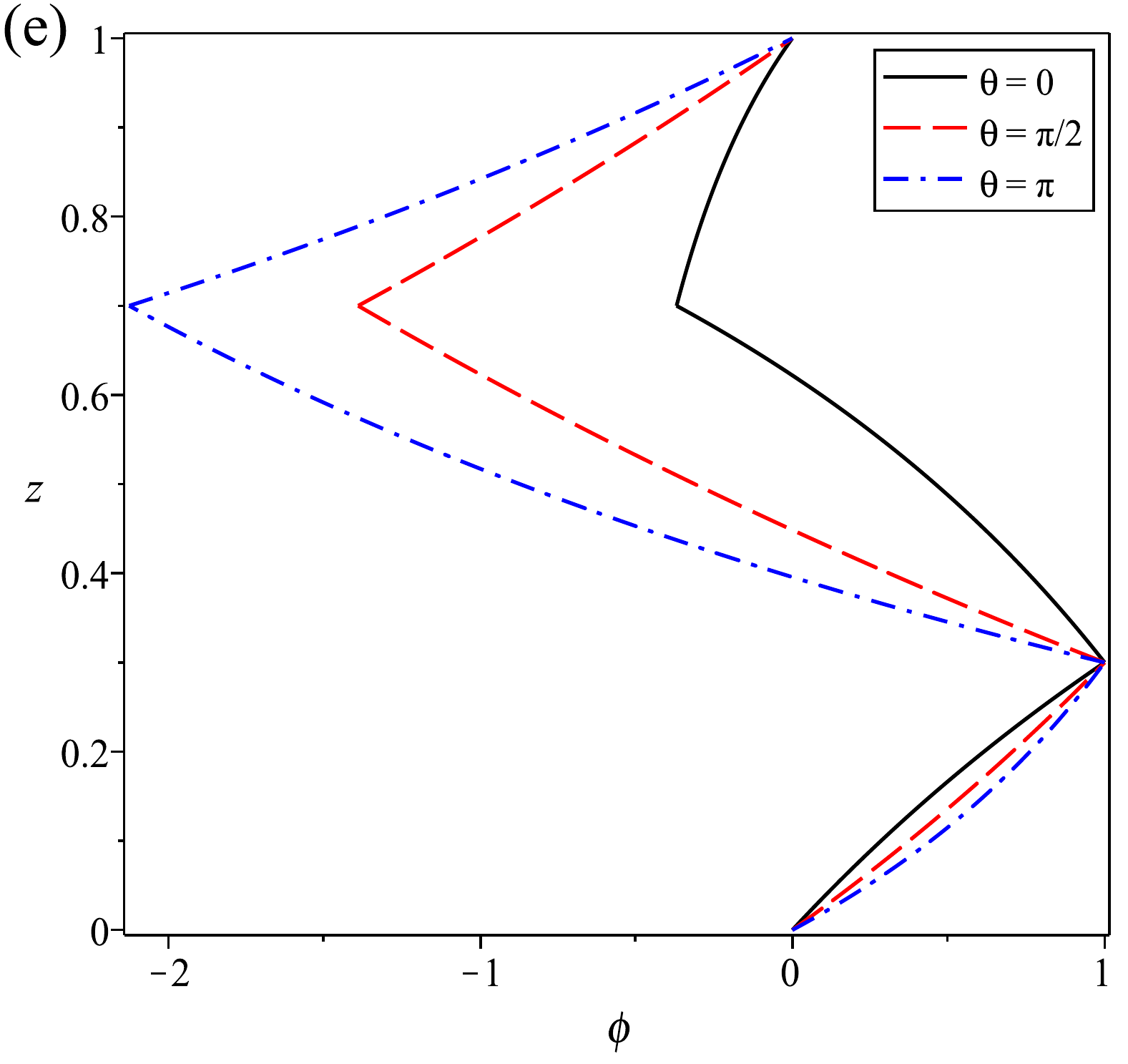} 
      \includegraphics[width=0.3\linewidth]{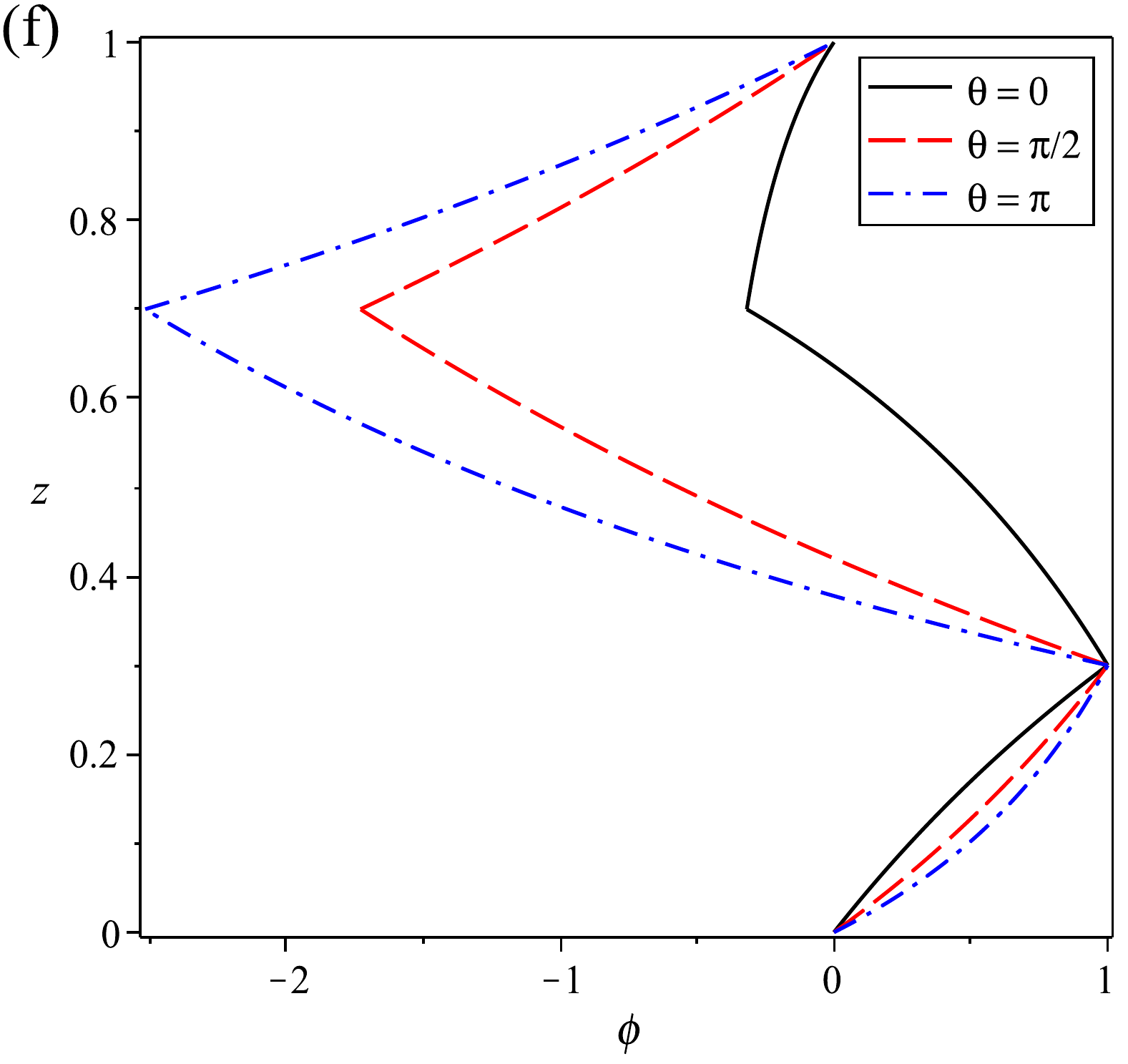} 
%\vspace{-0.5cm}
       \caption{%\footnotesize 
       Modal function $\phi(z)$ for mode-1 (top row) and mode-2 (bottom) when $\rho_1=1, \rho_2=1.1, \rho_3 = 1.2$ and $d_{1}= 0.3$ and $d_{2}= 0.7$. The first, second, and third columns correspond to $\gamma=0.01$, $\gamma=0.1$ and $\gamma=0.12$, respectively. Different angles are considered, $\theta = 0, {\pi}/{2}, \pi$ (see solid black, dashed red and dash-dotted blue lines, respectively).}
       \label{fig:PicS56_6_13}
\vspace{0.5cm}
%\end{figure}
%\begin{figure}%[!p!t] %hbt! !p!t
      \centering
     \includegraphics[width=0.3\linewidth]{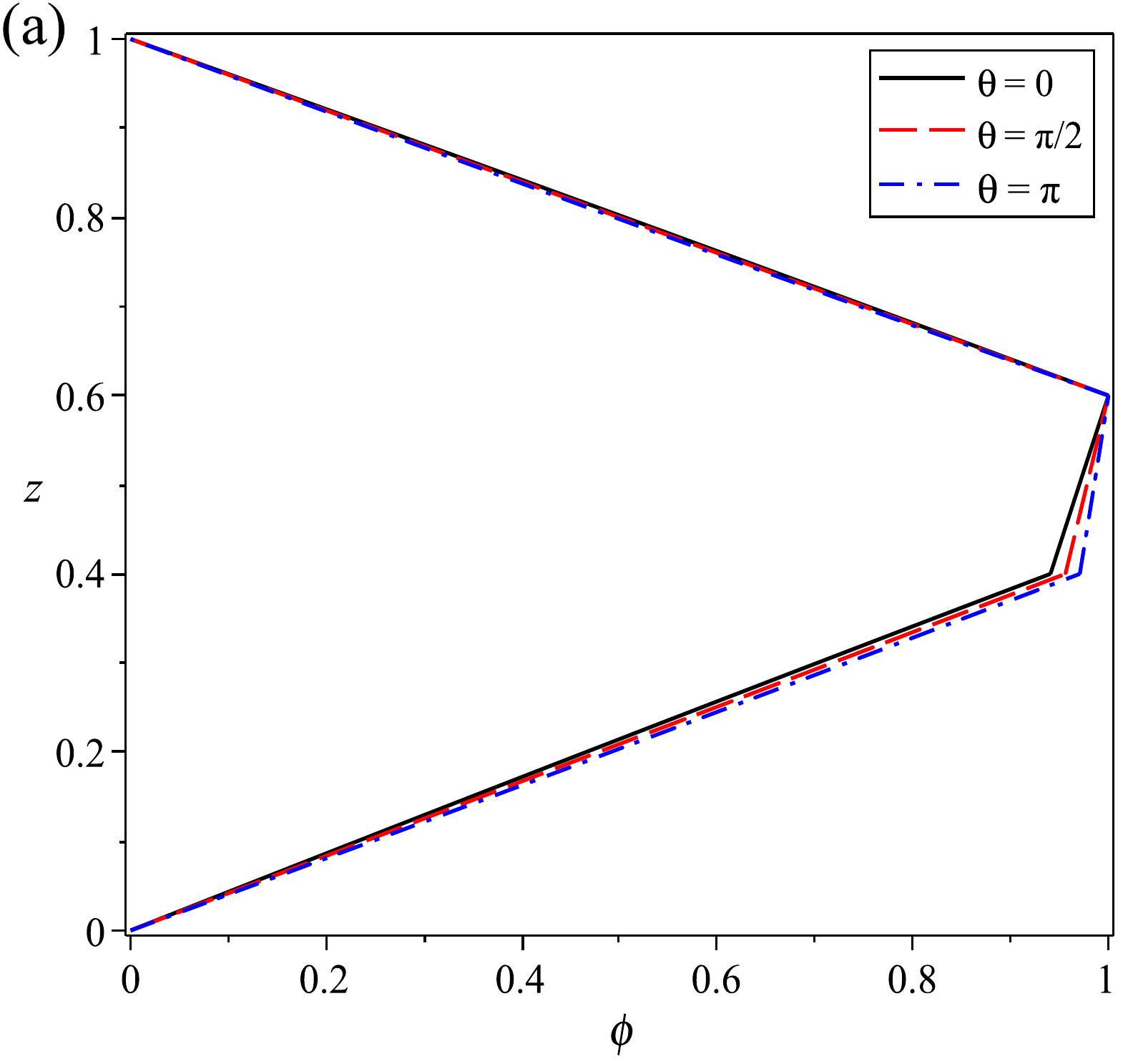} 
     \includegraphics[width=0.3\linewidth]{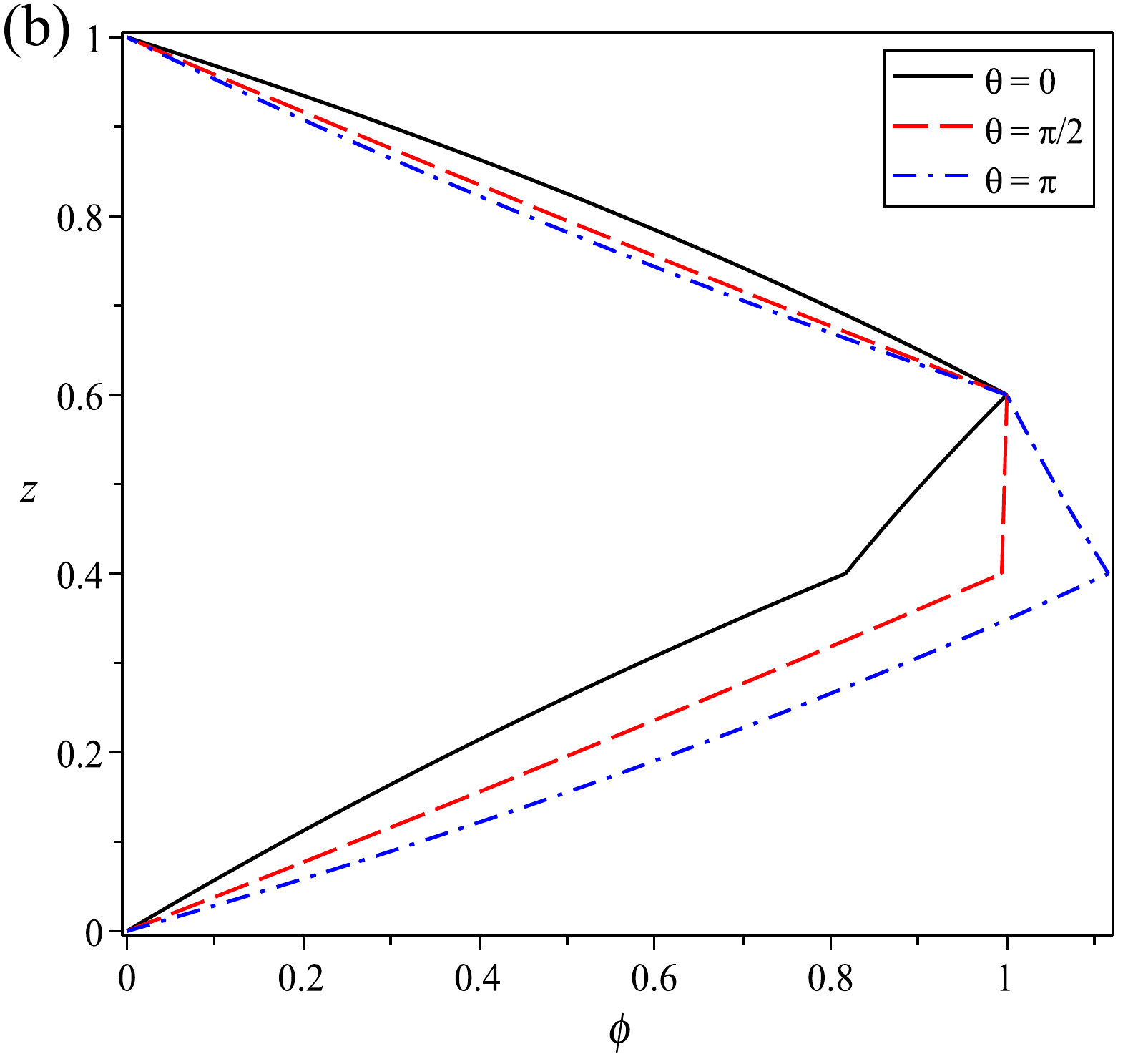}
     \includegraphics[width=0.3\linewidth]{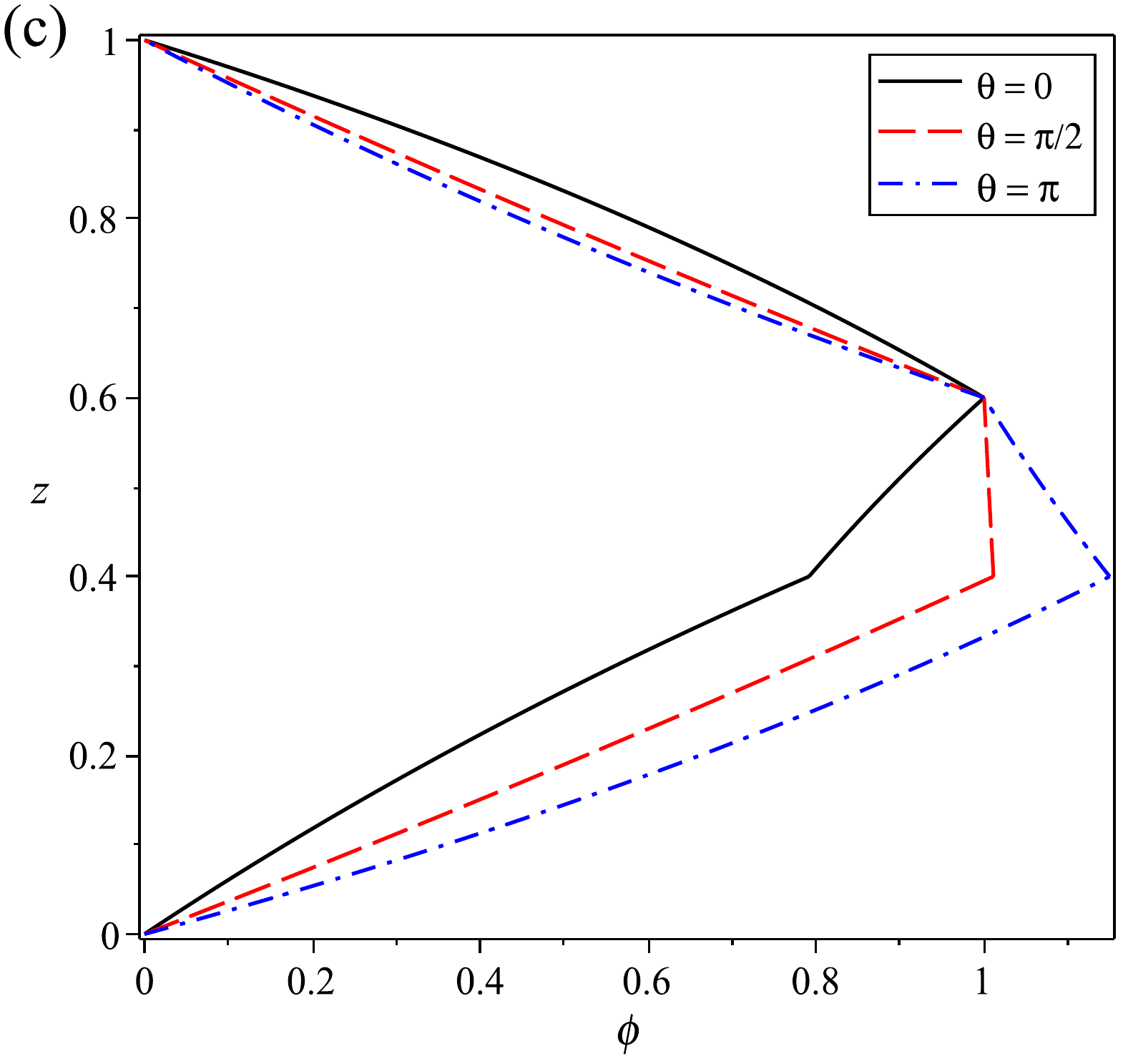} 
     \includegraphics[width=0.3\linewidth]{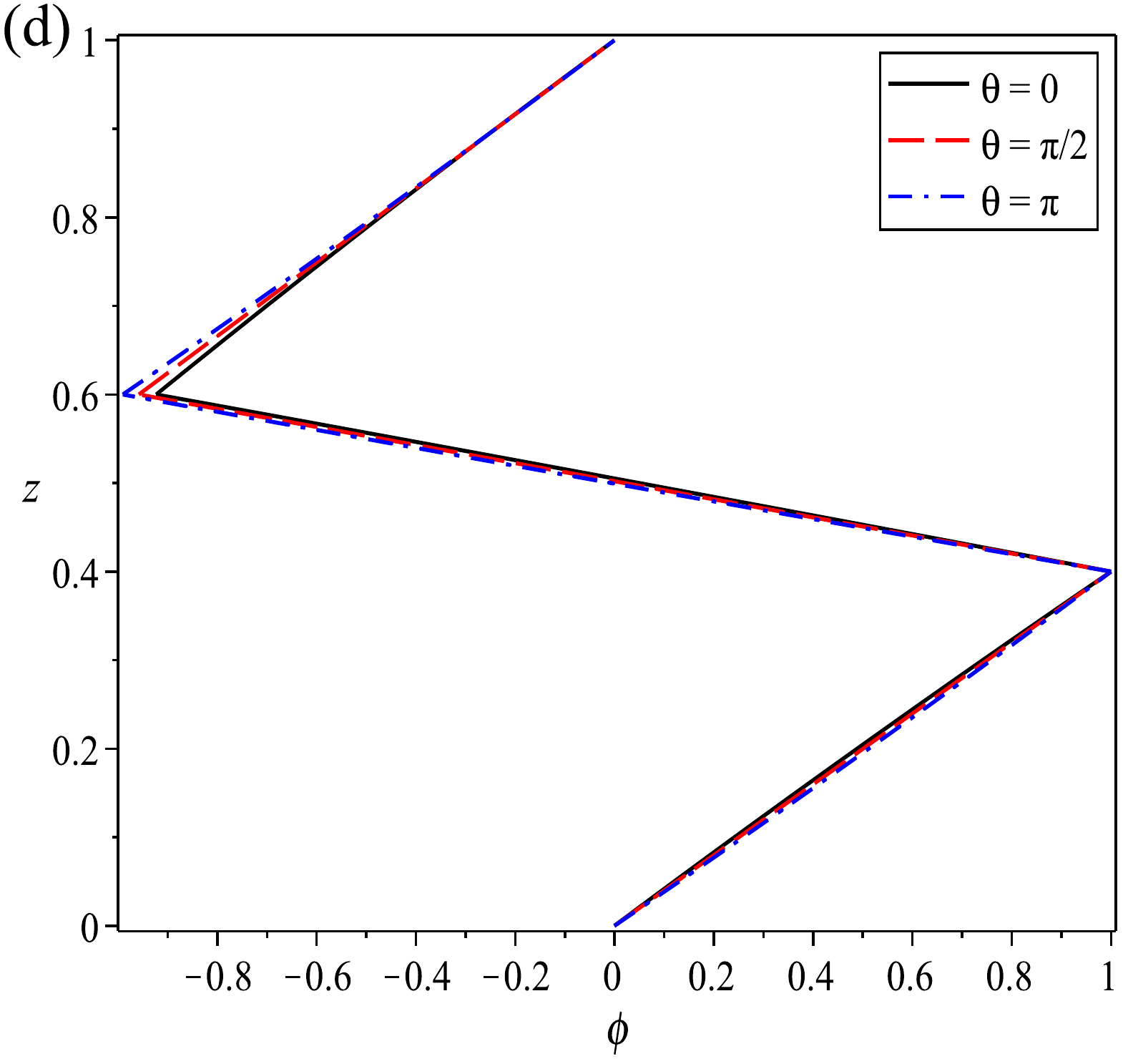} 
     \includegraphics[width=0.3\linewidth]{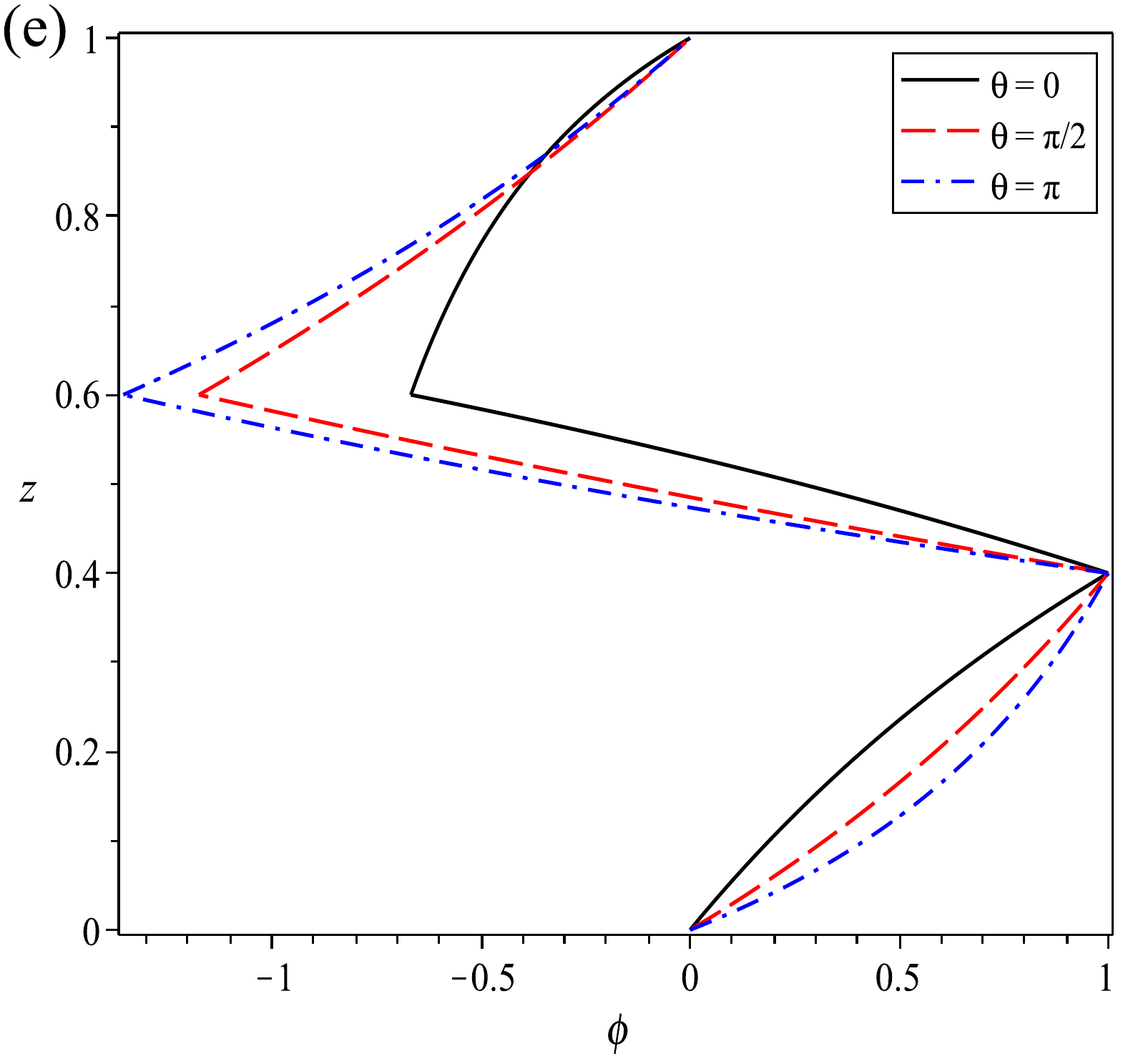} 
     \includegraphics[width=0.3\linewidth]{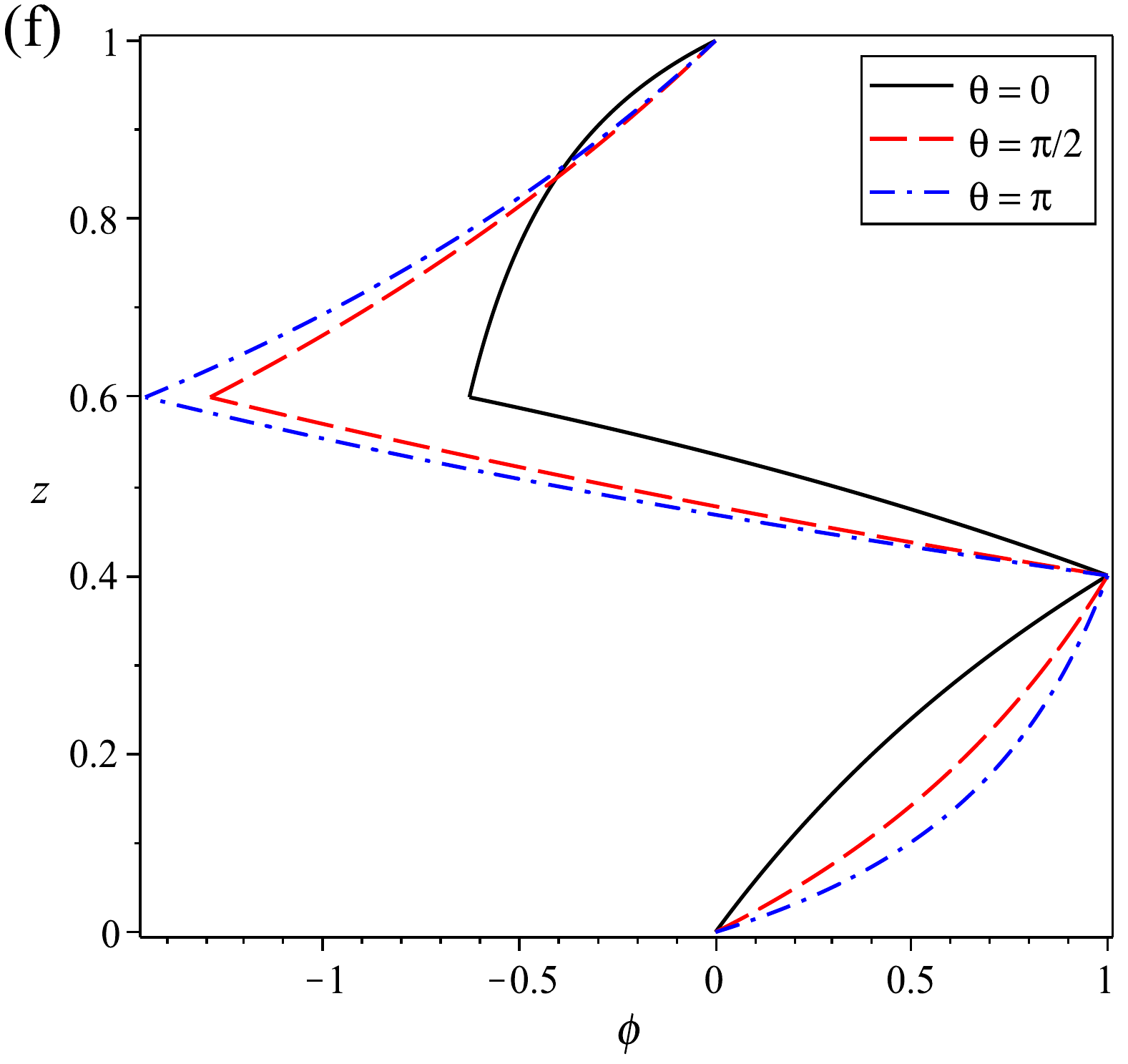} 
%\vspace{-0.5cm}
       \caption{%\footnotesize 
       Modal function $\phi(z)$ for mode-1 (top row) and mode-2 (bottom) when $\rho_1=1, \rho_2=1.1, \rho_3 = 1.2$ and $d_{1}= 0.4$ and $d_{2}= 0.6$. The first, second, and third columns correspond to $\gamma=0.01$, $\gamma=0.1$ and $\gamma=0.12$, respectively. Different angles are considered, $\theta = 0, {\pi}/{2}, \pi$ (see black solid, dashed red and dash-dotted blue lines, respectively).}
       \label{fig:PicS56_6_15}
\end{figure}
%%%%%%%%%%%%%%%

%%%%%%%%%%%%%%%%%%%%%%%%%%%%%%%%%%%%%%%%%%%%%%
%%%%%%%%%%%%%%%%%%%%%%%%%%%%%%%%%%%%%%%%%%%%%%
%%%%%%%%%%%%%%%%%%%%%%%%%%%%%%%%%%%%%%%%%%%%%%

%\newpage
%\section{Vertical structure}
%{\color{red} START READING FROM HERE}

Next, we illustrate the influence of the linear shear flow on the modal function $\phi^{(i)}$, $i=1,2$ defining the vertical structure of the internal wave field, where the superscripts $1$ and $2$ are used to denote the modal functions for the first and second modes, respectively. We normalise the modal functions for the first/second mode so that they are equal to unity on the upper/lower interface, respectively,  i.e. 
%\vspace{0.3cm}
%\begin{eqnarray}
$\phi^{(1)} = { \widetilde{\phi}^{(1)}}/{ \widetilde{\phi}^{(1)}(d_{2}, \theta)},         \ \mbox{and}\            \phi^{(2)} = { \widetilde{\phi}^{(2)}}/{ \widetilde{\phi}^{(2)}(d_{1}, \theta)}, $
%\label{modal_func1}
%\end{eqnarray}
where tildes denote unnormalised modal functions. This normalisation is chosen in accordance with the maxima of these functions in the absence of any current, or for a very weak current (see Figure~\ref{fig:PicS56_6_13}).
In Figures~\ref{fig:PicS56_6_13} and \ref{fig:PicS56_6_15}, we show the modal functions for mode-1 (top panels) and mode-2 (bottom panels) when different layer thicknesses ($d_1=0.3$, $d_2=0.7$ in Figure~\ref{fig:PicS56_6_13} and $d_1=0.4$, $d_2=0.6$ in Figure~\ref{fig:PicS56_6_15}) and vorticity values $\gamma$ are considered.  
\noindent We consider the angles $\theta=0$, $\theta={\pi}/{2}$ and $\theta=\pi$.
%\newpage
%\vspace{0.3cm}
%\noindent 
It is observed that for a very weak current ($\gamma = 0.01$), $\phi^{(1)}$ has a maximum at the upper interface $z=d_{2}$, while $\phi^{(2)}$ has a maximum at the lower interface $z=d_{1}$ for all directions $\theta$, 
%and the vertical structure is close to that in the absence of any current
%they are significantly close in the shape due to the 
justifying the choice for our normalisation.
%The modal function $\phi$ in Figure 6.14 are squeezed; while in Figure 6.12 are elongated in all the direction of %the shear flow. %(because of depth difference of internal position waves).
%\vspace{0.5cm}
%\noindent  
With increasing values of $\gamma$, we observe that 
%It is then interesting to study the effect of an increasing strength of the shear flow on the vertical structure of the first and second interfacial ring modes. This is also shown in Figure~\ref{fig:PicS56_6_13}. Here, there is an interesting effect that 
the maximum of $\phi^{(1)}$ 
%is always on the upper interface in the downstream direction, it 
shifts to the lower interface in the upstream direction. Similarly, the maximum of $\phi^{(2)}$ shifts to the upper interface in the upstream direction. Thus, the vertical structure of the ring waves on a shear flow is strongly three-dimensional.

%\vspace{0.5cm}
%\noindent 
The results for the thick (Figure~\ref{fig:PicS56_6_13}) and thin (Figure~\ref{fig:PicS56_6_15}) intermediate layers are qualitatively similar. However, there are significant quantitative differences. Overall, the effect of the shear flow on the vertical structure is stronger in the first case, similarly to the effect on the deformation of the wavefronts, discussed above.
\begin{figure}
\centering
\includegraphics[height=6.5cm]{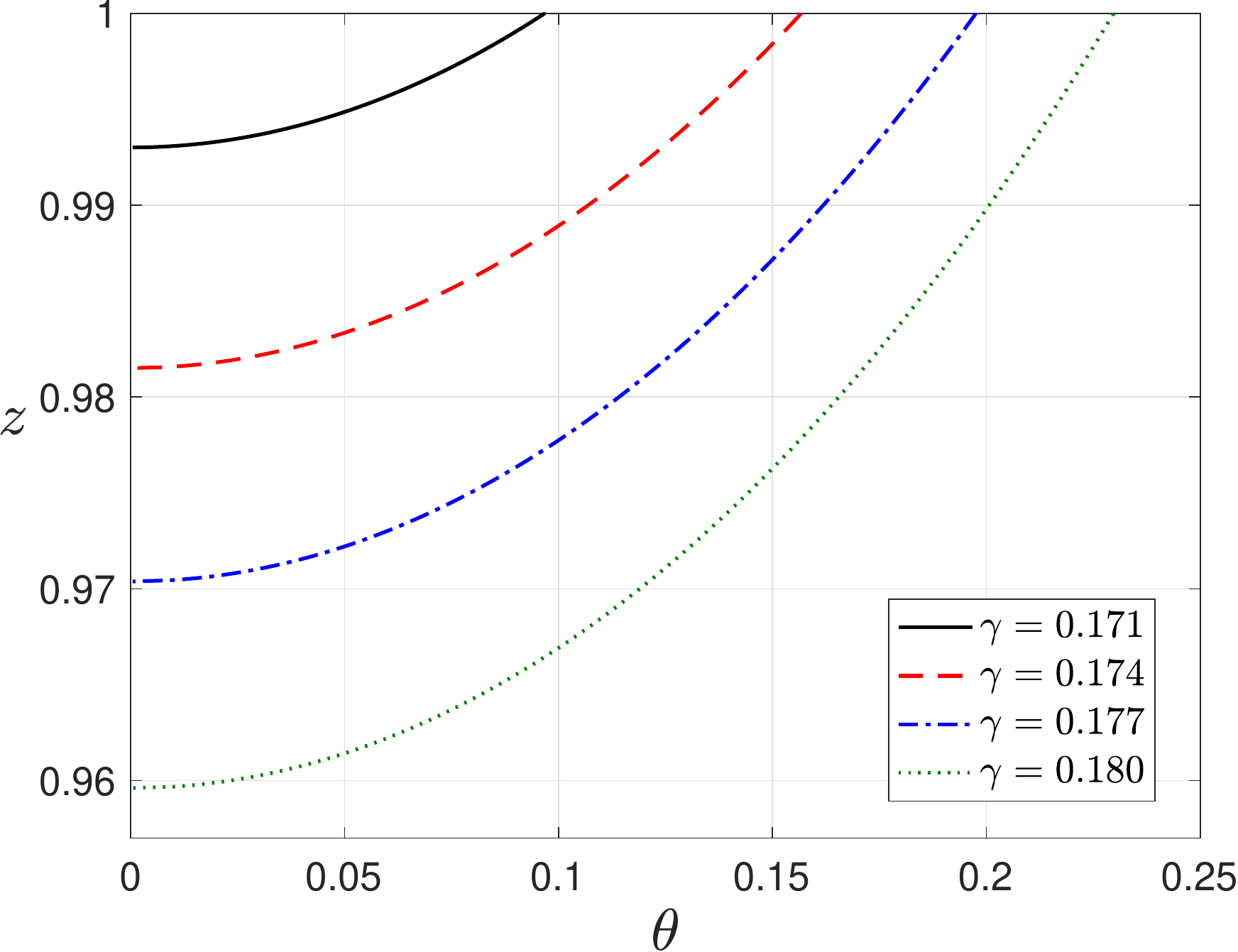}
\caption{%\footnotesize 
The location of the {\it critical surface} $z=z_c(\theta)$ as a function of $\theta$ for $\rho_1=1$, $\rho_2=1.1$, $\rho_3=1.2$ and $d_{1} = 0.3$, $d_{2} = 0.7$ for several values of $\gamma$ as indicated in the legend.}
\label{fig:critical_layer_d1_0_3_d2_0_7}
\end{figure}

When $\gamma$ is large enough, we notice that given $\theta$, $\phi$ may no longer be defined at certain values of $z$ as the spectral problem becomes singular -- the coefficient in front of the second derivative is equal to zero, which corresponds to the emergence of a {\it critical surface}, replacing a {\it critical level} known in the case of plane waves. In particular, it follows from (\ref{phi_all}) 
that $\phi_1$ is not defined when $F(1)=0$, which amounts to consider
$$
F(1)=-s+\gamma (k \cos\theta-k' \sin\theta)=-s+\gamma a,  
$$ 
taking into account (\ref{ss}). When $a<0$, clearly $F(1)$ does not vanish, since $\gamma>0$ by assumption. However, as seen in Figure~\ref{fig:Q_a2}, the range of values of $a$ defined by the condition $H-a^2>0$ always contains positive values, regardless of the regime being elliptic, hyperbolic, or parabolic. Considering first the elliptic regime we assert that the minimum value of $\gamma$ for which $F(1)=0$ is attained when $a=a_{\max}$, for which $\theta(a_{\max})=0$. Let $\gamma^*$ denote this critical value of $\gamma$. Then, $\gamma^*=s/k(0)$, which is precisely when the wave speed in the downstream direction matches %the vorticity strength, 
the speed of the background current at the top surface, i.e.~$u_0(1)$,
as depicted in Figure~\ref{fig:speeds}. For the parameters in Figure~\ref{fig:speeds}, we find $\gamma^*\approx 0.169226$ for mode-2 waves and $\gamma^*\approx 0.464706$ for mode-1 waves. Moreover, as the value of $\gamma$ increases further, we observe that the critical surfaces persist. More precisely, $F(z)=0$ on the geometrical locus defined by $z=z_c(\theta)$ for a certain range of values of $\theta$ ({\it cf.} Figure~\ref{fig:critical_layer_d1_0_3_d2_0_7}).  
%As a consequence, as the values of $\gamma$ are increased 
%
%For the special value of $a$ 
%From (\ref{phi_all}) it follows that 
%Note that as $\gamma$ increases, there appears a critical layer. For mode~2, the critical layer appears first in the streamwise direction  (i.e. for $\theta=0$) at $\gamma=\gamma_c\approx 0.169$. For $\gamma>\gamma_c$, the critical layer exists for $\theta\in [0,\tilde\theta]\cup[2\pi,2\pi-\tilde\theta]$, where $\tilde\theta=0$ for $\gamma=\gamma_c$ and $\tilde\theta$ increases as $\gamma$ increases. Figure~\ref{fig:critical_layer_d1_0_3_d2_0_7} shows the location of the critical layer $z=z_c(\theta)$ as a function of $\theta$.
%
%it is useful to point out that the values $a_{\max}$, $a_{\min}$, and $a^*$ introduced above correspond to assigning the values $0$ or $\pi$ to $\theta$. More precisely: in the elliptic regime $\theta(a_{\min})=\pi$, $\theta(a_{\max})=0$; in the hyperbolic regime $\theta(a_{\min})=\theta(a_{\max})=0$; in the parabolic regime $\theta(a^*)=0$. Note that in the hyperbolic regime $a_{\min}$ 
%
We notice that the solution for the equation for $k$ as a function of $\theta$ still formally exists, even after the appearance of a critical surface. 
%Moreover, at about $\gamma=0.181192$, there is a transition from the ``elliptic'' regime to the ``hyperbolic'' regime for mode~2, see Figure~\ref{fig:Q_a2}. 
%The ``elliptic'' regime corresponds to the case when $Q- a^2>0$  for $ a\in( a_\mathrm{min},\, a_\mathrm{max})$ and the ``hyperbolic'' regime corresponds to the case when $Q- a^2<0$ for $ a\in( a_\mathrm{min},\, a_\mathrm{max})$. Here, $ a_\mathrm{min}$ and $ a_\mathrm{max}$ are the smallest and the largest roots of $Q- a^2$. (Note that in all the cases considered, we found that $Q- a^2$ has two roots, except at the transition from the ``elliptic'' regime to the ``hyperbolic'' regime, at which there exists only one root.) So, in the ``elliptic'' regime, the range for $ a$ considered is $ a\in[ a_\mathrm{min},\, a_\mathrm{max}]$, while in the ``hyperbolic'' regime, the range for $ a$ considered is $ a\in(-\infty,\, a_\mathrm{min}]\cup[ a_\mathrm{max},\,\infty)$. {\bf DT: All this paragraph is probably not needed.}}

%The counterparts of the results shown in Figures~\ref{fig:PicS56_6_13} for the case when the interfaces are positioned at $d_1 = 0.3$ and $d_2 = 0.7$, are shown in Figure~\ref{fig:PicS56_6_15} for $d_1 = 0.4$ and $d_2 = 0.6$. The results are qualitatively similar, but there are significant quantitative differences. Overall, the effect of the shear flow on the vertical structure is stronger in the first case, similarly to the effect on the deformation of the wavefronts, discussed above.
%

\begin{figure}
\centering
\includegraphics[height=6.5cm]{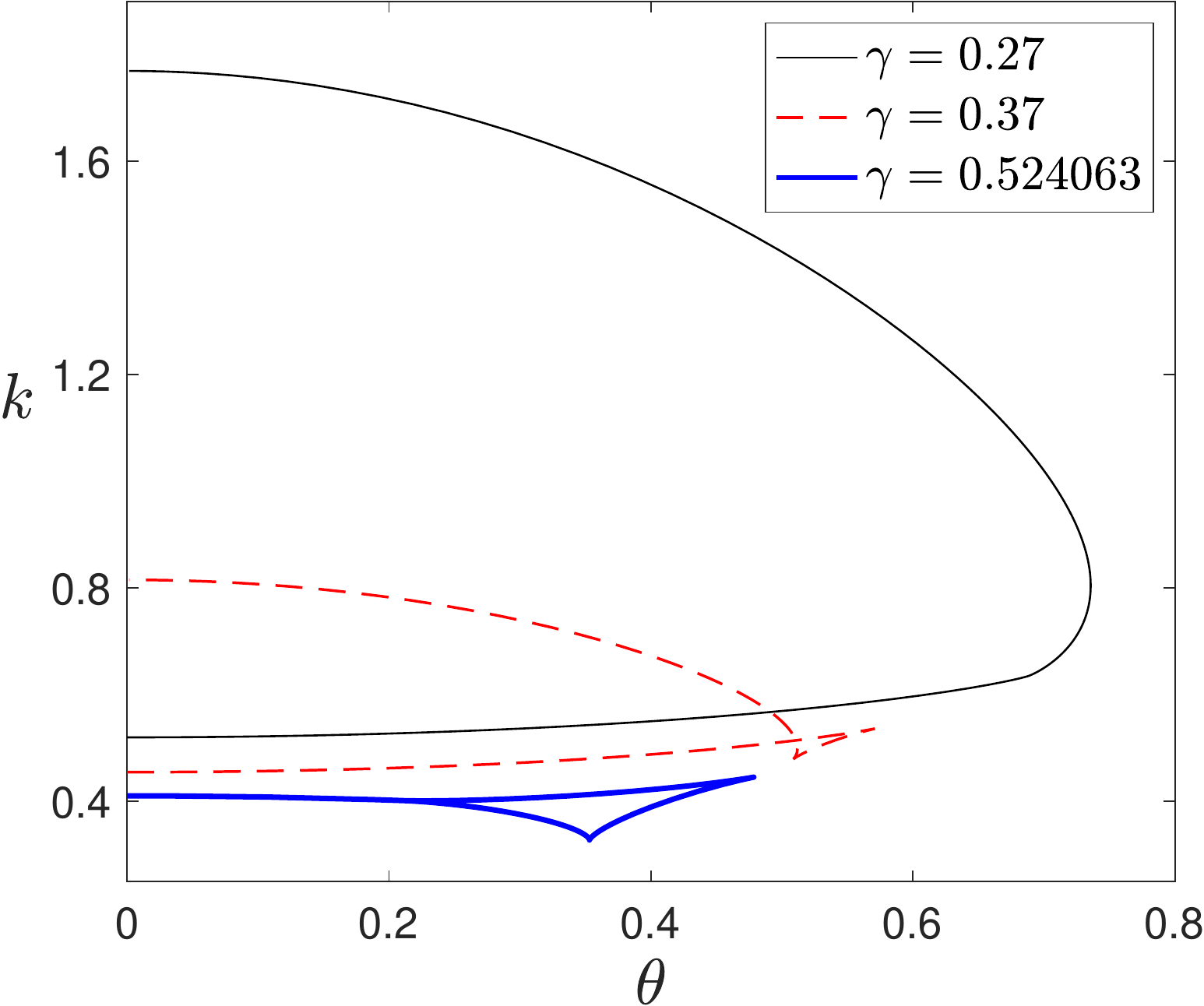}
\caption{%\footnotesize 
$k(\theta)$ of the mode-2 internal ring waves for $\rho_1=1$, $\rho_2=1.1$, $\rho_3=1.2$ and $d_{1} = 0.3, d_{2} = 0.7$ in the hyperbolic regime when $\gamma=0.27,\, 0.37,\, 0.524063$.}
\label{fig:k_theta_hyperbolic}
\vspace{1cm}
\centering
\includegraphics[height=6.5cm]{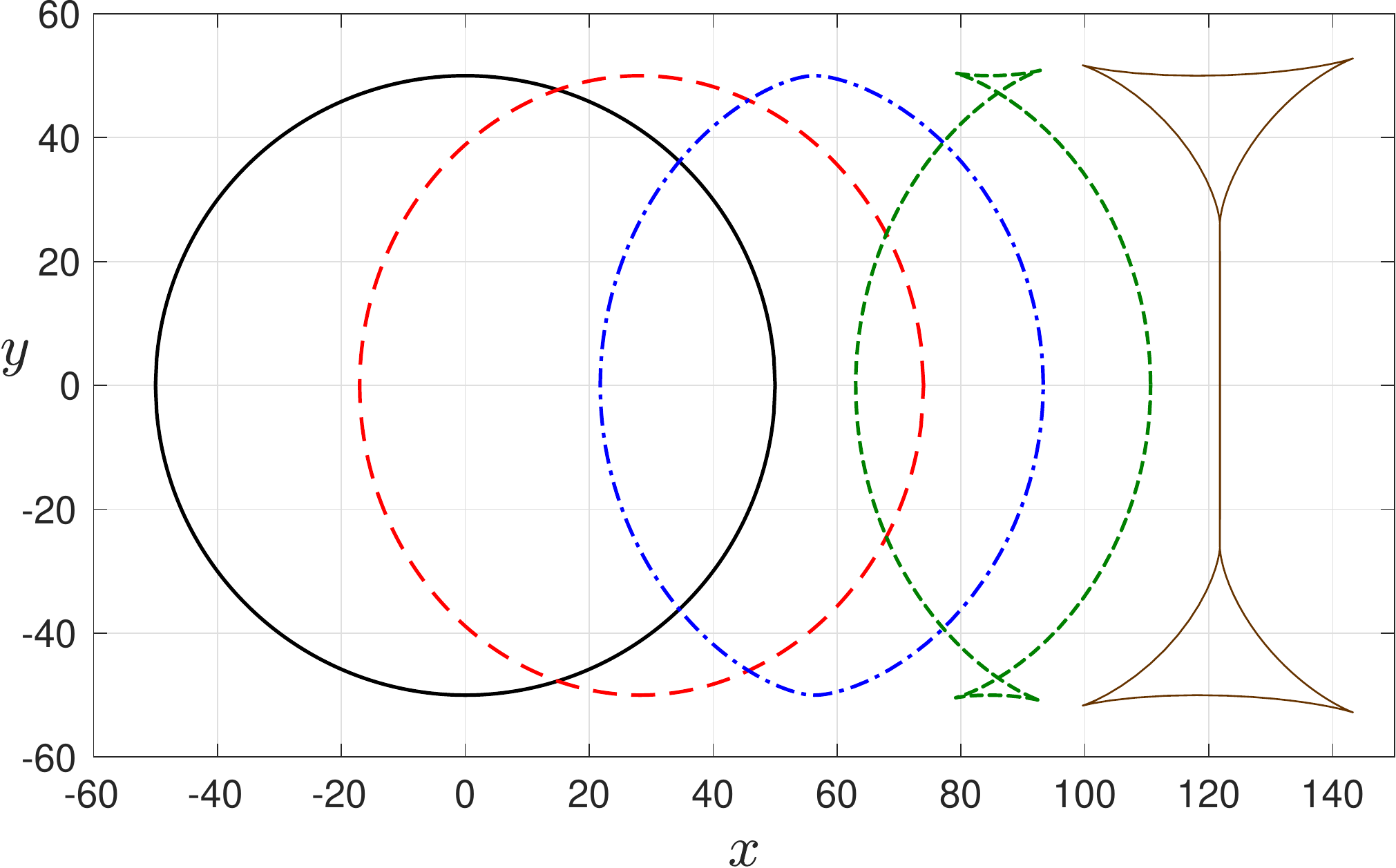}
\caption{%\footnotesize 
The wavefronts of the mode-2 ring waves for $\rho_1=1$, $\rho_2=1.1$, $\rho_3=1.2$ and $d_{1} = 0.3$, $d_{2} = 0.7$, described by $k(\theta)r = 50$, showing the transition from the elliptic to hyperbolic regime with $\gamma=0$ (thick black solid line), $\gamma =0.125$ (red long-dashed line), $\gamma =0.25$ (blue dash-dotted line), $\gamma =0.375$ (green short-dashed line) and $\gamma =0.524063$ (thin brown solid line), respectively.}
\label{fig:wavefronts_elliptic_hyperbolic}
\end{figure}

\begin{figure}
\centering
\includegraphics[height=6.5cm]{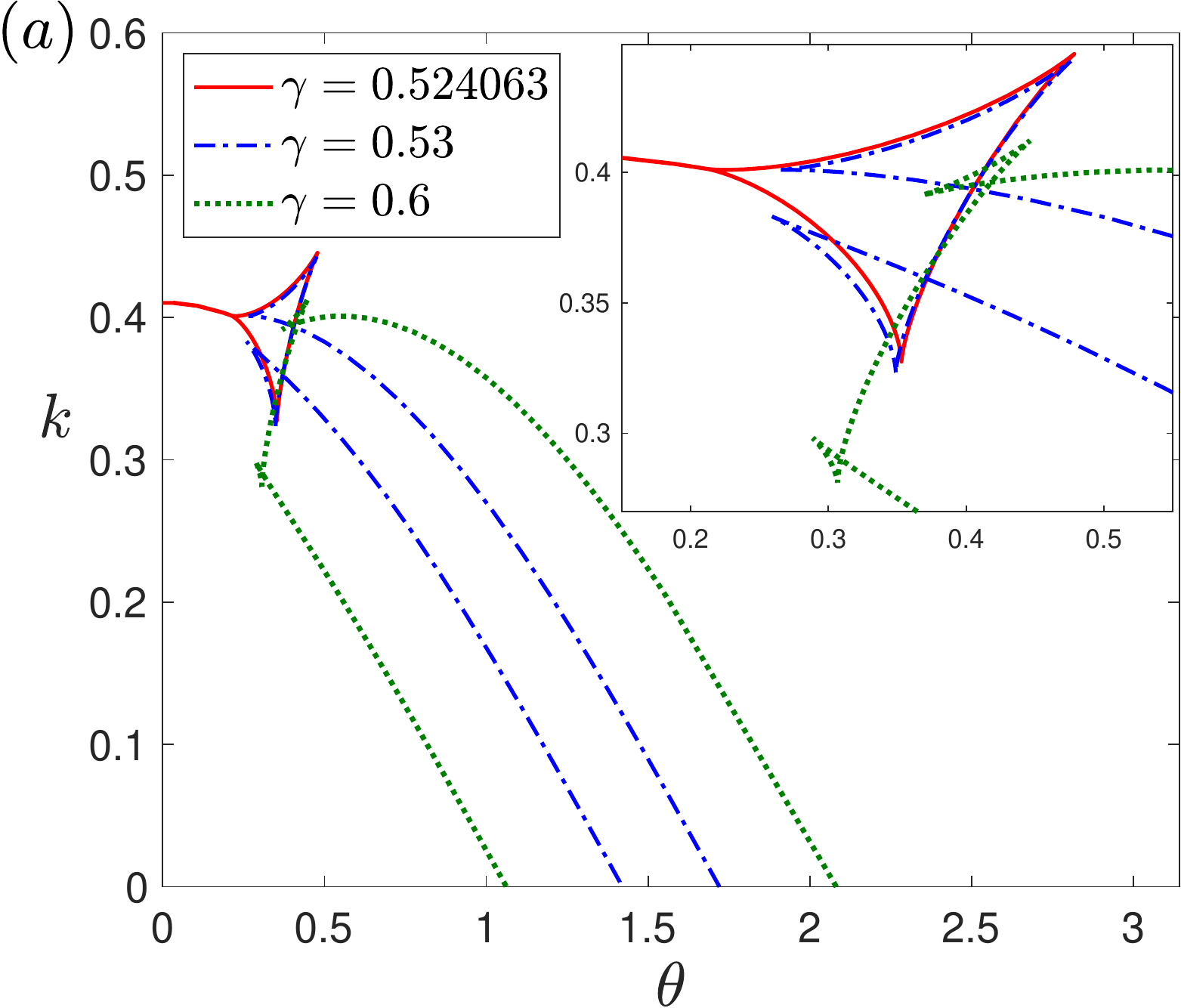}
%\vspace{1cm}
\centering
\includegraphics[height=6.5cm]{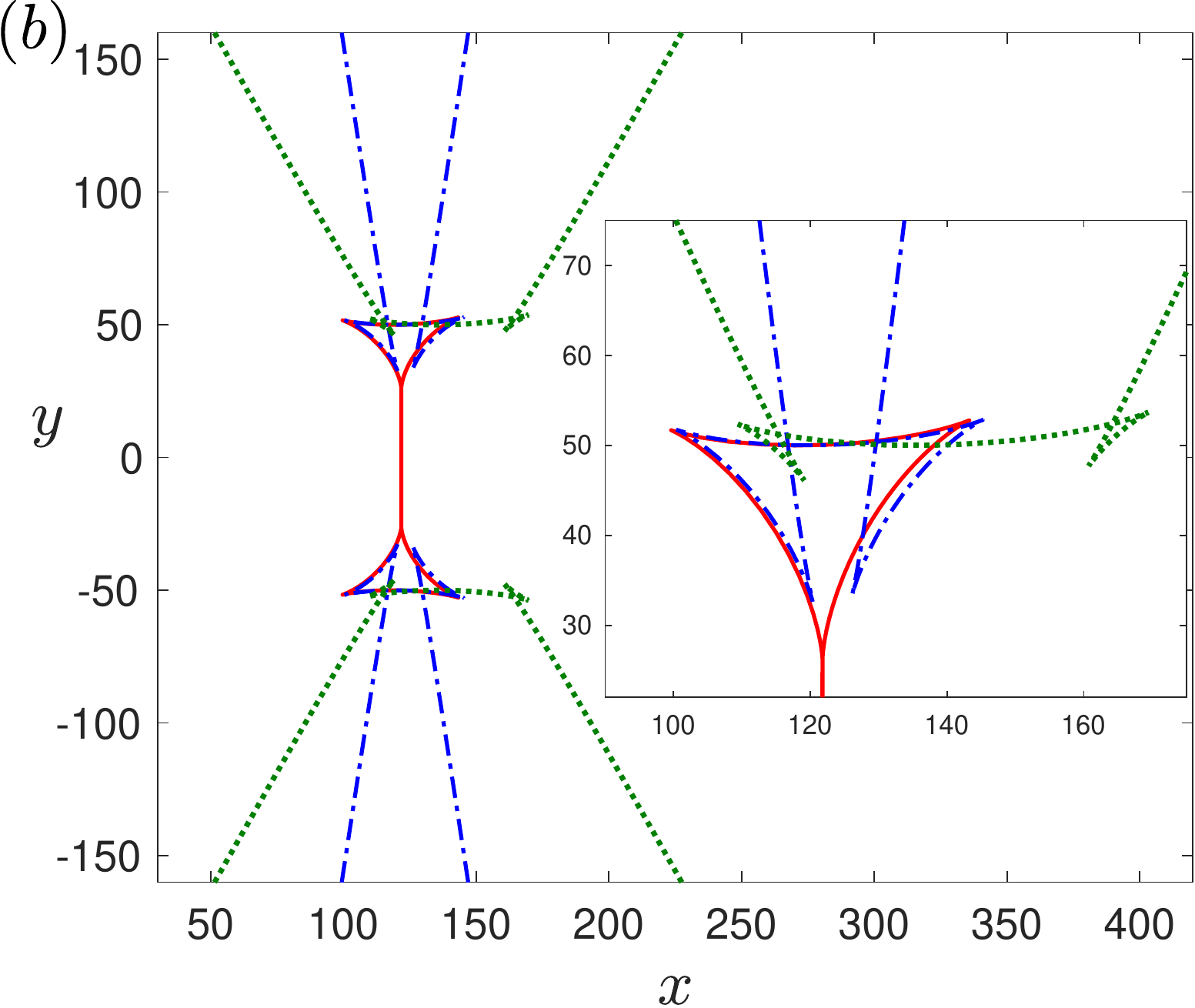}
\caption{%\footnotesize 
Shown are (a) $k(\theta)$ and (b) the wavefronts of the $\mbox{mode-2}$ internal waves for $\rho_1=1$, $\rho_2=1.1$, $\rho_3=1.2$ and  $d_{1} = 0.3, d_{2} = 0.7$ in the hyperbolic regime when $\gamma=0.524063,\,0.53,\,0.6$. The wavefronts are described by $k(\theta)r = 50$. }
\label{fig:mode2_restructure}
\end{figure}

\subsection{Hyperbolic regime}
\label{sec:hyperbolic_regine}

As discussed above, when $\gamma>\gamma_{p2}$ the mode-2 ring waves will transition to the hyperbolic regime. When this happens, we observe that the relationship between $k$ and $\theta$ is no longer one-to-one as shown in Figure~\ref{fig:k_theta_hyperbolic}. Moreover, with increasing values of $\gamma$   
we start observing the formation of a ``swallowtail" singularity (e.g.~\cite{Arnold}) at about $\gamma=\gamma_s\approx 0.274$. For $\gamma>\gamma_s$, the curve in the $(k,\theta)$-plane becomes self-intersecting with two cusps formed, as shown in Figure~\ref{fig:k_theta_hyperbolic} for $\gamma=0.37$. We recall that at $\gamma=\gamma^-\approx 0.524063$ the speeds of the mode-2 wave fronts along the current coincide and {mode-2} ring waves with compact wavefronts crossing the $x$-axis cease to exist (see Section~\ref{sec:lw_instability}).  This is reflected in the corresponding curve for $\gamma=0.524063$ in the $(k,\theta)$-plane by a coalescence of some of its branches.  {\color{red}}

%For this limiting value of $\gamma$ the corresponding curve on the $(k,\theta)$-plane     
The transition from the elliptic to the hyperbolic regime is depicted in Figure~\ref{fig:wavefronts_elliptic_hyperbolic} where the wavefronts are shown for different values of $\gamma$ up to $\gamma=\gamma^-$, and where the formation of the swallowtail singularity is well patent. We note that such feature has been  documented e.g. in the context of optics \cite{Berry1} and tsunamis \cite{Berry2}.

Finally, we note that although {mode-2} ring waves with compact wavefronts crossing the $x$-axis do not exist for $\gamma\in(\gamma^-,\gamma^+)$, where $\gamma^-\approx 0.524063$ and $\gamma^+\approx0.829932$, other types of solutions with non-compact wavefronts extending to infinity appear for such values of $\gamma$. Such a restructuring of the $\mbox{mode-2}$ solutions is demonstrated in Figure~\ref{fig:mode2_restructure}, where the solutions for mode-2 are shown for $\gamma=0.524063,\,0.53$ and $0.6$ (the red solid, blue dash-dotted and green dotted lines respectively). It is apparent from panel (a) that for $\gamma>\gamma^-$ there appear two branches in the solutions for $k(\theta)$  which converge to $k=0$ for certain values of $\theta$, say $\theta_a$ and $\theta_b$ with $\theta_a<\theta_b$. We have verified that at $\gamma=\gamma^-$, $\theta_a$ and $\theta_b$ apparently emerge from $\pi/2$, and as $\gamma$ increases, $\theta_a$ decreases while $\theta_b$ increases. Convergence of $k$ to $0$ as $\theta\rightarrow\theta_{a,b}$ implies that $r\rightarrow\infty$ as $\theta\rightarrow\theta_{a,b}$, i.e. the corresponding wavefronts extend to infinity. This is confirmed in panel (b), where it can be observed that for $\gamma>\gamma^-$ the wavefronts do not cross the $x$-axis and consist of two disconnected parts (symmetric with respect to the $x$-axis). Each of the parts apparently has two oblique asymptotes, one with a positive slope and another one with a negative slope.

%each contour possesses two cusps and one self-intersection
%
%In geometrical language (Arnold 1986), the double-cusped wavefront is a swallowtail singularity 

\section{Long-wave instability}\label{sec:lw_instability}

In this section we consider the linear stability analysis of an inviscid, incompressible, stratified shear flow, with the ambient density stratification $\rho_0(z)$ and background velocity $u_0(z)$. The behaviour of a small two-dimensional, monochromatic disturbance of wavenumber $\alpha$ and wave speed $c$ is governed by (see \cite{DR}) 
\begin{equation}\label{Euler_linearised}
\hat{\psi}'' +\frac{\rho_0'}{\rho_0}\left (\hat{\psi}'-{u_0'\over u_0-c} \hat{\psi}\right )+ \left[-\frac{g\,\rho_0'}{\rho_0\, (u_0-c)^2}  - \frac{u_0''}{u_0-c} - \alpha^2 \right] \hat{\psi}=0,
\end{equation}
where the prime indicates differentiation with respect to $z$, $g$ is the gravitational acceleration, and $\hat{\psi}$ is the complex amplitude of the stream function $\psi$ defined by $\psi(x,z,t)=\hat{\psi}(z) {\rm exp}[i\alpha(x-ct)]$ at each point $(x,z)$ and time $t$. The wave speed $c$ may be complex, and such a wave is said to be unstable if ${\rm Im}(c)>0$.  

As before, linear velocity and piecewise-constant density profiles are adopted (see Figure~\ref{fig:Pic20}),
$$
u_0(z) = \gamma z, \quad \rho_0 (z)= \rho_{3}H(z) + (\rho_{2}-\rho_{3})H(z-d_{1}) + (\rho_{1}-\rho_{2})H(z-d_{2}).
$$

As a consequence, $u_0''=0$, and so, in each subdomain where $\rho_0=\rm{constant}$, equation (\ref{Euler_linearised}) can be solved explicitly as linear combinations of $\exp (\pm \alpha z)$. More precisely:
$$
\hat{\psi}(z) = \left\{ 
\begin{array}{rcl}
A_3 \,e^{\alpha z} + B_3 \,e^{-\alpha z} & \mbox{if } & d_2<z<h_0\\
A_2 \,e^{\alpha z} + B_2 \,e^{-\alpha z} & \mbox{if } & d_1 < z < d_2\\
A_1 \,e^{\alpha z} + B_1 \,e^{-\alpha z} & \mbox{if } & 0<z < d_1\\
\end{array}  
\right.,
$$
for arbitrary constants $A_1, A_2, A_3, B_1, B_2, B_3$. Then, at the levels $z=d_1$ and $z=d_2$ where $\rho_0(z)$ is discontinuous, the continuity of pressure and normal velocity at each one of these interfaces requires the following jump conditions: 
\begin{equation}\label{jump_conditions}
\left\llbracket  \rho_0 \left[ (u_0-c) \, \hat{\psi}^\prime - \left( u_0^\prime + \frac{g}{u_0-c} \right) \hat{\psi} \right]  \right\rrbracket =0, \quad \left\llbracket \hat{\psi} \right\rrbracket =0,
\end{equation}
respectively. Here we have used $\llbracket \cdot \rrbracket$ to denote a jump across the interface. By imposing these jump conditions along with no flux conditions at the rigid boundaries, the system can be reduced to two equations
\begin{eqnarray}
&&\hspace{-2cm}\Big\{ (\gamma d_1-c) \alpha \left[ \rho_3 \,\coth(\alpha d_1) + \rho_2 \,\coth(\alpha(d_2-d_1)) \right] -(\rho_3-\rho_2) \left( \gamma+ \frac{g}{\gamma d_1-c}\right)  \Big\} \,b_1 -  \nonumber \\
&&- \rho_2 (\gamma d_1-c) \alpha \, {\rm csch} (\alpha(d_2-d_1)) \, b_2 = 0,\label{disprel1}
\end{eqnarray}
\begin{eqnarray}
 &&\hspace{-2cm}- \rho_2 (\gamma d_2-c) \alpha \, {\rm csch} (\alpha(d_2-d_1)) \, b_1 + \nonumber \\
&&\hspace{-1cm}+\Big\{ (\gamma d_2-c) \alpha \left[ \rho_1 \,\coth(\alpha(h_0-d_2)) + \rho_2 \,\coth(\alpha(d_2-d_1)) \right] + \nonumber \\
&&+(\rho_1-\rho_2) \left( \gamma+ \frac{g}{\gamma d_2-c}\right) \Big\} \,b_2 = 0, \label{disprel2}
\end{eqnarray}
with $b_1 \equiv A_2 \,e^{\alpha d_1} + B_2 \,e^{-\alpha d_1}$ and $b_2 \equiv  A_2 \,e^{\alpha d_2} + B_2 \,e^{-\alpha d_2}$, from which it follows that the dispersion relation between the wave speed $c$ and the wavenumber $\alpha$ is obtained as a polynomial equation (of degree 4) for $c$. To examine the long-wave instability, we consider the equations (\ref{disprel1}),~(\ref{disprel2}) in the limit when $\alpha \rightarrow 0$,
{\small
\begin{eqnarray*}
&&\hspace{-3cm}\Big\{ (\gamma d_1-c)^2 \left[ \rho_3(d_2-d_1) + \rho_2 d_1\right] - (\rho_3-\rho_2)d_1 (d_2-d_1) \left[ \gamma (\gamma d_1 -c)+g \right] \Big\} \,b_1 
- \rho_2 d_1 (\gamma d_1-c)^2 \, b_2 = 0, \nonumber \\
&&\hspace{-2cm}- \rho_2 (h_0-d_2) (\gamma d_2-c)^2 \, b_1 + \nonumber \\
&&\hspace{-3cm}+\Big\{ (\gamma d_2-c)^2 \left[ \rho_1 (d_2-d_1) + \rho_2 (h_0-d_2)\right]+(\rho_1-\rho_2)(d_2-d_1)(h_0-d_2)\left[ \gamma (\gamma d_2 -c)+g \right] \Big\} \,b_2 = 0.
\end{eqnarray*}
}
Requiring the determinant of this linear system to vanish, and non-dimensionalising the variables as in Section 1, we obtain the long-wave speeds as the roots of the quartic equation for $c$, which exactly coincides with the equation (\ref{lin_lw_speeds}) previously obtained in Section~3 as a reduction of the angular adjustment equation (\ref{dr}).
%\begin{eqnarray}
%&&\hspace{-2cm}\Big\{ (\gamma d_1-c)^2 \left[ \rho_3(d_2-d_1) + \rho_2 d_1\right] - (\rho_3-\rho_2)d_1 (d_2-d_1) \left[ \gamma (\gamma d_1 -c)+1 \right] \Big\}  \nonumber \\
%&&\hspace{-2cm}\times \Big\{ (\gamma d_2-c)^2 \left[ \rho_1 (d_2-d_1) + \rho_2 (1-d_2)\right]+(\rho_1-\rho_2)(d_2-d_1)(1-d_2)\left[ \gamma (\gamma d_2 -c)+1 \right] \Big\}= \nonumber \\
%&&\hspace{-2cm}=\rho_2^2 d_1 (1-d_2) (\gamma d_1-c)^2 (\gamma d_2 -c)^2. \label{lin_lw_speeds}
%\end{eqnarray}
%
%We can work in the non-dimensional variables introduced in Section 1 simply by dividing each one of the two equations by $g \rho^* h_0^2$, with $\rho^*$ the dimensional reference density of the fluid.} 
%%If we scale time such that the acceleration due to gravity g is also taken to be unity, 
%Then, linear longwave speeds are the roots of the following quartic equation for $c$:
For brevity, the same symbols are used for non-dimensional quantities here. It was proved in \cite{BC} that regardless of the physical parameters used, there is always a limited range of values $(\gamma^-, \gamma^+)$ at which two of the four roots of (\ref{lin_lw_speeds}) are complex, and long waves are unstable. Outside this range, all four roots are real and long waves are stable. 

In the absence of shear current ($\gamma=0$), (\ref{lin_lw_speeds}) reduces to 
$$
a_4 c^4 + a_2 c^2 + a_0 =0,
$$
where the coefficients $a_0, a_2, a_4$ are precisely those found in (\ref{lw_no_shear}). When $\gamma> 0$, the critical values $\gamma^-, ~\gamma^+$ between which long waves are unstable can be found by computing the roots of the {\it discriminant}, with respect to $c$, to the quartic equation (\ref{lin_lw_speeds}). This yields a polynomial equation for $\gamma$ (of degree $12$) whose roots can be computed numerically. 
%that has two positive roots $\gamma^-$ and $\gamma^+$, with $\gamma^-<\gamma^+$. 
However, simple explicit estimates to such critical values can be obtained by adopting the Boussinesq approximation, commonly used in the study of weakly stratified fluids. To do so, it is convenient to go back again to dimensional variables. Under the Boussinesq approximation, the linear long-wave speeds, given in dimensional form, are the roots of
\begin{eqnarray}
&&\hspace{-2cm}\left[ d_2 (\gamma d_1-c)^2 - g\Delta_2 d_1 (d_2-d_1) \right] \left[ (h_0-d_1)(\gamma d_2-c)^2 - g \Delta_1 (d_2-d_1)(h_0-d_2) \right]= \nonumber \\
&&=d_1 (h_0-d_2) (\gamma d_1-c)^2 (\gamma d_2 -c)^2. \label{lw_speeds_Boussinesq} 
\end{eqnarray}
Here, $\Delta_1$ and $\Delta_2$ are the density increments defined as $\Delta_1 = (\rho_2-\rho_1)/\rho_2$ and $\Delta_2 = (\rho_3-\rho_2)/\rho_2$, and assumed to be small. By taking $\Delta_1=\Delta_2$ and $d_1=h_0-d_2$, we recover the so-called {\it symmetric configuration}, for which the equation (\ref{lw_speeds_Boussinesq}) is considerably reduced:
\begin{eqnarray}
&&\hspace{-2cm}\left[ (h_0-d_1) (\gamma d_1-c)^2 - g^\prime d_1 (h_0-2d_1) \right] \left[ (h_0-d_1)(\gamma d_2-c)^2 - g^\prime d_1(h_0-2d_1)\right]= \nonumber \\
&&=d_1^2 (\gamma d_1-c)^2 (\gamma d_2 -c)^2, \label{lw_speeds_sym_Boussinesq} 
\end{eqnarray}
\begin{figure}
\centering
\includegraphics[width = 0.5\textwidth] {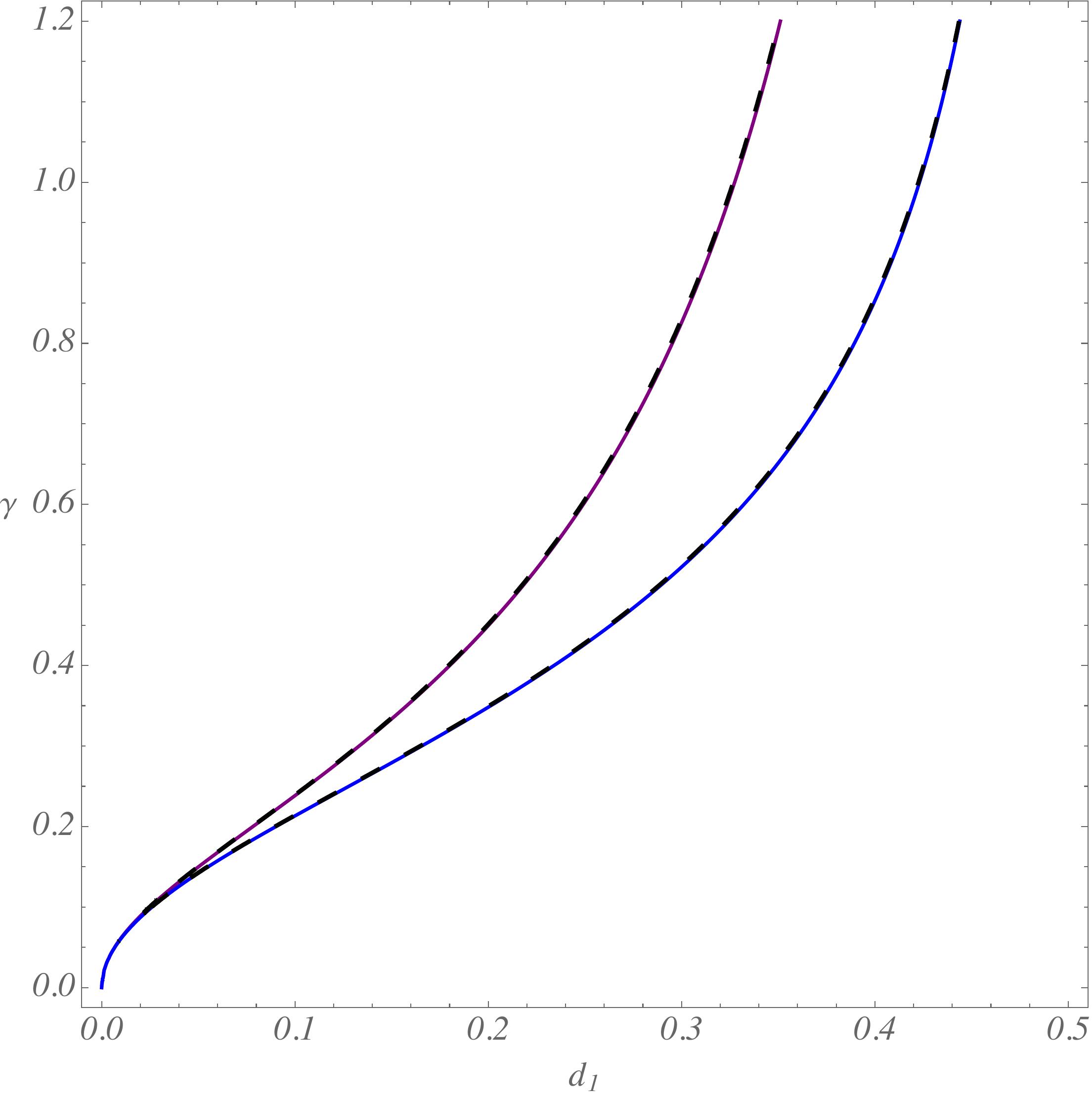}
\caption{Values of $\gamma^-$ and $\gamma^+$ at which the discriminant of (\ref{lin_lw_speeds}) vanishes (black dashed lines), and comparison with the values ${\gamma_B^-}, ~{\gamma_B^+}$ obtained under Boussinesq approximation (red and blue solid lines). Here, the top and bottom layers have the same thickness in rest, i.e., $d_1=1-d_2$, and we set equal density increments $\Delta_1=\Delta_2=1/11$ with $\rho_1=1$. We recall that when 
$\gamma^-< \gamma< \gamma^+$, mode-2 ring waves do not exist.} 
\label{fig:values_gamma_c}
\end{figure}
where $g^\prime=g \Delta_1$ is the reduced gravity. The significance of this form is that the quartic equation for $c$ can be rewritten as a biquadratic for a special speed $\tilde{c}$. To do so, notice that the equation can be cast into the form: 
\begin{eqnarray}
&&\hspace{-2cm} h_0(h_0-2d_1)\left[ (\gamma d_1 -c)(\gamma d_2-c) \right]^2 - g^\prime d_1 (h_0-d_1)(h_0-2d_1) \left[ (\gamma d_1-c)^2 +(\gamma d_2-c)^2 \right]+ \nonumber\\
&&+ {g^\prime}^2 d_1^2 (h_0-2d_1)^2=0\nonumber.
\end{eqnarray}
Let $\tilde{u}_0=\gamma h_0/2$ be the average velocity. Then 
$$
\gamma d_1 -c = \tilde{u}_0-c-\delta, \quad \gamma d_2 -c = \tilde{u}_0-c+\delta,
$$
with $\delta=\gamma (h_0-2d_1)/2$. If we define $\tilde{c}$ as the wave speed relative to the mean flow, {\it i.e.,} $\tilde{c} =c-\tilde{u}_0$, then 
$$
(\gamma d_1-c)^2 + (\gamma d_2-c)^2 = 2( \tilde{c}^2 + \delta^2), \quad  (\gamma d_1-c) (\gamma d_2-c) =\tilde{c}^2 - \delta^2,
$$ 
which shows that (\ref{lw_speeds_sym_Boussinesq}) is indeed a biquadratic form for $\tilde{c}$. 

Using once again non-dimensional variables we find that the {\it discriminant} is given by a polynomial on $\gamma$ (of degree 8), which is proportional to  
$$
\left[ (1-2d_1)^2 \gamma^2 - 4 d_1 \Delta_1 \right] \left[ (1-2d_1)\gamma^2 - 4d_1 \Delta_1 \right] \left[ (1-2d_1)^2 (1-d_1) \gamma^2 + d_1^3 \Delta_1 \right]^2. 
$$ 
The last term of this expression is clearly always positive, so the discriminant vanishes only when 
$$
\gamma^2=  ({\gamma_B^-})^2 \equiv 4d_1 \Delta_1 /(1-2d_1), \quad \gamma^2= ({\gamma_B^+})^2 \equiv 4d_1 \Delta_1 /(1-2d_1)^2.
$$
%We conclude that $\gamma^- = 2 \sqrt{d_1 \Delta_1 /(1-2d_1)}$ is the value above which long waves become unstable. 
The remarkable agreement of these estimates with the values $\gamma^-$, $\gamma^+$, computed by considering the non-Boussinesq effects, is shown in Figure~\ref{fig:values_gamma_c}.

%on the density stratified fluid 
%%approximation with the density values 
%adopted in this study: $\rho_1=1, \rho_2=1.1, \rho_3=1.2$.   
%In Figure~\ref{fig:values_gamma_c} it is shown 

\section{Singular solutions and the $p$-discriminant method}

In Sections \ref{sec:angular_eq} and \ref{sec:wave_fronts}, wavefronts of ring waves were obtained from singular solutions to the angular adjustment equation (\ref{dr}). To find such solutions the envelope of the family of integral curves to the equation was examined (see Section \ref{sec:angular_eq}). We will show in this Section that there is another method to find singular solutions to the differential equation (\ref{dr}). 

In general terms, given a differential equation $\varphi(x,y,y')=0$, with $x$ denoting the independent variable and $y(x)$ the dependent variable, a solution is called a singular solution if uniqueness is violated at each point of the domain of the equation. Geometrically this means that more than one integral curve with the common tangent line passes through each point $(x_0, y_0)$.
%A solution of a differential equation $\varphi(x,y,y')=0$ is called a singular solution if uniqueness of solution is violated at each point of the domain of the equation. Geometrically this means that more than one integral curve with the common tangent line passes through each point $(x_0, y_0)$.
One of the ways to find a singular solution is by examining the envelope of the family of integral curves, based on using what is known as the {\it $C$-discriminant} (as in Section \ref{sec:angular_eq}). 
Another way to find a singular solution, proposed by Darboux \cite{Darboux}, consists on investigating the so-called {\it $p$-discriminant} of the differential equation. If the function $\varphi(x,y,y')$ and its partial derivatives $\frac{\partial \varphi}{\partial y}$, $\frac{\partial \varphi}{\partial y'}$ are continuous in the domain of the differential equation, the singular solution can be found from the system of equations: 
$$
\varphi(x,y,y')=0, \quad \frac{\partial \varphi(x,y,y')}{\partial y'}=0.
$$   
Upon finding the $p$-discriminant curve (a necessary condition for $y(x)$ to be a solution of this pair of equations), one should check whether it is a solution of the differential equation, and whether it is a singular solution, that is whether there are any other integral curves of the differential equation that touch the $p$-discriminant curve at each point.

\begin{figure}
\centering
\includegraphics[height=6.5cm]{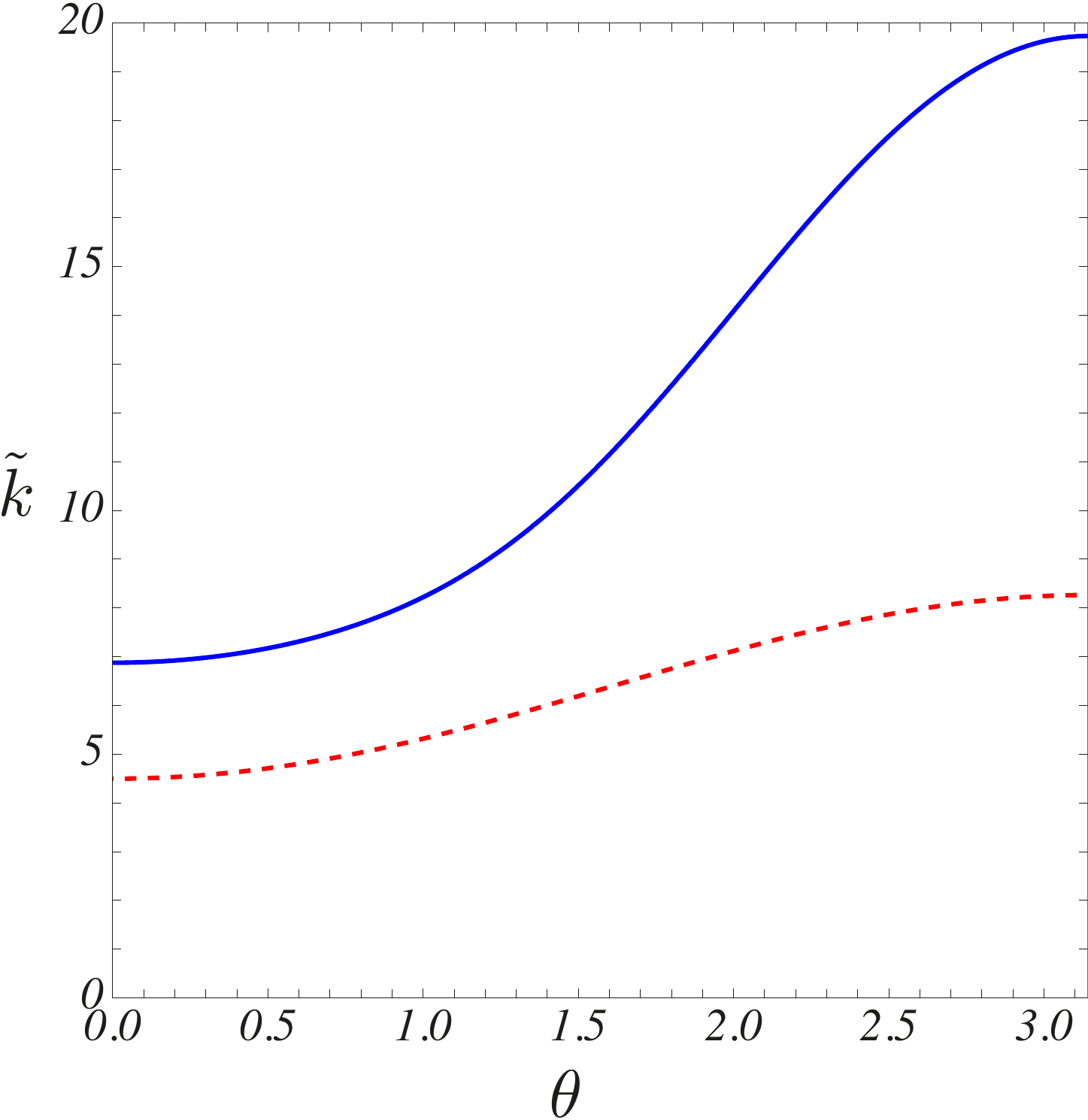}
\caption{%\footnotesize 
The geometrical locus on the $(\theta,\tilde{k})$-plane where the $p-$discriminant curve vanishes for  $\rho_1=1$, $\rho_2=1.1$, $\rho_3=1.2$ and $d_{1} = 0.3, d_{2} = 0.7$, and $\gamma=0.1$.}
\label{fig:p_discriminant_curve}
\end{figure}

We now go back to the angular adjustment equation (\ref{dr}) and use the $p$-discriminant to find the singular solution pertinent to ring waves. We remark that (\ref{dr}) can be rescaled, by introducing a new variable $\tilde{k}$ defined as $\tilde{k}=k/s$. Then, by writing the differential equation as $\varphi (\theta,\tilde{k}, \tilde{k}^\prime)=0$ and finding the discriminant of $\varphi$ with respect to $\tilde{k}^\prime$, we obtain a 12th degree polynomial in $\tilde{k}$ with the coefficients depending on $\theta$. The zero set of this expression is what is known as the $p$-discriminant curve. It should be stressed that this is not a plane algebraic curve, since its expression (too cumbersome to be presented here) is not a polynomial in both $\tilde{k}$ and $\theta$. Using a mathematical software such as {\sc Mathematica} \cite{Mathematica}  we can easily visualise in the $(\theta,\tilde{k})$-plane the geometrical locus at which the discriminant vanishes, once all the physical parameters are fixed. In Figure~\ref{fig:p_discriminant_curve}, we set $d_1=0.3$ and $d_2=0.7$, as in Figure \ref{fig:combined_6_9}, and let $\gamma=0.1$. 
%What is striking here is that 
Two solution branches are obtained (solid and dashed lines in the figure), each one corresponding to a particular mode. In this specific example, we can show that the plots of $k(\theta)$ in Figure~\ref{fig:combined_6_9} obtained for mode-1 and -2 in panels $(a)$, $(b)$, with $\gamma=0.1$, can be recovered from Figure~\ref{fig:p_discriminant_curve} by a simple rescaling of each solution branch. More precisely, mode-1 solution is obtained by multiplying the solution branch in (red) dashed line by $s=c_0^+$, and mode-2 solution is obtained by multiplying the solution branch in (blue) solid line by $s=c_0^-$. 

\begin{figure}
\centering
\includegraphics[height=6.3cm]{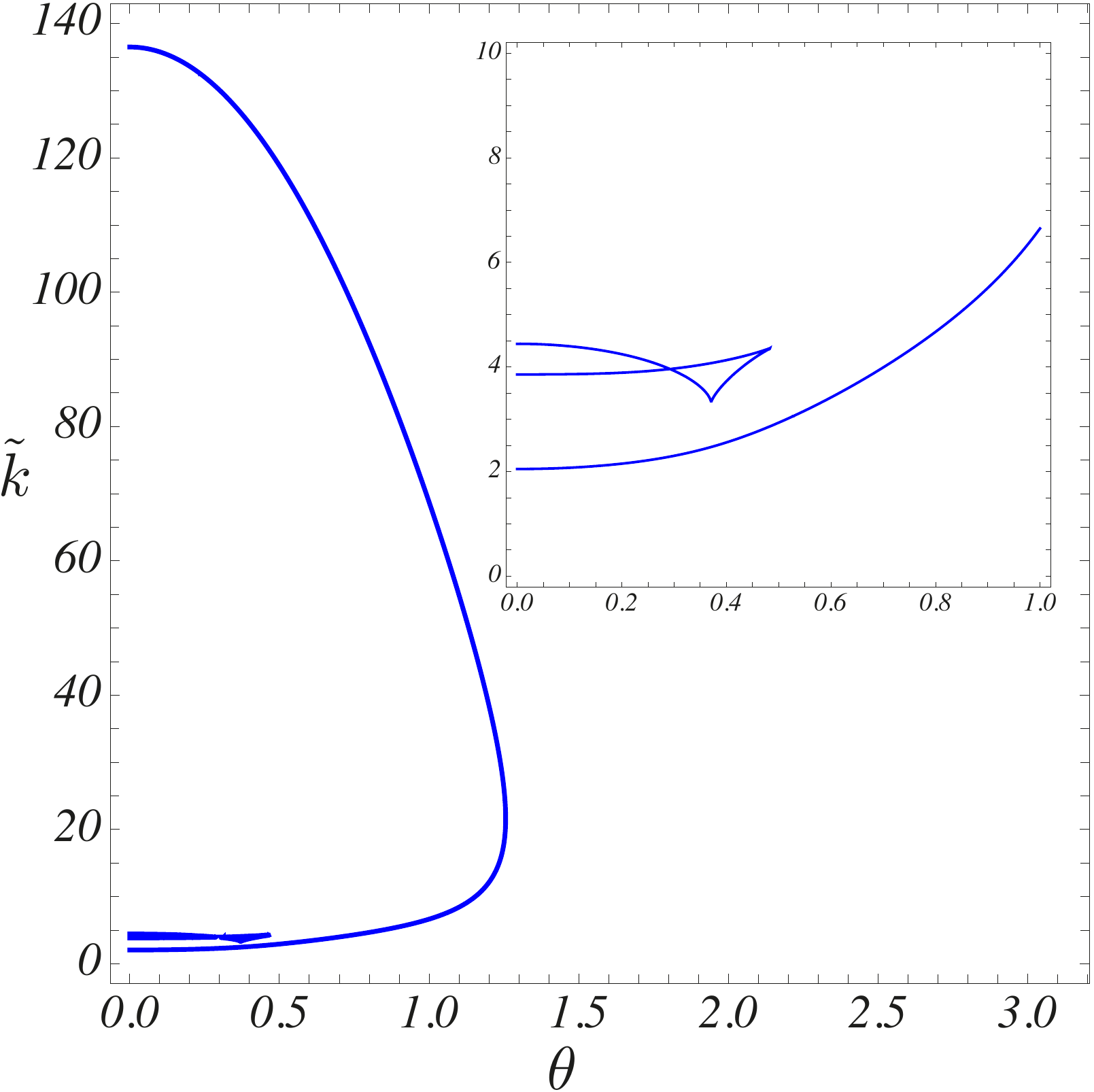}
\qquad 
\includegraphics[height=6.3cm]{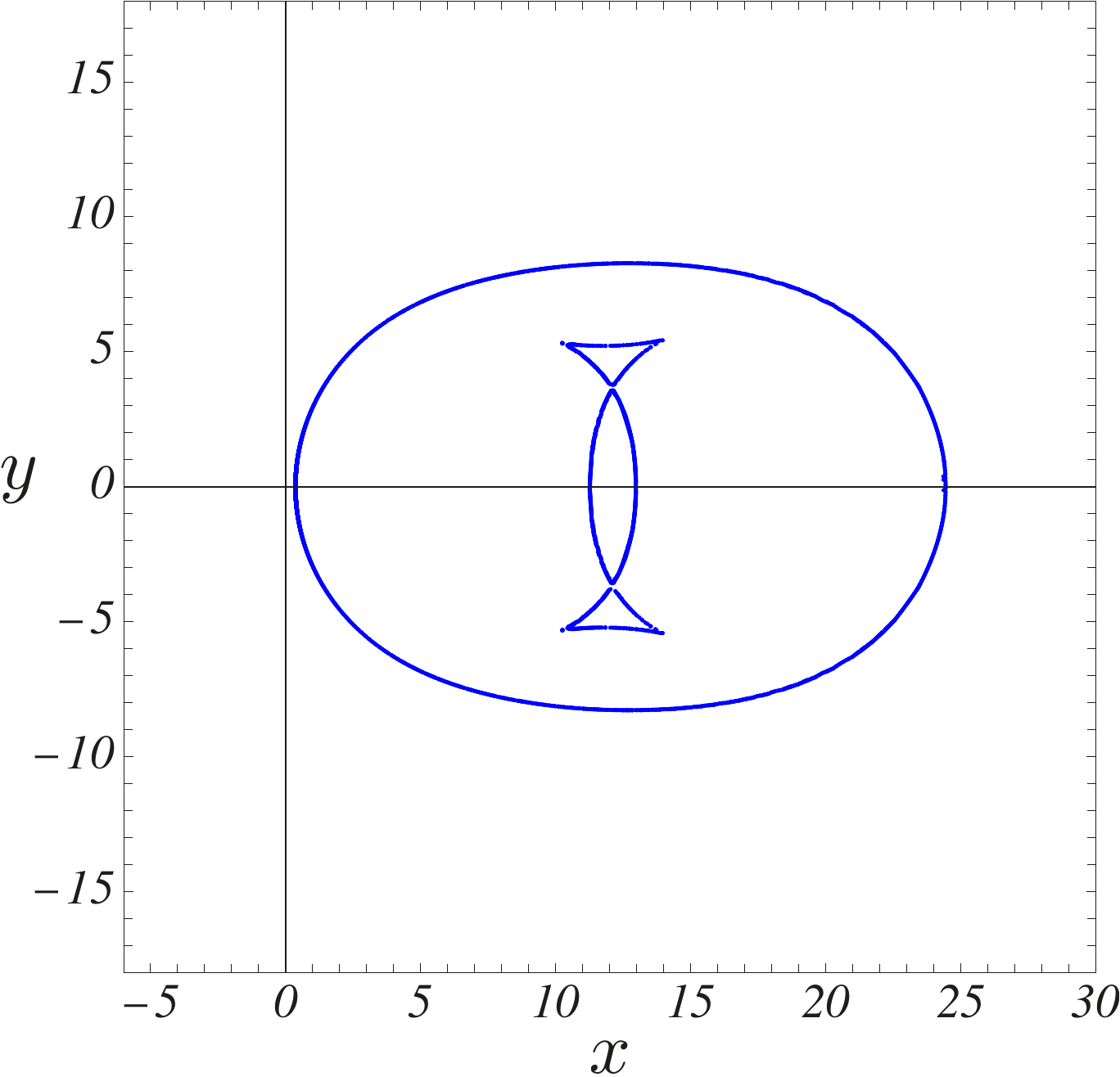}\\
{\footnotesize $(a)$ \hspace{5cm}  $(b)$}
\caption{The geometric locus on the $(\theta,\tilde{k})$-plane where the $p-$discriminant curve vanishes for  $\rho_1=1$, $\rho_2=1.1$, $\rho_3=1.2$ and $d_{1} = 0.3, d_{2} = 0.7$,  $\gamma=0.5$ is shown in panel $(a)$. The inset shows a close-up view where the swallowtail singularity is clearly visible in the solution branch corresponding to the mode-2. Wavefronts of ring waves described by $\tilde{k}(\theta)r = 50$ are shown in panel $(b)$.}
\label{fig:p_discriminant_curve_ex_2}
\end{figure}

With this method, we do not need to know $k(\theta)$ in a parametric form in order to plot the wavefronts. In particular, all the subtleties involved with the transition from the elliptic regime to the hyperbolic regime can be entirely avoided. The exact same procedure is taken regardless the value of $\gamma$ being small or large. However, the solution is implicit, and the parametric form obtained earlier is useful for computing the coefficients of the amplitude equation.  In Figure~\ref{fig:p_discriminant_curve_ex_2} we keep the same values of $d_1$, $d_2$ as in Figure~\ref{fig:p_discriminant_curve}, but  
increase the strength of the current to $\gamma=0.5$. The geometrical locus in the $(\theta,\tilde{k})$-plane where the $p-$discriminant curve vanishes is shown in panel $(a)$. A swallowtail singularity can be easily detected in one of the solution branches (see the inset in panel $(a)$). From Section~\ref{sec:wave_fronts} and Figures~\ref{fig:k_theta_hyperbolic}, \ref{fig:wavefronts_elliptic_hyperbolic} we know that this corresponds to the mode-2 solution. The other branch then corresponds to the mode-1 solution. The wavefronts of the ring waves described by $\tilde{k}(\theta)r = 50$ are shown in panel $(b)$. The outer ring, elongated in the direction of the shear flow, corresponds to the mode-1 ring waves, and the inner wavefront is akin to what was found in Figure~\ref{fig:wavefronts_elliptic_hyperbolic} in green short-dashed line for $\gamma=0.375$.

\begin{figure}
\centering
\includegraphics[height=6.3cm]{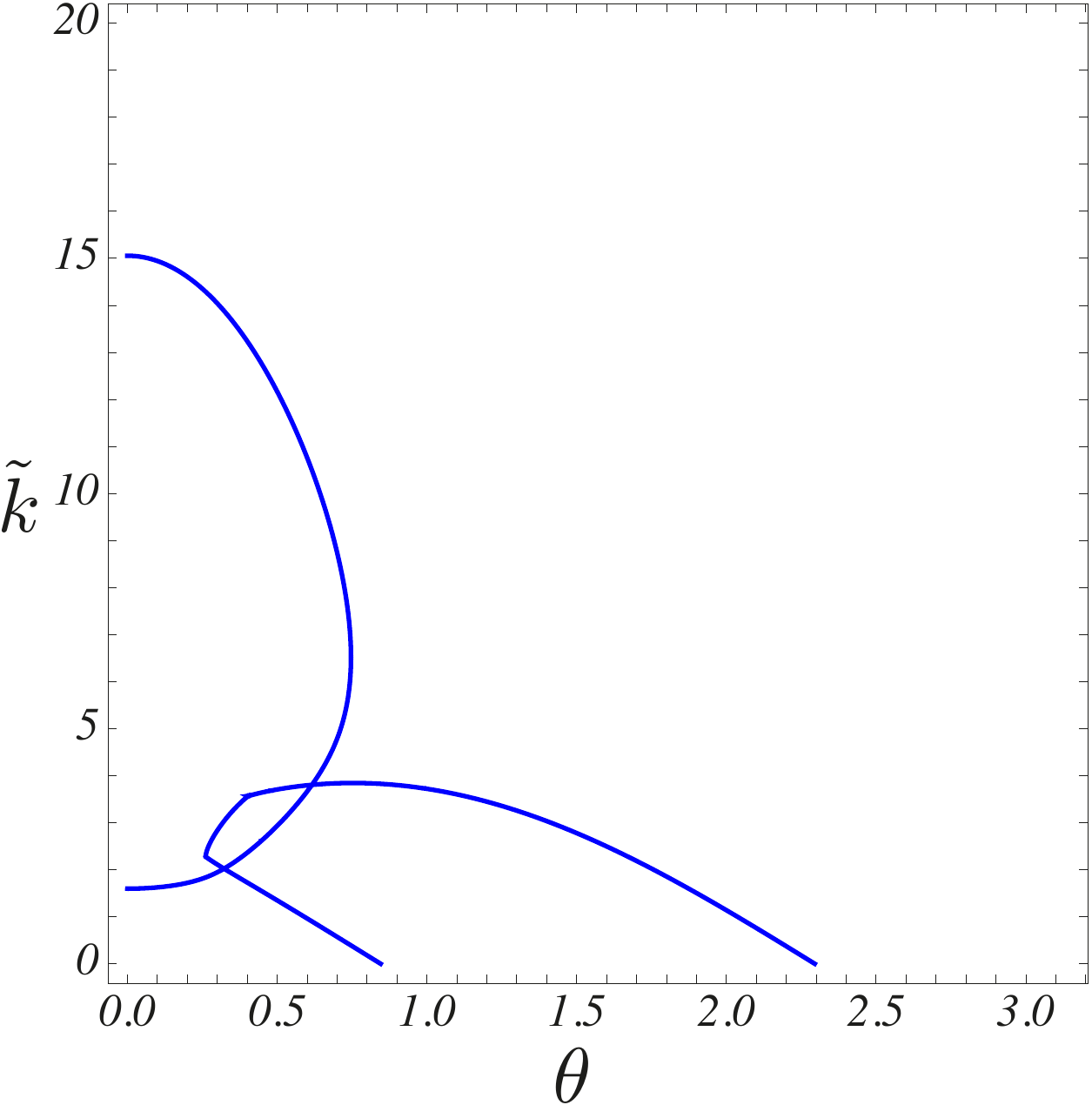}
\qquad 
\includegraphics[height=6.3cm]{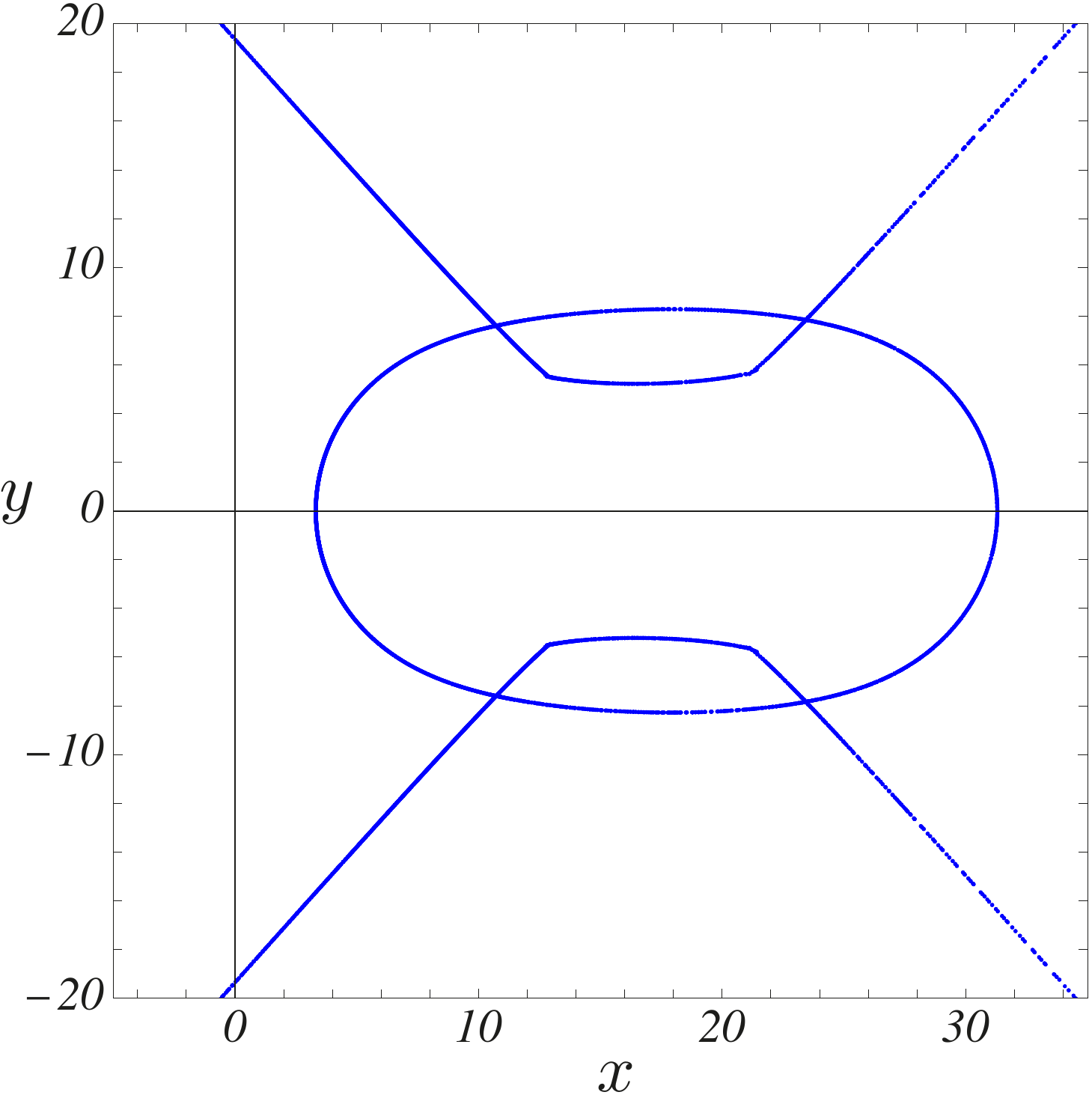}\\
{\footnotesize $(a)$ \hspace{5cm}  $(b)$}
\caption{The geometric locus on the $(\theta,\tilde{k})$-plane where the $p-$discriminant curve vanishes for  $\rho_1=1$, $\rho_2=1.1$, $\rho_3=1.2$ and $d_{1} = 0.3, d_{2} = 0.7$, $\gamma=0.7$ is shown in panel $(a)$. Wavefronts of waves described by $\tilde{k}(\theta)r = 50$ are shown in panel $(b)$.}
\label{fig:p_discriminant_curve_ex_3}
\end{figure}

\begin{figure}
\centering
\includegraphics[height=6.3cm]{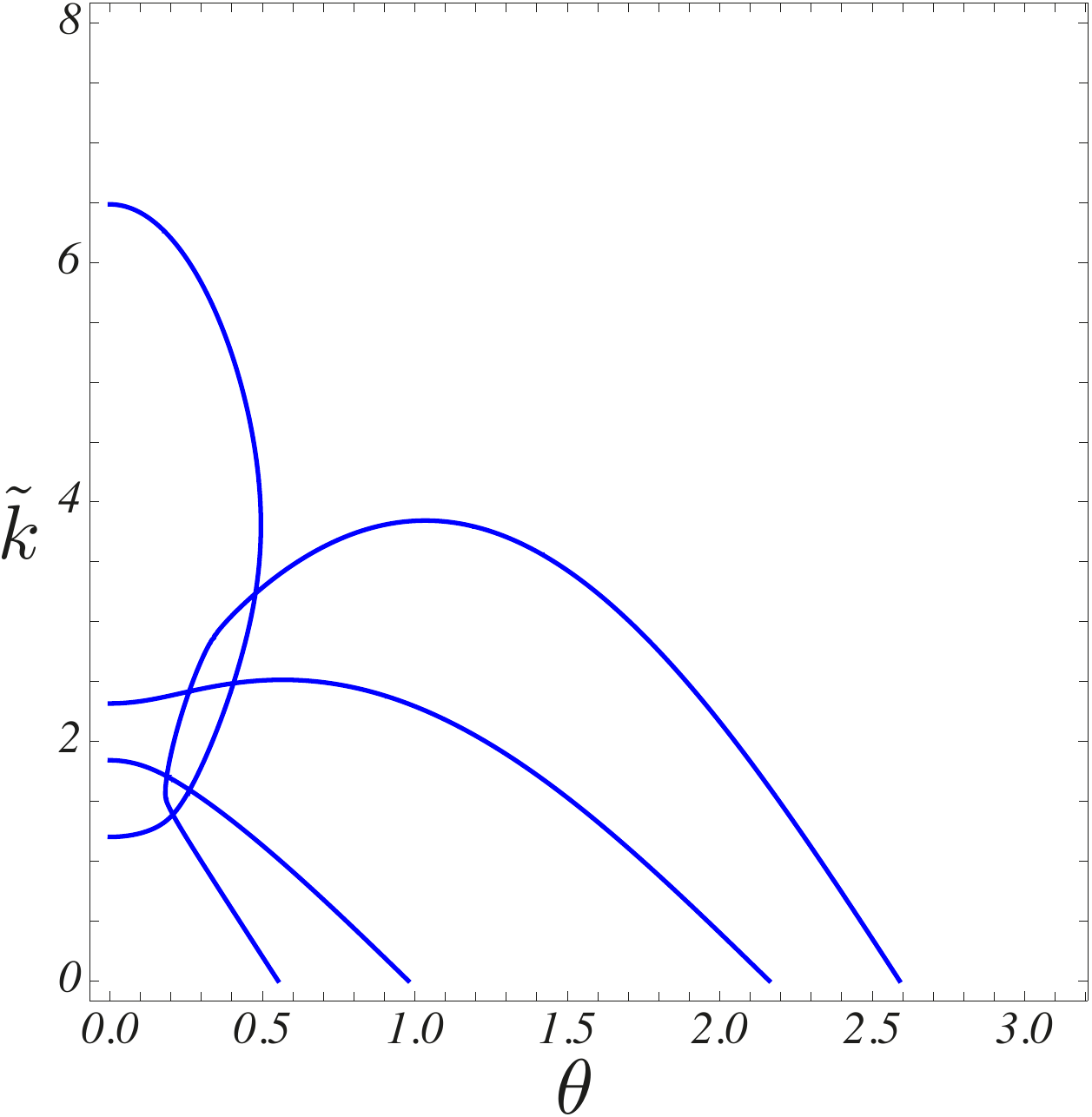}
\qquad 
\includegraphics[height=6.3cm]{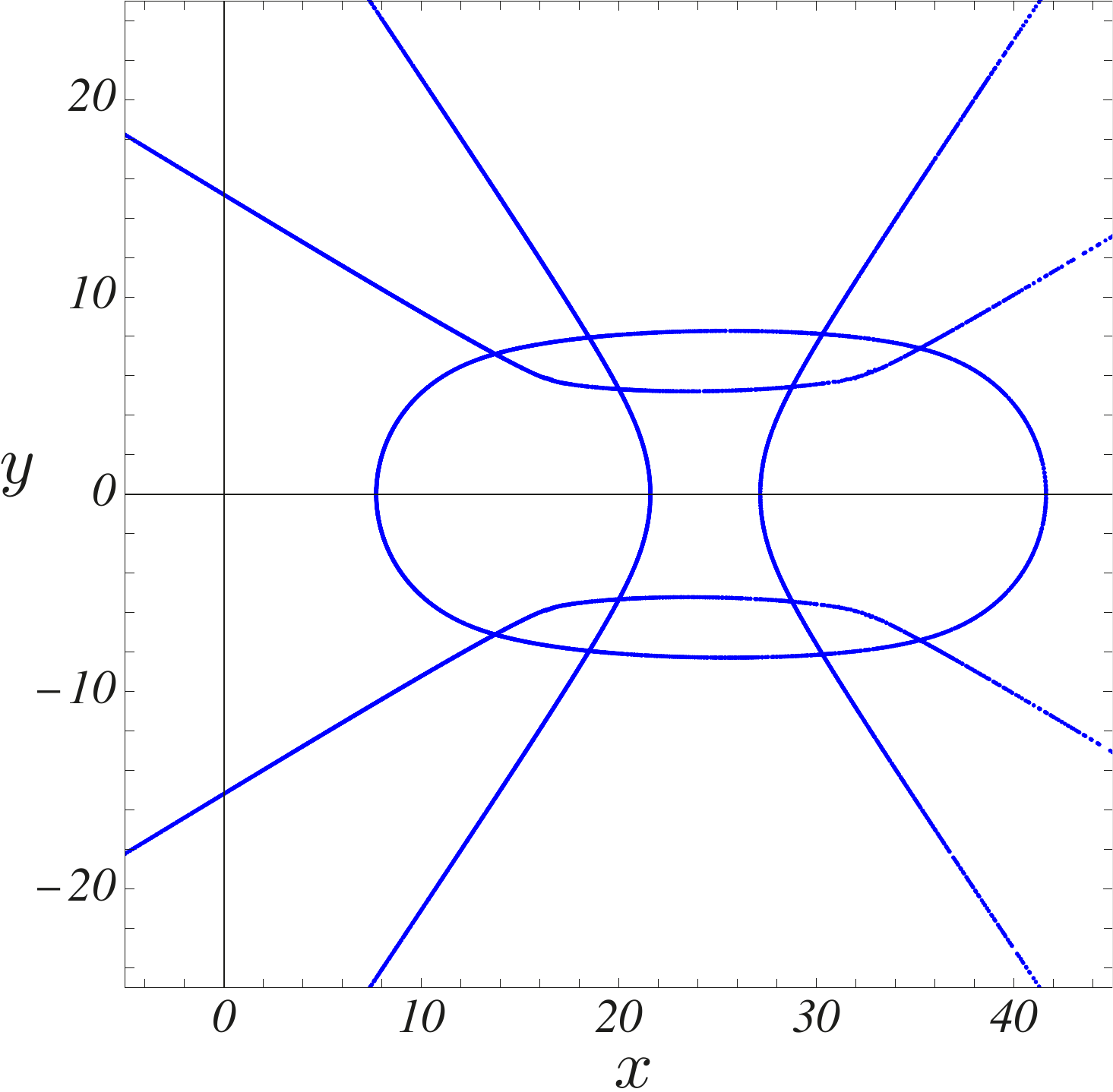}\\
{\footnotesize $(a)$ \hspace{5cm}  $(b)$}
\caption{The geometric locus on the $(\theta,\tilde{k})$-plane where the $p-$discriminant curve vanishes for  $\rho_1=1$, $\rho_2=1.1$, $\rho_3=1.2$ and $d_{1} = 0.3, d_{2} = 0.7$, and $\gamma=1$ is shown in panel $(a)$. Wavefronts of waves described by $\tilde{k}(\theta)r = 50$ are shown in panel $(b)$.}
\label{fig:p_discriminant_curve_ex_4}
\end{figure}

We can keep increasing the value of $\gamma$, namely beyond the value $\gamma^- \approx 0.524063$ at which the linear long-wave speeds of mode-2 in the flow direction cease to exist. The results obtained for $\gamma=0.7$ are shown in Figure~\ref{fig:p_discriminant_curve_ex_3}. Here, we observe that the corresponding wavefronts clearly include the mode-1 ring  elongated in the direction of the shear flow. Another component consists of two branches which converge to $\tilde k=0$ at certain values of $\theta$. The corresponding mode-2 wavefront does not cross the $x$-axis and extends to infinity. This is in agreement with the discussion in Section~\ref{sec:hyperbolic_regine} (cf. Figure~\ref{fig:mode2_restructure}). 
 A further example is shown in Figure~\ref{fig:p_discriminant_curve_ex_4} for an even larger value of $\gamma$. In this Figure, we set $\gamma=1$, for which linear long-wave speeds of mode-2 in the flow direction exist once again. A more detailed investigation of the wavefronts for such large values of $\gamma$ is left out of scope of our present study.

%Re-scale the problem so that $s$ disappears. Explain the method and add figure showing the two solutions at once}

%{\color{blue} Please, add the singular solution and wavefronts obtained with the old method and compare.}

%\newpage
\section{Numerical results for the amplitude equation}

%
%{\color{red}
%We then obtained the analytical expressions for the coefficients (\ref{mu}) of the amplitude equation (\ref{cKdV}) which are found using the solution of the modal equations. These coefficients are given in Appendix A. They are used and discussed in Section 5. 
%There will be no critical levels, i.e. $F(z, \theta) \ne 0$, provided $\gamma z < s < \frac{s}{k(0)}$, see \cite{HKG} for details. Continuation of the constructed solutions for $k(\theta)$ for stronger currents is discussed in Section 5. 
%}

\noindent In this section, we discuss the numerical solutions of the amplitude equation for the propagation of the two interfacial ring modes in the specific case of a three-layer linear shear current.
%in a three-layer fluid over the linear shear current, 
% (i.e. the shear flow with the constant vorticity $\gamma$). 
%\vspace{0.3cm}
%\noindent
%\noindent   
%Both modes will be considered and .
%\begin{eqnarray}
%\mu_1A_R+\mu_2AA_{\xi}+\mu_3A_{\xi\xi\xi}+\mu_4\frac{A}{R}+\mu_5\frac{A_\theta}{R}=0,
%\end{eqnarray}
%where $A=A(R,\xi,\theta)$. 
Here, the coefficients $\mu_1,\ldots, \mu_5$ of (\ref{cKdV}) are given in Appendix A. 
%We note that the cKdV equation is recovered when $\mu_5=0$. 
It is known (see \cite{KZ}) that for a homogeneous free surface flow with general shear current, $\mu_5=0$ and the cKdV equation is recovered for each value of $\theta$ (although we note that $\mu_5 \ne 0$ for surface waves in a two-layer fluid  \cite{KZ1}). For interfacial waves in our three-layer rigid-lid configuration with linear shear current, we have strong numerical evidence that $\mu_5 = 0$ as well. This justifies solving the $1+1$-dimensional cKdV-type amplitude equation with the coefficients parametrised by $\theta$, instead of (\ref{cKdV}):
\begin{eqnarray}
\mu_1A_R+\mu_2AA_{\xi}+\mu_3A_{\xi\xi\xi}+\mu_4\frac{A}{R}=0.
\label{cKdV1}
\end{eqnarray}
%for each fixed value of $\theta$.
%$\mu_5$ is also zero. 
 %analytical coefficients (\ref{mu})  of this amplitude equation for the three-layer stratification and a linear current are given 
% in the Appendix A. 
%\vspace{0.3cm}
%\noindent
%We note that since the computation of the coefficients of the weakly-nonlinear equation indicates that $\mu_5 = 0$ in the case under study, we are solving the $1+1$-dimensional cKdV-type amplitude equation 
%\begin{eqnarray}
%\mu_1A_R+\mu_2AA_{\xi}+\mu_3A_{\xi\xi\xi}+\mu_4\frac{A}{R}=0
%\label{cKdV1}
%\end{eqnarray}
%for each fixed value of $\theta$. 
For our numerical runs we use an extension of the efficient  implicit finite-difference scheme developed in \cite{FM} (see \cite{KZ1} and Appendix C, where we fixed some typos).

To describe the initial evolution of the weakly-nonlinear waves, we use the two-dimensional linear wave equation for each mode, assuming that a shear flow is initially negligible (see \cite{KZ1} and references therein): 
\begin{eqnarray}
A_{tt}-s^2 (A_{xx}+A_{yy})=0,
\end{eqnarray}
which has the following exact solution 
\begin{eqnarray}
A(x,y,t)=Q\,\mathrm{Re}\left(\frac{1+\mathrm{i}st/\nu}{\bigl[(1+\mathrm{i}st/\nu)^2+(x^2+y^2)/\nu^2\bigr]^{3/2}}\right),
\end{eqnarray}
\noindent describing waves from a localised initial condition at $t=0$. Here, $s$ is the wave speed in the absence of a shear flow, and $Q$ and $\nu$ are arbitrary constants (see \cite{DS} and references therein). 
%The value of $Q$ corresponds to the amplitude of the initial condition 

%\vspace{0.3cm}
%\noindent
%The variables $\xi$, $R$ and $\theta$ in the amplitude equation (\ref{cKdV})  are related to the original variables $t$, $r$ and $\theta$ (where $r$ and $\theta$ are the polar coordinates so that $x=r\cos\theta$ and $y=r\sin\theta$) via the following expressions:
%\begin{eqnarray}
%\xi=rk(\theta)-st, \qquad R=\varepsilon r k(\theta),\qquad \theta=\theta.
%\end{eqnarray}
Since
%\begin{eqnarray}
$t=\frac{R/\varepsilon-\xi}{s}, \ r=\frac{R}{\varepsilon k(\theta)},\ \theta=\theta,$
%\end{eqnarray}
then, assuming that 
%\begin{eqnarray}
$\xi\in[\xi_{\min},\,\xi_{\max}],\  R\in[R_0,\,R_{\max}],\  \theta\in[0,\,2\pi], $
%\end{eqnarray}
we have
%\begin{eqnarray}
$$t\in\left[\frac{R_0/\varepsilon-\xi_{\max}}{s},\,\frac{R_{\max}/\varepsilon-\xi_{\min}}{s}\right],$$
%\end{eqnarray}
and, for each $\theta\in[0,\,2\pi]$, the range of values of $r$ is
%\begin{eqnarray}
$$r\in\left[\frac{R_0}{\varepsilon k(\theta)},\,\frac{R_{\max}}{\varepsilon k(\theta)}\right].$$
%\end{eqnarray}
To specify the initial condition for the weakly nonlinear equation at $R=R_0$, we need the data from the linear wave equation at 
%\begin{eqnarray}
$r=\frac{R_0}{\varepsilon k(\theta)},$
%\end{eqnarray}
for each $\theta\in[0,\,2\pi]$ and for the time interval
%\begin{eqnarray}
$t\in\left[\frac{R_0/\varepsilon-\xi_{\max}}{s},\,\frac{R_0/\varepsilon-\xi_{\min}}{s}\right].$
%\end{eqnarray}
The initial condition for the weakly-nonlinear equation written in the $(R,\xi,\theta)$ coordinates takes the form 
\begin{eqnarray}
A(R_0,\xi,\theta)=Q\,\mathrm{Re}\left(\frac{1+\mathrm{i}\left(\frac{R_0}{\varepsilon\nu}-\frac{\xi}{\nu}\right)}{\biggl[\left(1+\mathrm{i}\left(\frac{R_0}{\varepsilon\nu}-\frac{\xi}{\nu}\right)\right)^2+\frac{R_0^2}{\varepsilon^2\nu^2k^2(\theta)}\biggr]^{3/2}}\right).
\end{eqnarray}

%\vspace{0.3cm}
%\noindent
%\noindent   
For the computational examples presented in this section, we assume that $\varepsilon=0.02$ and $\nu=0.5$ and use $Q=\pm5$. The same initial condition is set for both modes, and two configurations are considered in the following subsections: a symmetric configuration with  $d_1 = 0.3, d_2 = 0.7$, and an asymmetric configuration with $d_1 = 0.2, d_2 = 0.9$.

\begin{figure}%[!p!t]
      \centering
      \includegraphics[width=0.3\linewidth]{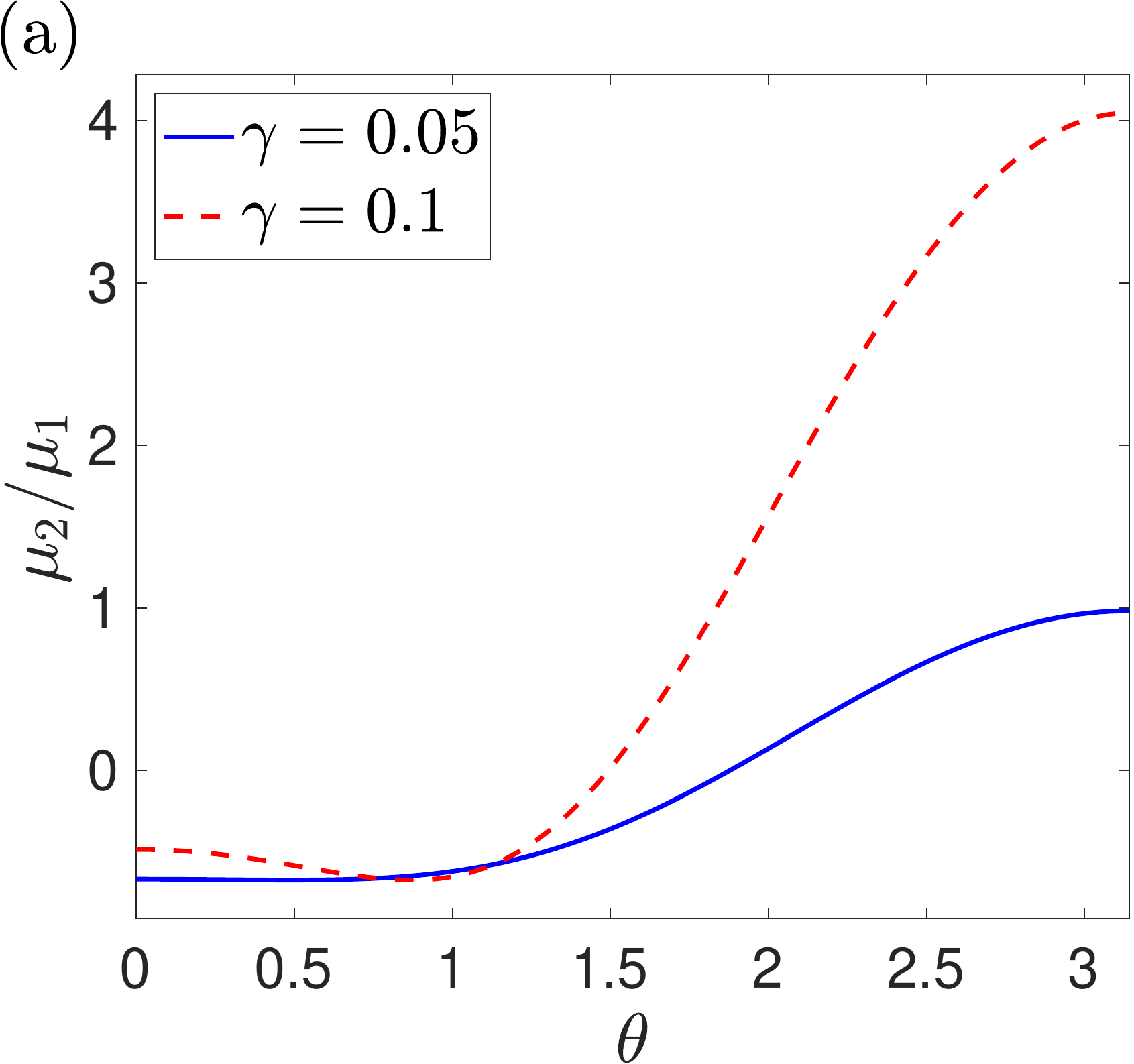} \hfill
      \includegraphics[width=0.3\linewidth]{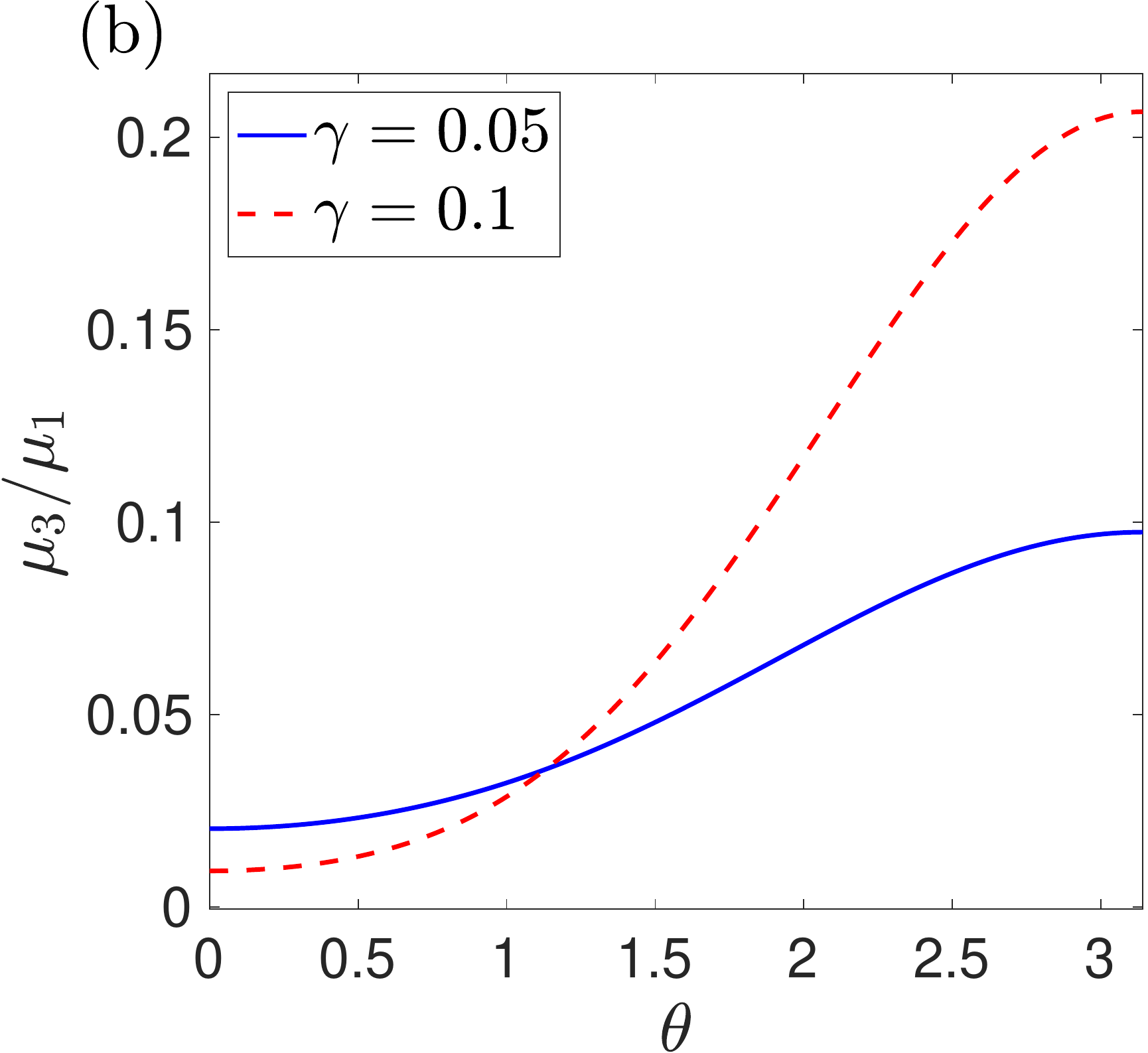} \hfill
      \includegraphics[width=0.3\linewidth]{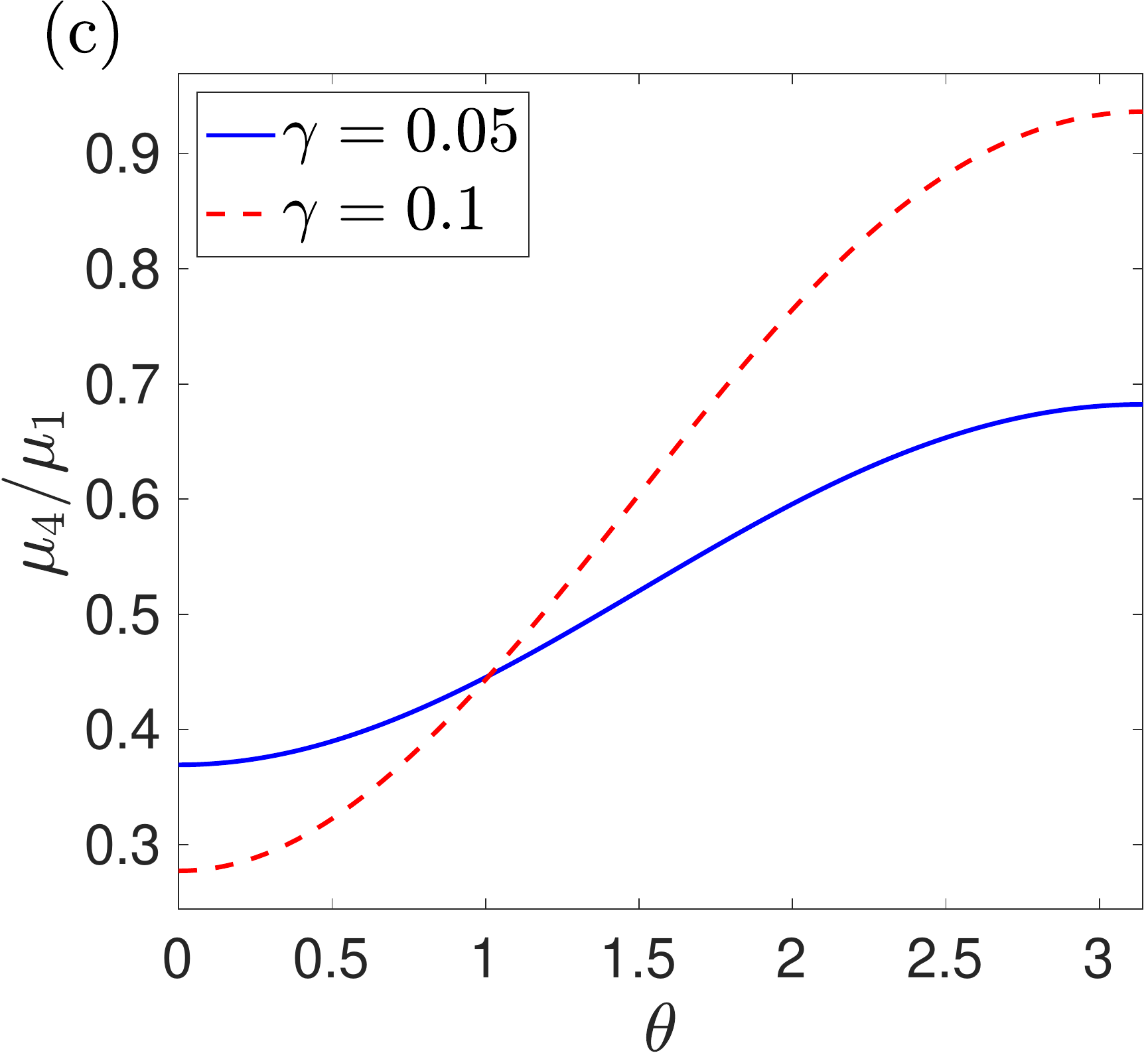} 
\vspace{-0.2cm}
       \caption{Coefficients $\mu_i(\theta)/ \mu_1(\theta), i = 2, 3, 4,$  for the mode-1 ring waves for  $\rho_1=1, \rho_2=1.1, \rho_3 = 1.2$ and $d_{1} = 0.3, d_{2} = 0.7$, when $\gamma=0.05$ (blue solid lines) and $\gamma=0.1$ (red dashed lines).}
       \label{fig:PicS56_005_mod1}
\vspace{0.5cm}       
      \includegraphics[width=0.3\linewidth]{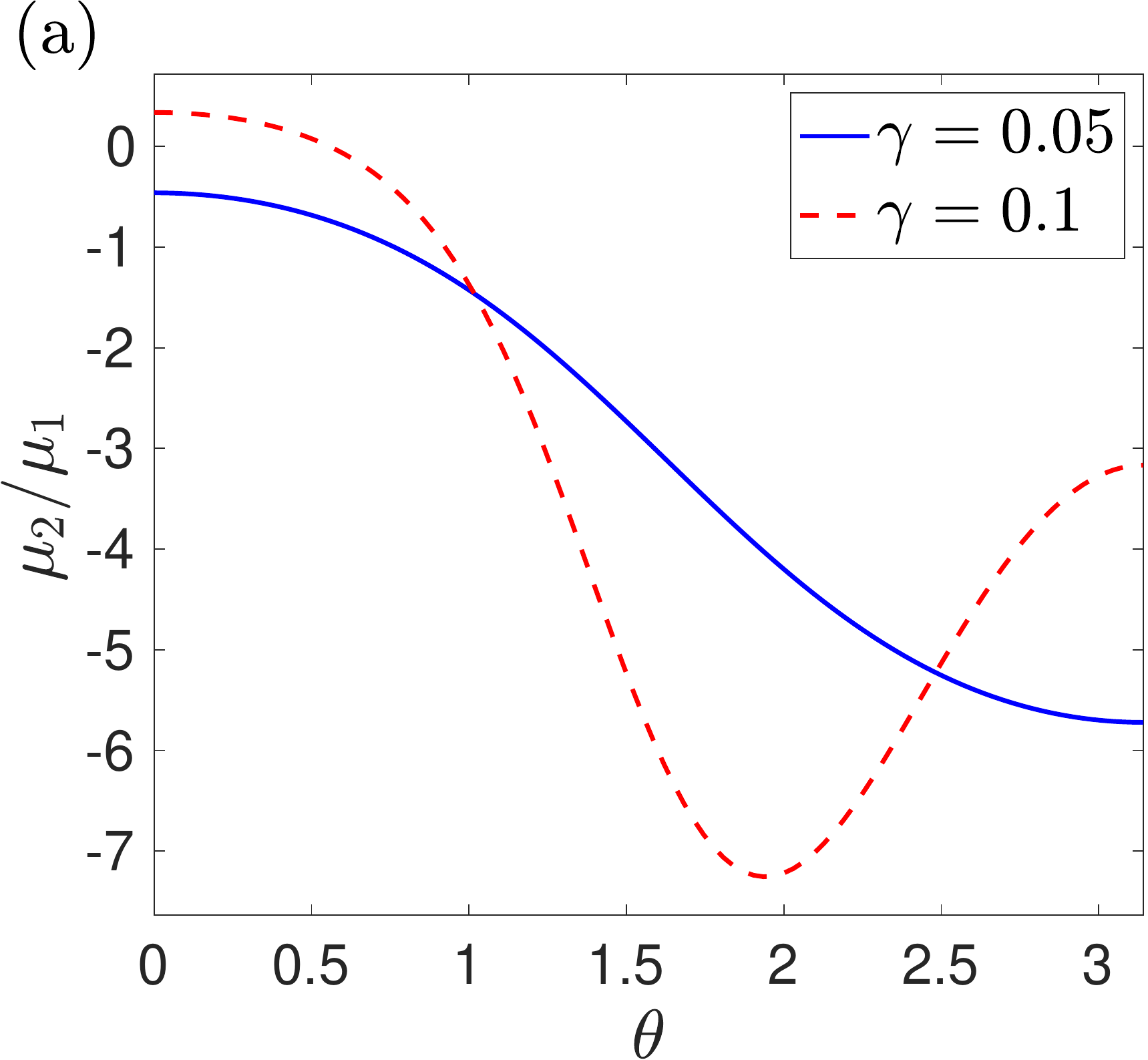} \hfill
      \includegraphics[width=0.3\linewidth]{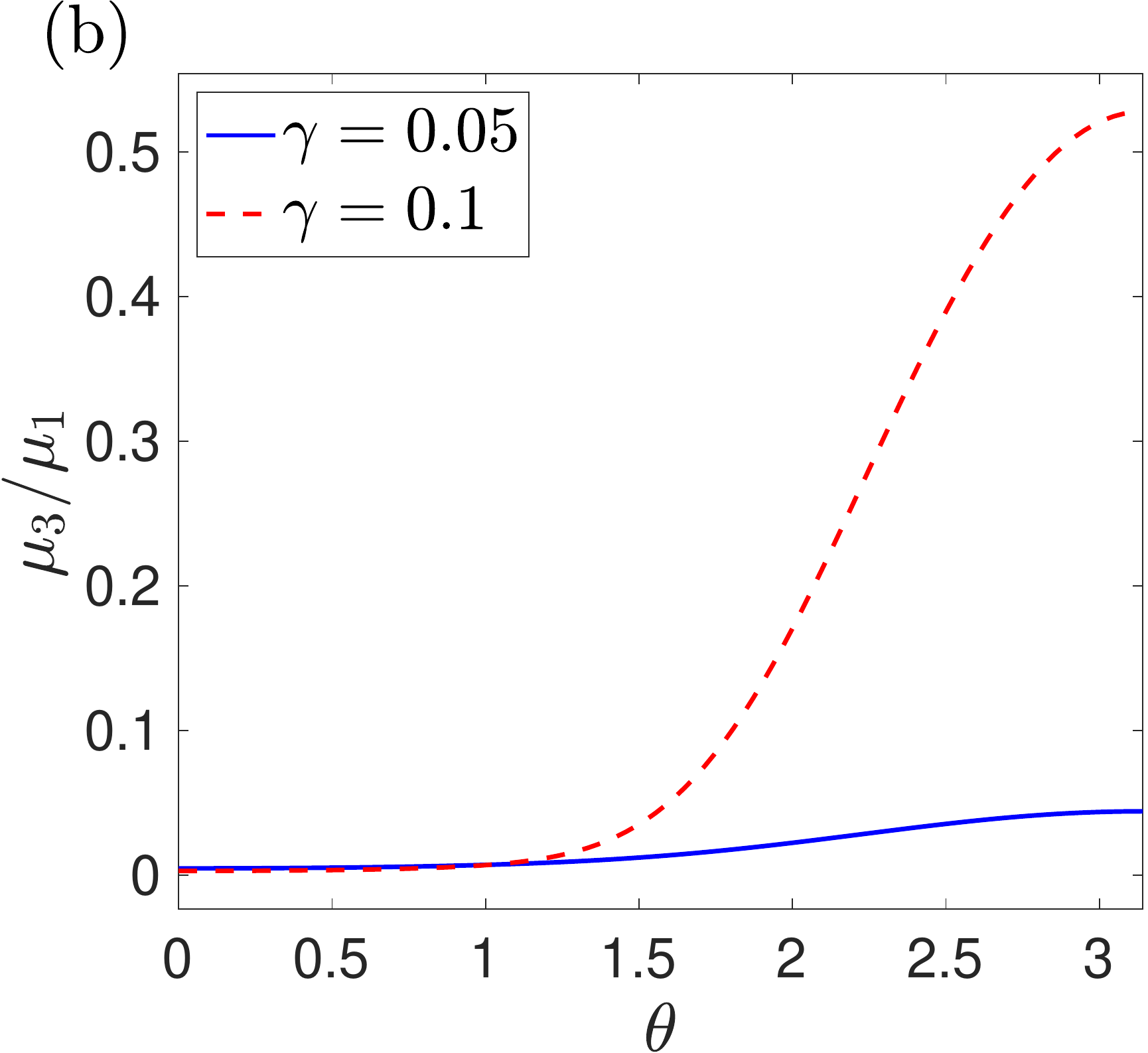} \hfill
      \includegraphics[width=0.3\linewidth]{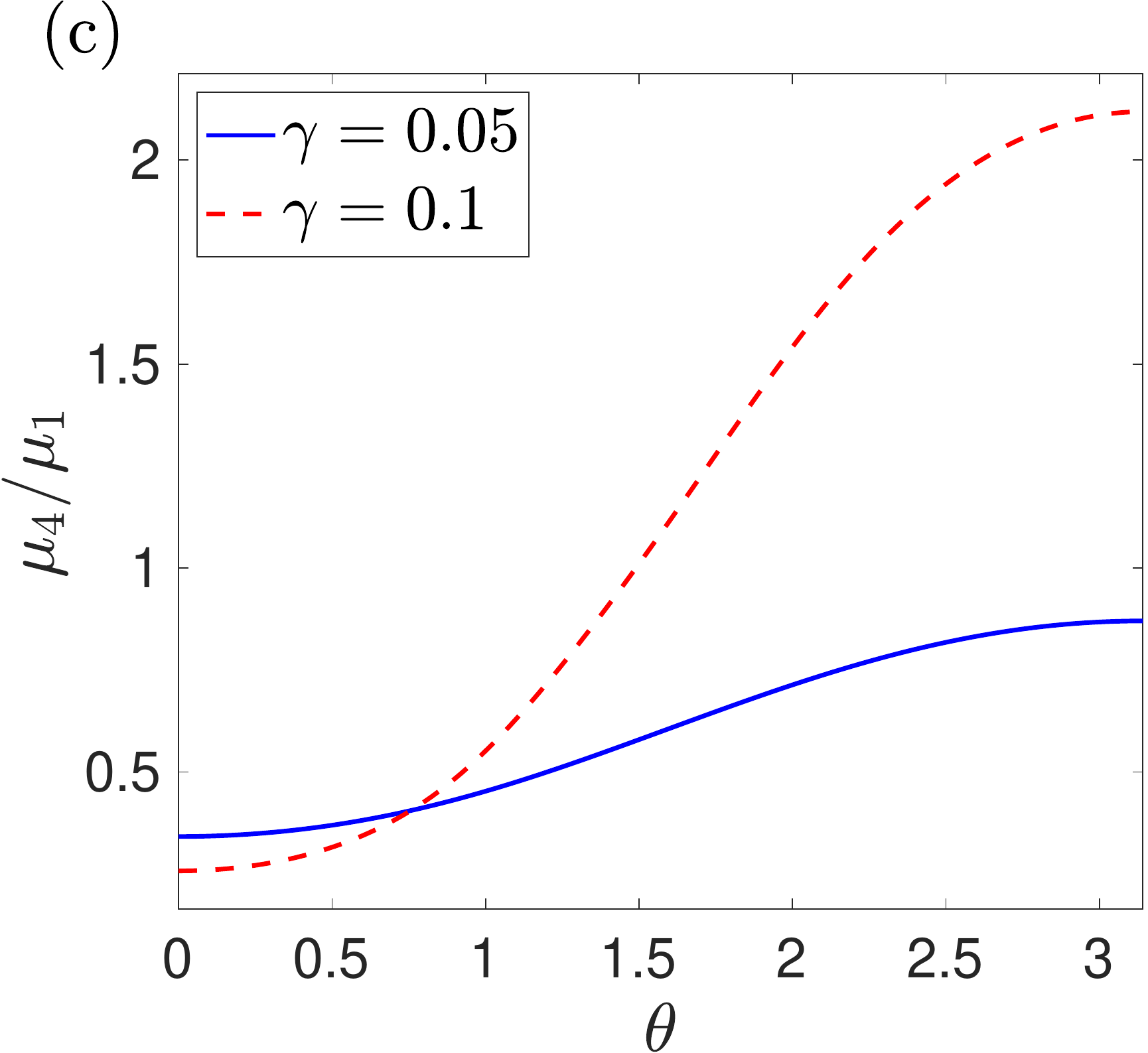} 
\vspace{-0.2cm}
       \caption{Coefficients $\mu_i(\theta)/ \mu_1(\theta), i = 2, 3, 4,$  for the mode-2 ring waves for  $\rho_1=1, \rho_2=1.1, \rho_3 = 1.2$ and $d_{1} = 0.3, d_{2} = 0.7$, when $\gamma=0.05$ (blue solid lines) and $\gamma=0.1$ (red dashed lines).}
       \label{fig:PicS56_005_mod2}
\vspace{0.5cm}
%\end{figure}
%
%\begin{figure}%[t!]
\end{figure}

\subsection{Symmetric configuration with $d_1=0.3$, $d_2=0.7$}

To solve (\ref{cKdV1}) we first need to compute the coefficients $\displaystyle \mu_i/\mu_1,\ i = {2, 3, 4}$. 
%Two values of $\gamma$ will be considered here: $\gamma = 0.05$ and $\gamma = 0.1$. With this choice of parameters, 
The behaviour of the coefficients is given in Figures \ref{fig:PicS56_005_mod1} and \ref{fig:PicS56_005_mod2} for the first and second mode, respectively, for a weaker current $\gamma = 0.05$ and a stronger current $\gamma = 0.1$.
 
%are shown for the first  and second interfacial ring modes in Figures \ref{fig:PicS56_005_mod1} and \ref{fig:PicS56_005_mod2}, respectively,  for $\gamma = 0.05$ and $\gamma = 0.1$. 
We note that for the first mode the nonlinearity coefficient $\mu_2/\mu_1$  in the downstream direction is small, while it is large in the upstream direction. The increase in the strength of the current results in a significant increase of all three coefficients in the upstream direction, and the strongest effect is on the nonlinearity coefficient.  For the second mode, although a similar qualitative behaviour is found for the dispersion coefficient $\mu_3/\mu_1$ and cylindrical divergence coefficient $\mu_4/\mu_1$, significant differences can be perceived for the nonlinearity coefficient $\mu_2/\mu_1$. More precisely, when the current is weak, $\mu_2/\mu_1$ is negative and monotonically decreasing for all $\theta \in [0, 2 \pi]$. However, for stronger currents it may change sign and lose monotonicity. As we will see, many features of the numerical solutions can be understood from the behaviour of these coefficients. 

%\begin{figure}%[t!]
%\begin{minipage}{.55\textwidth}
%\begin{center}
%\vspace{5mm}
%\includegraphics[width=\linewidth]{FIG/figure9a}
%\vspace{0.3cm}
%\end{center}
%\end{minipage}%\hspace{-0.04\textwidth}
%\begin{minipage}{.45\textwidth}
%\begin{center}
%\includegraphics[width=0.95\linewidth]{FIG/figure9b}\\[0.2cm]
%\includegraphics[width=0.95\linewidth]{FIG/figure9c}
%\end{center}
%\end{minipage}
%\vspace{-1cm}
%\caption{Nonlinear internal waves in a three-layer fluid over a shear flow with the vorticity $\gamma = 0.05$ ($Q=5$) for the first interfacial ring mode. The undisturbed interfaces are at $d_{1}= 0.3$ and $d_{2}= 0.7$. Panel (a) shows the wavefronts, with the colour scheme corresponding to the wave amplitude, while panels (b) and (c) show the wave profiles in the directions $y=0$ (i.e. the $x$ axis) and $x = 0$ (i.e. the $y$ axis), respectively, at $t=0$, $t=75$, $t=150$, $t=255$ and $t=300$.}
%       \label{fig:PicS56_6_26}
%\end{figure}
%
\begin{figure}%[t!]
\begin{minipage}{.55\textwidth}
\begin{center}
\vspace{5mm}
\includegraphics[width=\linewidth]{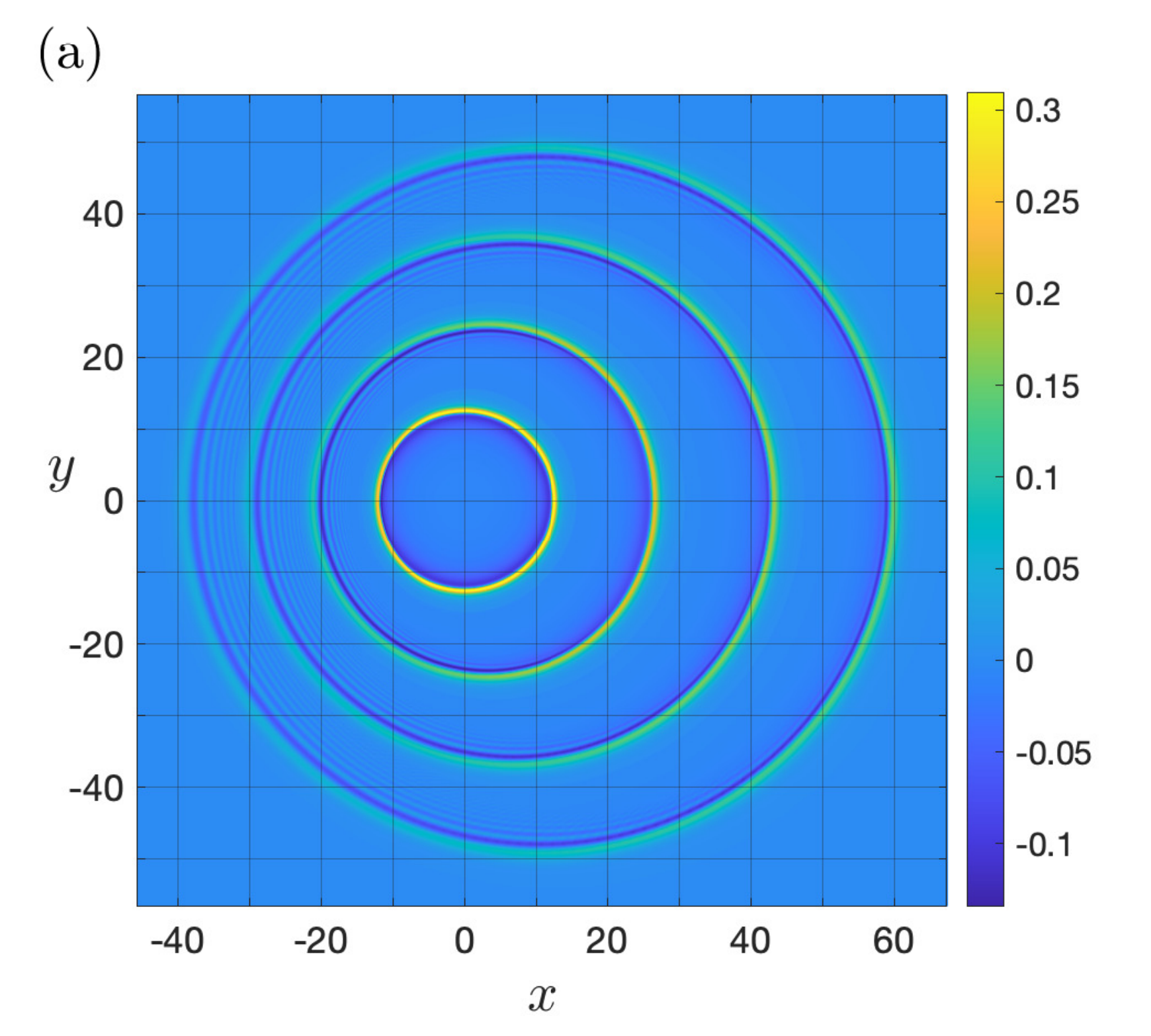}
\vspace{0.3cm}
\end{center}
\end{minipage}%\hspace{-0.04\textwidth}
\begin{minipage}{.45\textwidth}
\begin{center}
\includegraphics[width=0.95\linewidth]{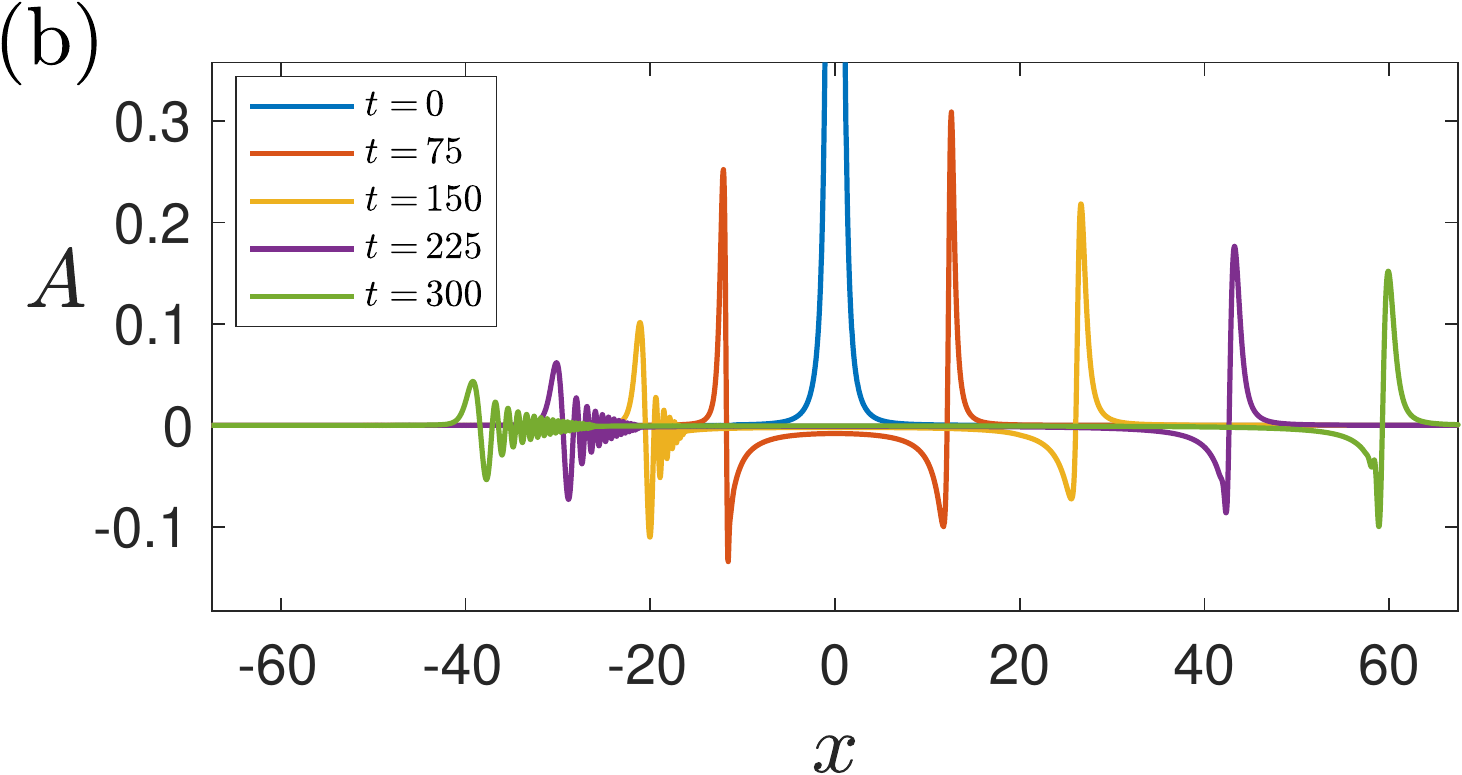}\\[0.2cm]
\includegraphics[width=0.95\linewidth]{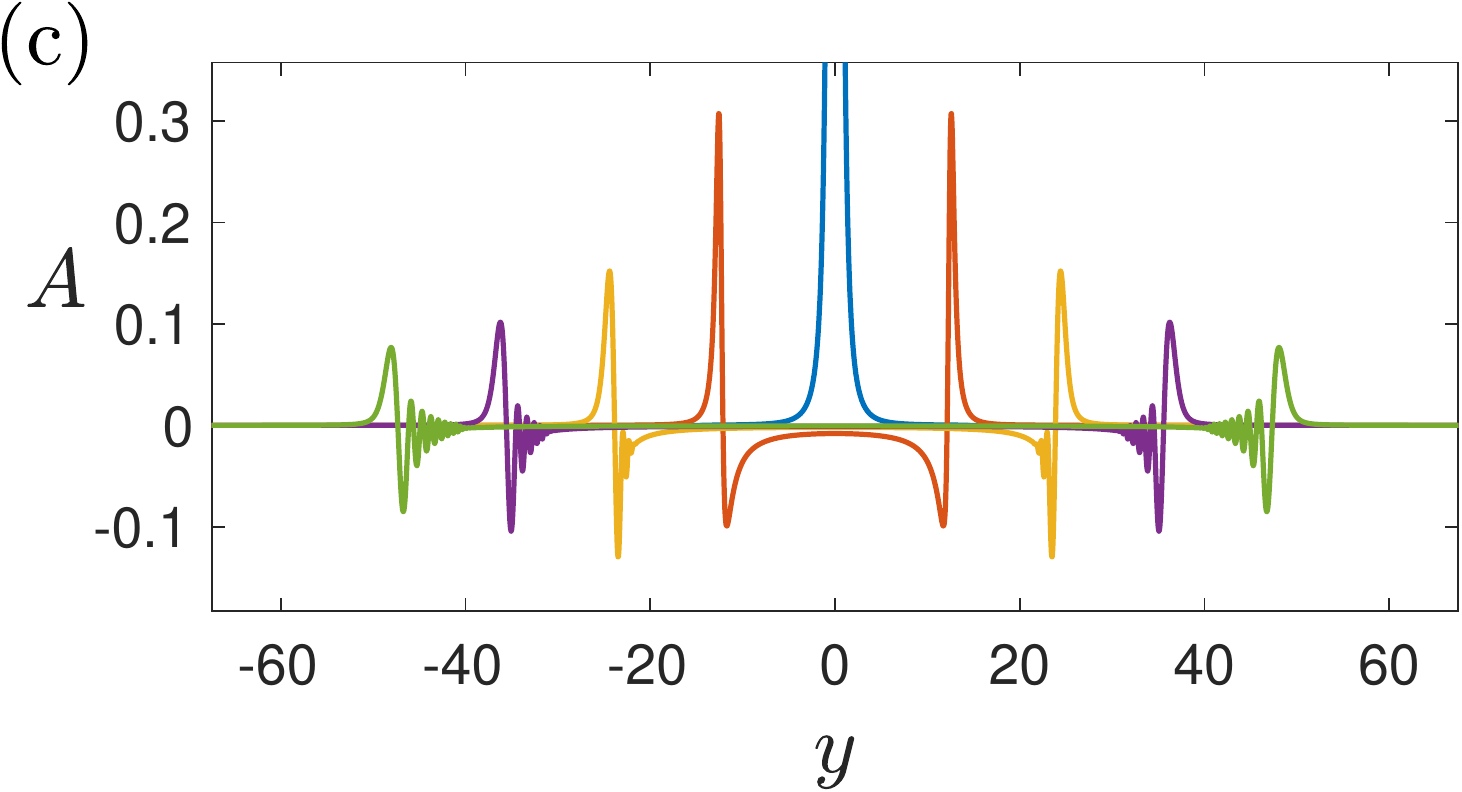}
\end{center}
\end{minipage}
\vspace{-1cm}
\caption{Mode-1 nonlinear ring waves for $\gamma = 0.1$, $Q = 5$, when  $\rho_1=1, \rho_2=1.1, \rho_3 = 1.2$ and $d_{1} = 0.3, d_{2} = 0.7$. Panel (a) shows the wavefronts, with the colour scheme corresponding to the wave amplitude, while panels (b) and (c) show the wave profiles in the directions $y=0$ (i.e. the $x$ axis) and $x = 0$ (i.e. the $y$ axis), respectively, at $t=0$, $t=75$, $t=150$, $t=255$ and $t=300$.}
       \label{fig:PicS56_6_30}
%\end{figure}
%
%\begin{figure}%[t!]
\begin{minipage}{.55\textwidth}
\begin{center}
\vspace{5mm}
\includegraphics[width=\linewidth]{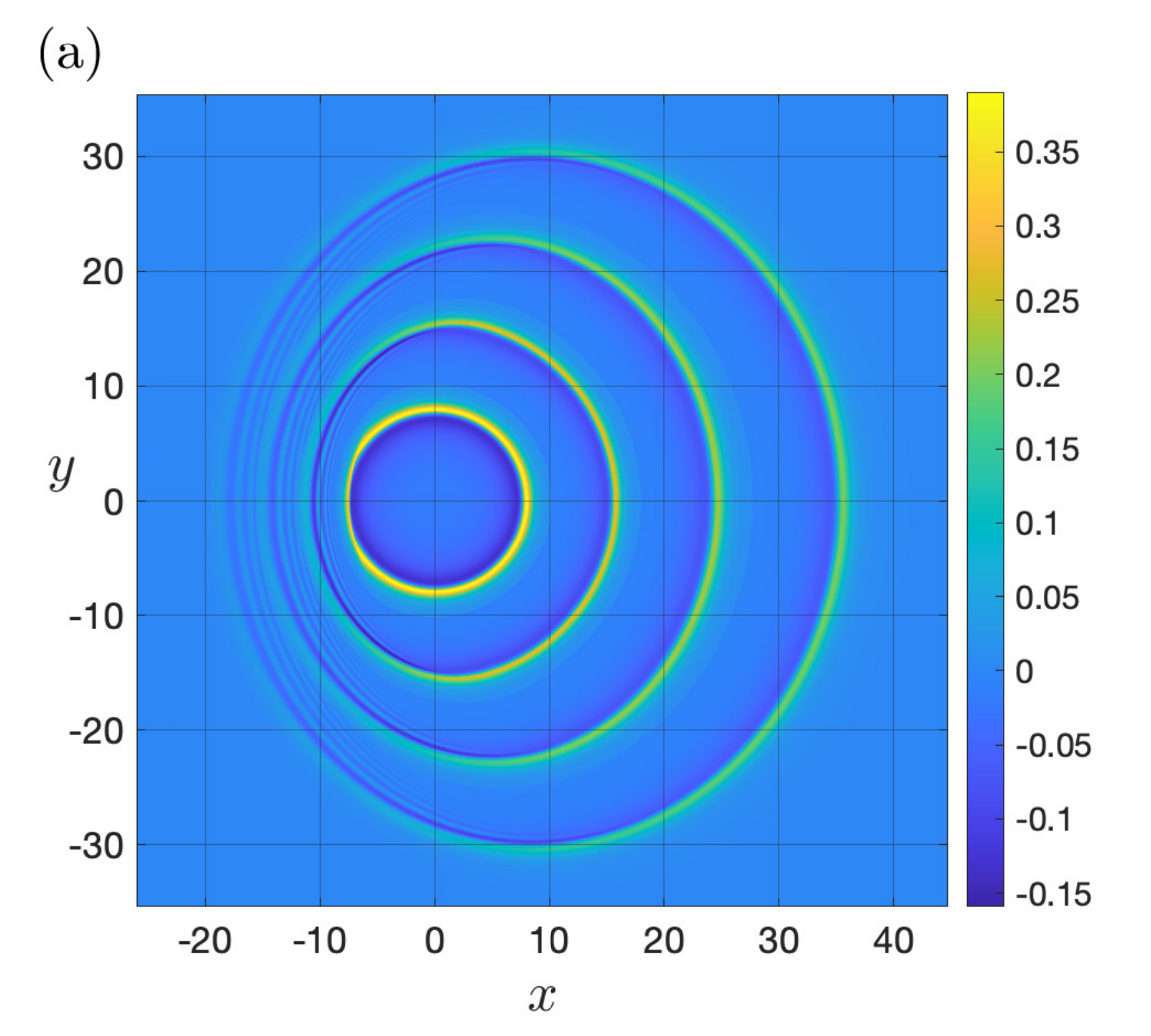}
\vspace{0.3cm}
\end{center}
\end{minipage}
\begin{minipage}{.45\textwidth}
\begin{center}
\includegraphics[width=0.95\linewidth]{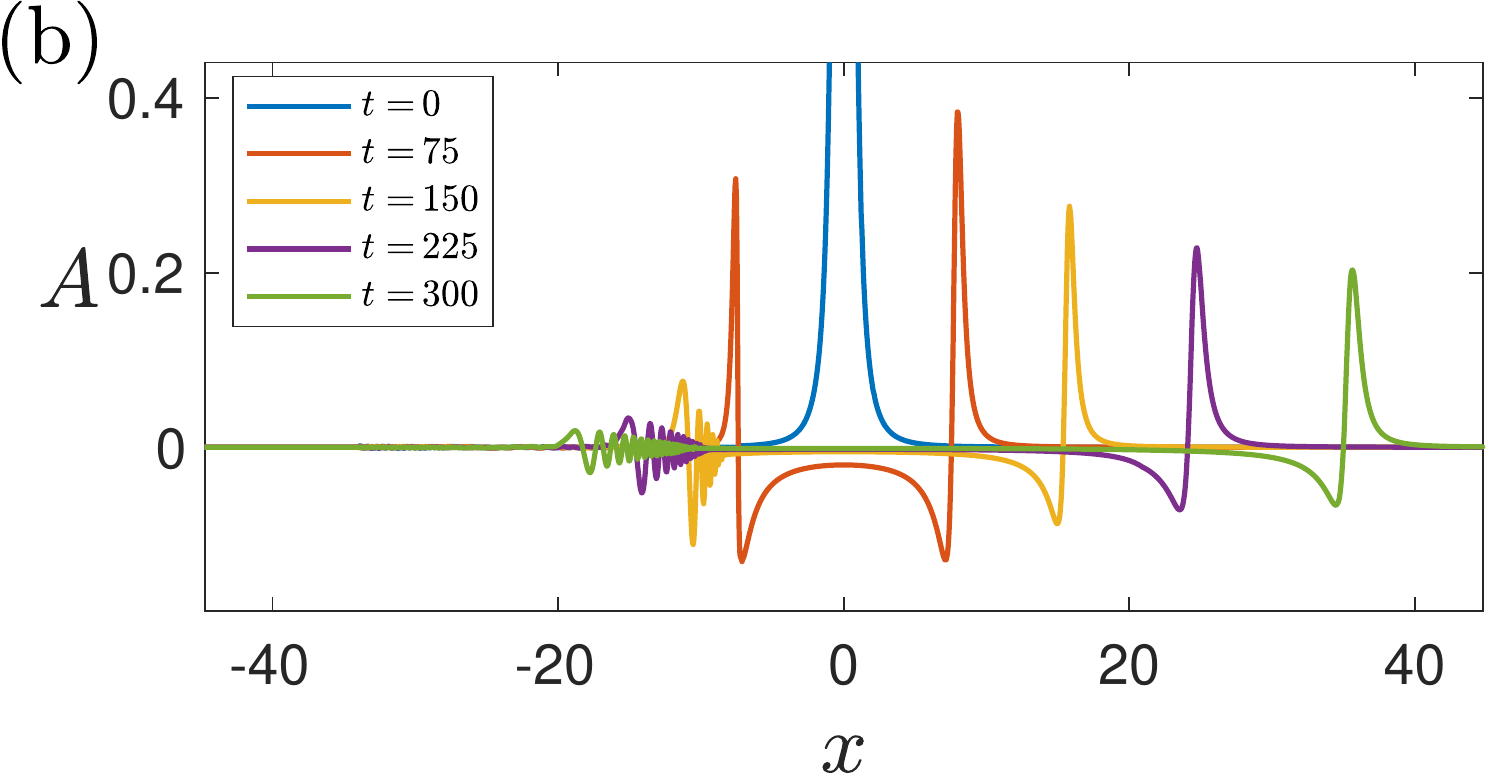}\\[0.2cm]
\includegraphics[width=0.95\linewidth]{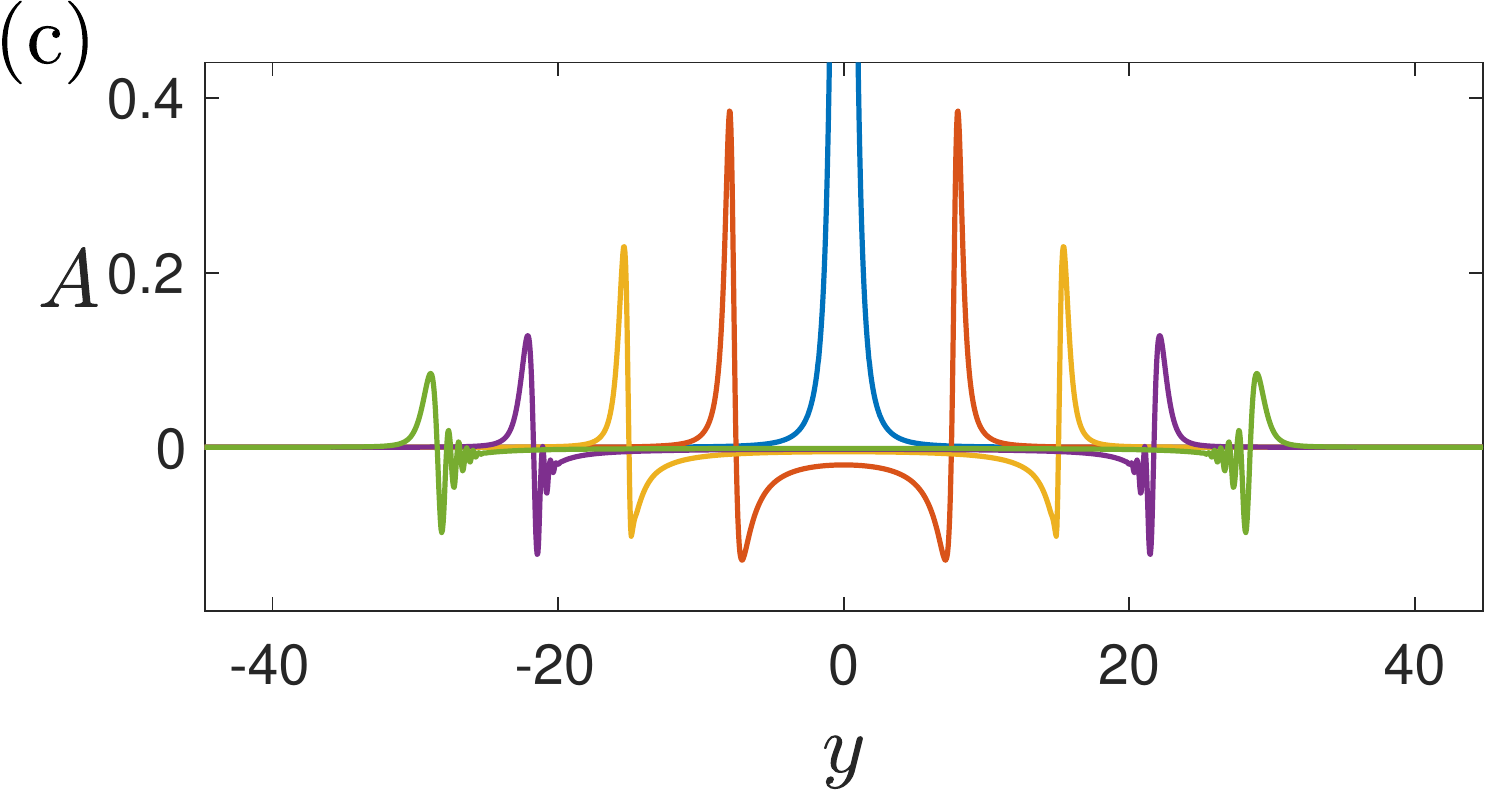}
\end{center}
\end{minipage}
\vspace{-0.7cm}
\caption{Mode-2 nonlinear ring waves for $\gamma = 0.1$, $Q = 5$, when  $\rho_1=1, \rho_2=1.1, \rho_3 = 1.2$ and $d_{1} = 0.3, d_{2} = 0.7$. Panel (a) shows the wavefronts, with the colour scheme corresponding to the wave amplitude, while panels (b) and (c) show the wave profiles in the directions $y=0$ (i.e. the $x$ axis) and $x = 0$ (i.e. the $y$ axis), respectively, at $t=0$, $t=75$, $t=150$, $t=255$ and $t=300$.}
       \label{fig:PicS56_6_31}
\end{figure}

\begin{figure}%[t!]
\begin{minipage}{.55\textwidth}
\begin{center}
\vspace{5mm}
\includegraphics[width=\linewidth]{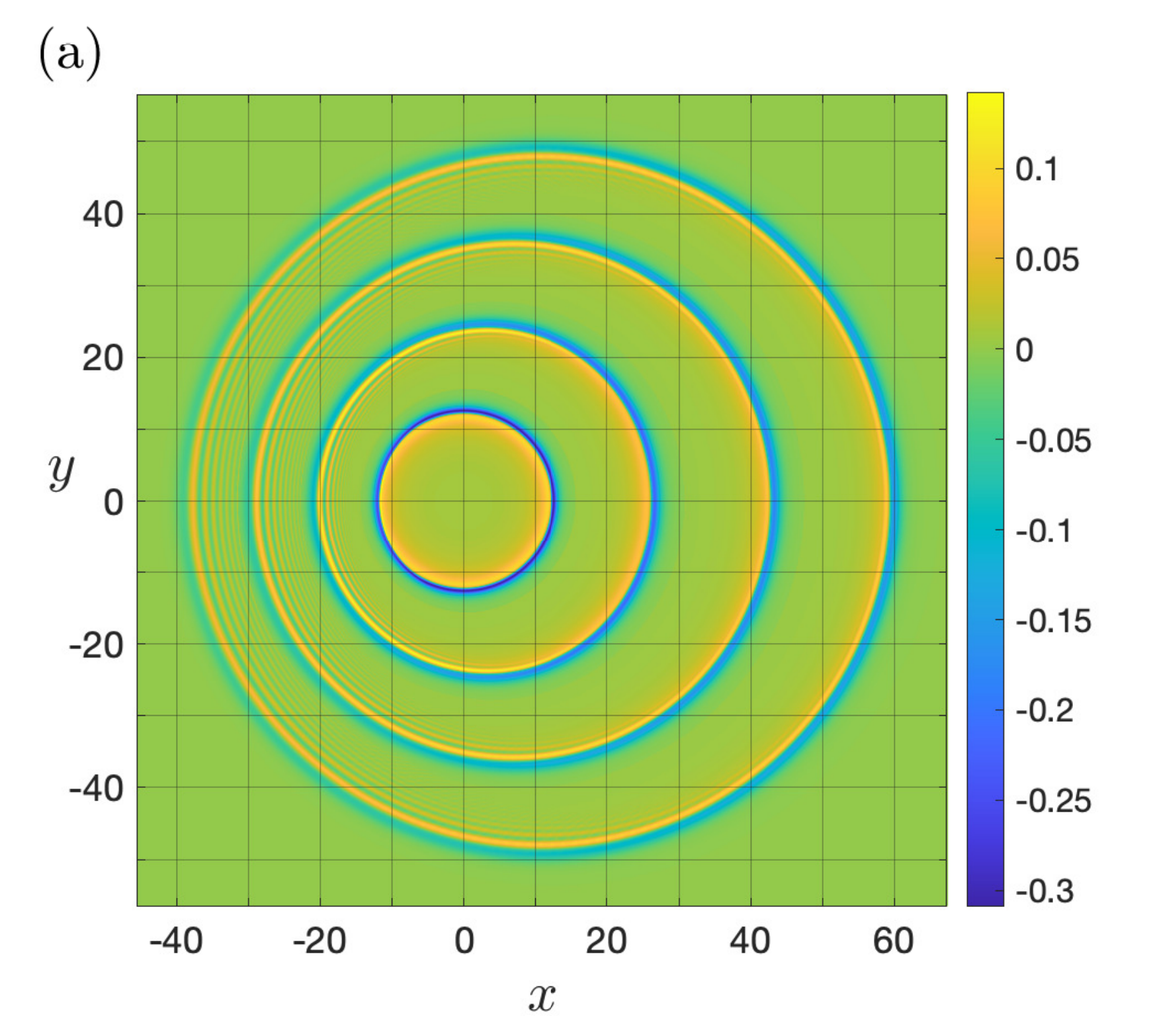}
\vspace{0.3cm}
\end{center}
\end{minipage}%\hspace{-0.04\textwidth}
\begin{minipage}{.45\textwidth}
\begin{center}
\includegraphics[width=0.95\linewidth]{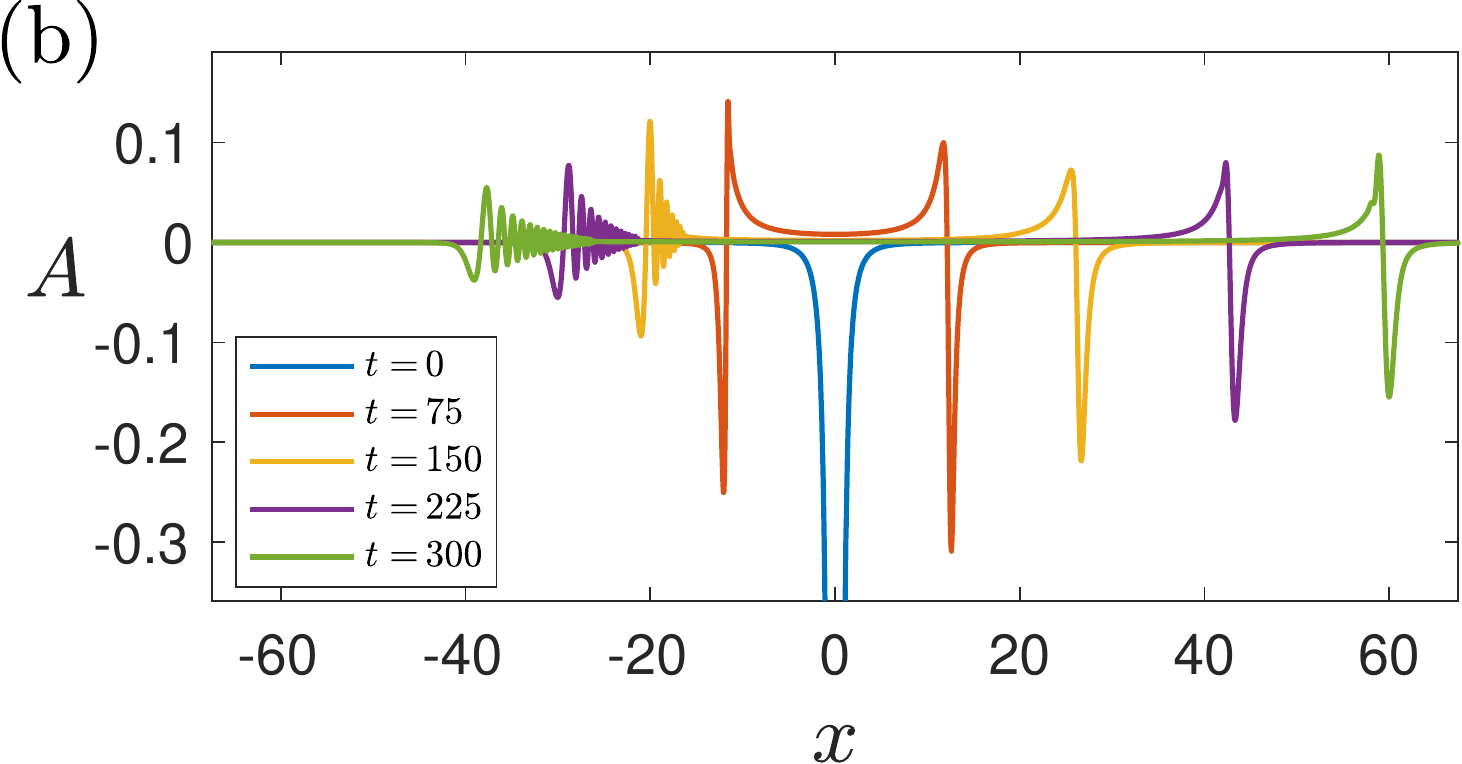}\\[0.2cm]
\includegraphics[width=0.95\linewidth]{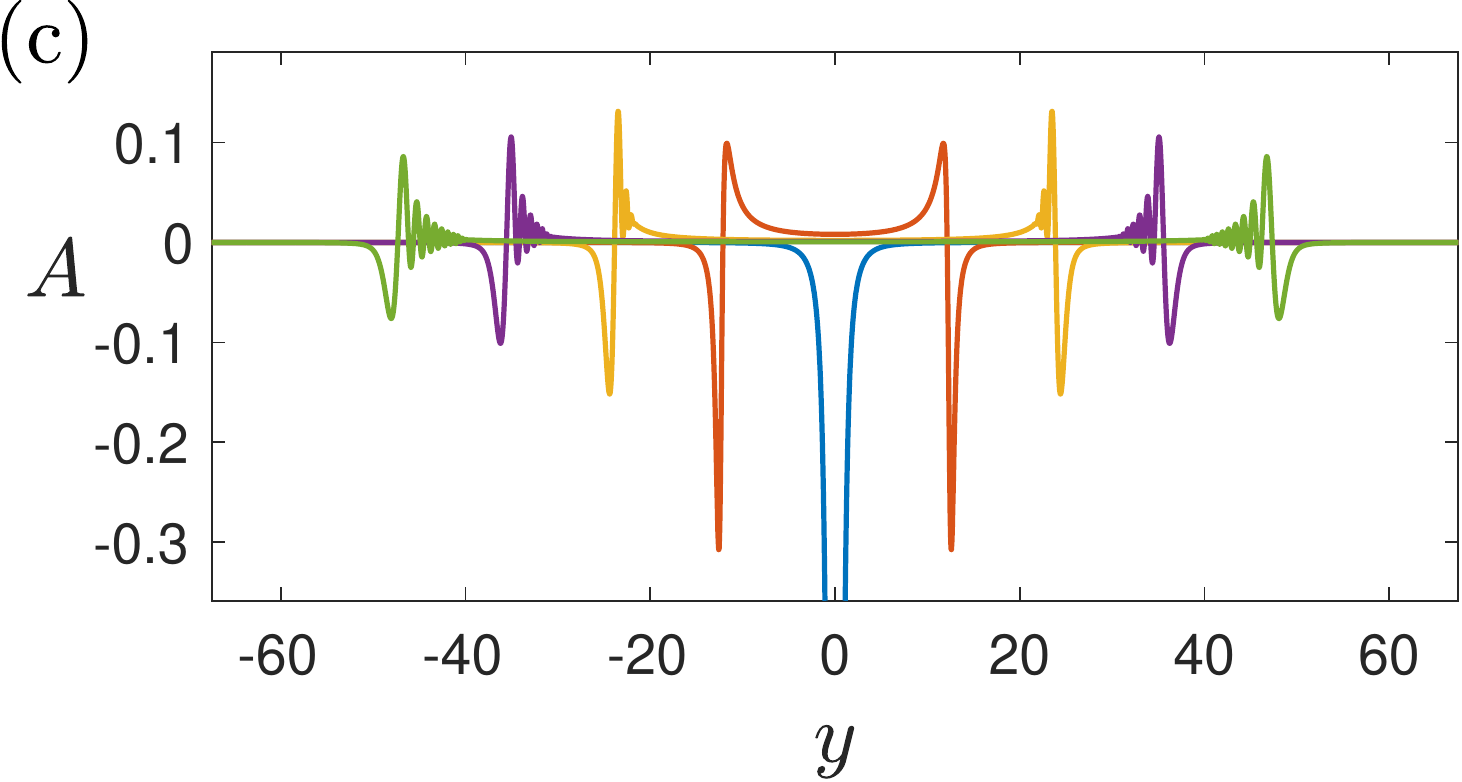}
\end{center}
\end{minipage}
\vspace{-0.7cm}
\caption{Mode-1 nonlinear ring waves for $\gamma = 0.1$, $Q = -5$, when $\rho_1=1, \rho_2=1.1, \rho_3 = 1.2$ and $d_{1} = 0.3, d_{2} = 0.7$. Panel (a) shows the wavefronts, with the colour scheme corresponding to the wave amplitude, while panels (b) and (c) show the wave profiles in the directions $y=0$ (i.e. the $x$ axis) and $x = 0$ (i.e. the $y$ axis), respectively, at $t=0$, $t=75$, $t=150$, $t=255$ and $t=300$.}
       \label{fig:PicS56_6_30b}
%\end{figure}
%
%\begin{figure}%[t!]
\begin{minipage}{.55\textwidth}
\begin{center}
\vspace{5mm}
\includegraphics[width=\linewidth]{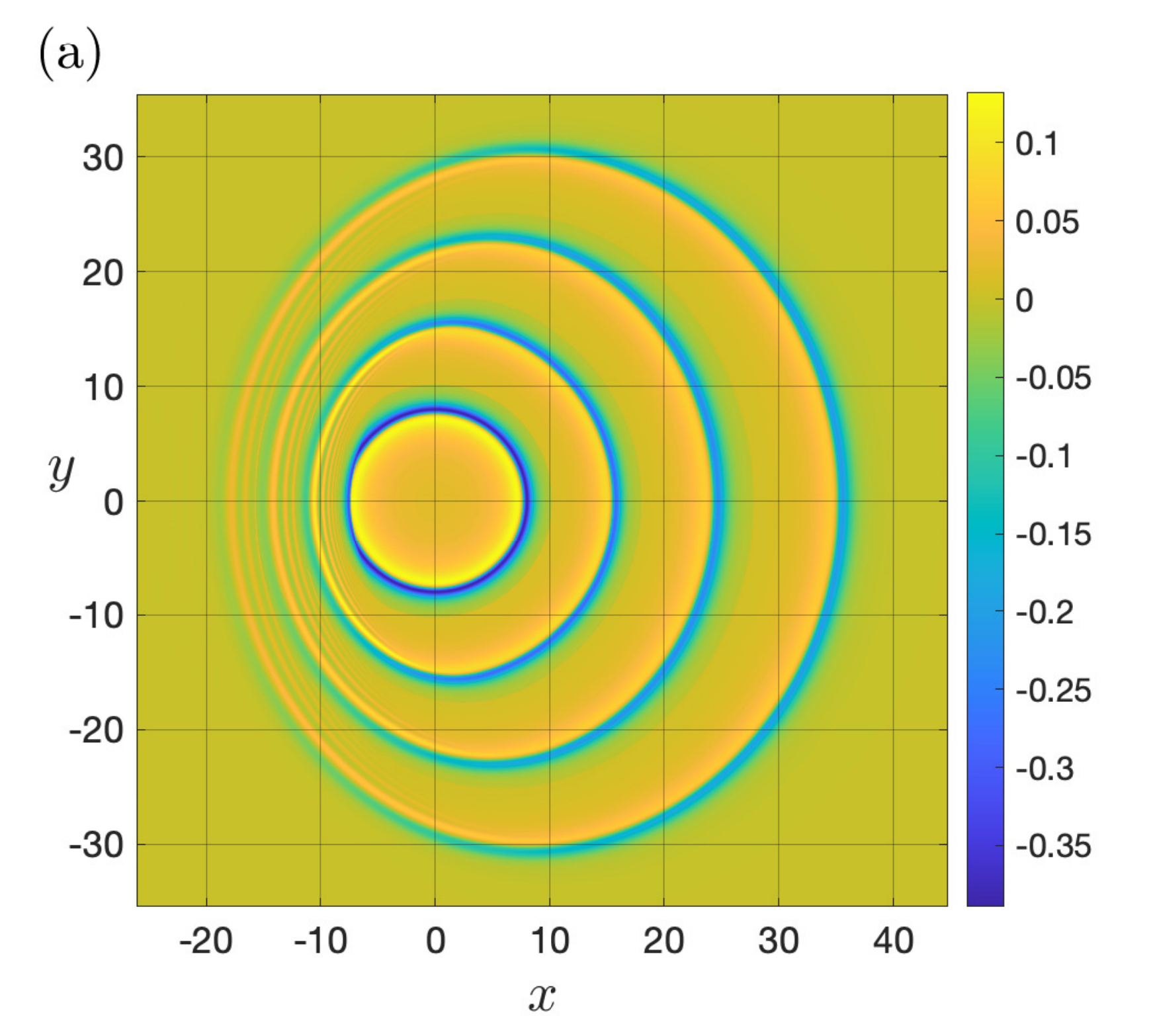}
\vspace{0.3cm}
\end{center}
\end{minipage}
\begin{minipage}{.45\textwidth}
\begin{center}
\includegraphics[width=0.95\linewidth]{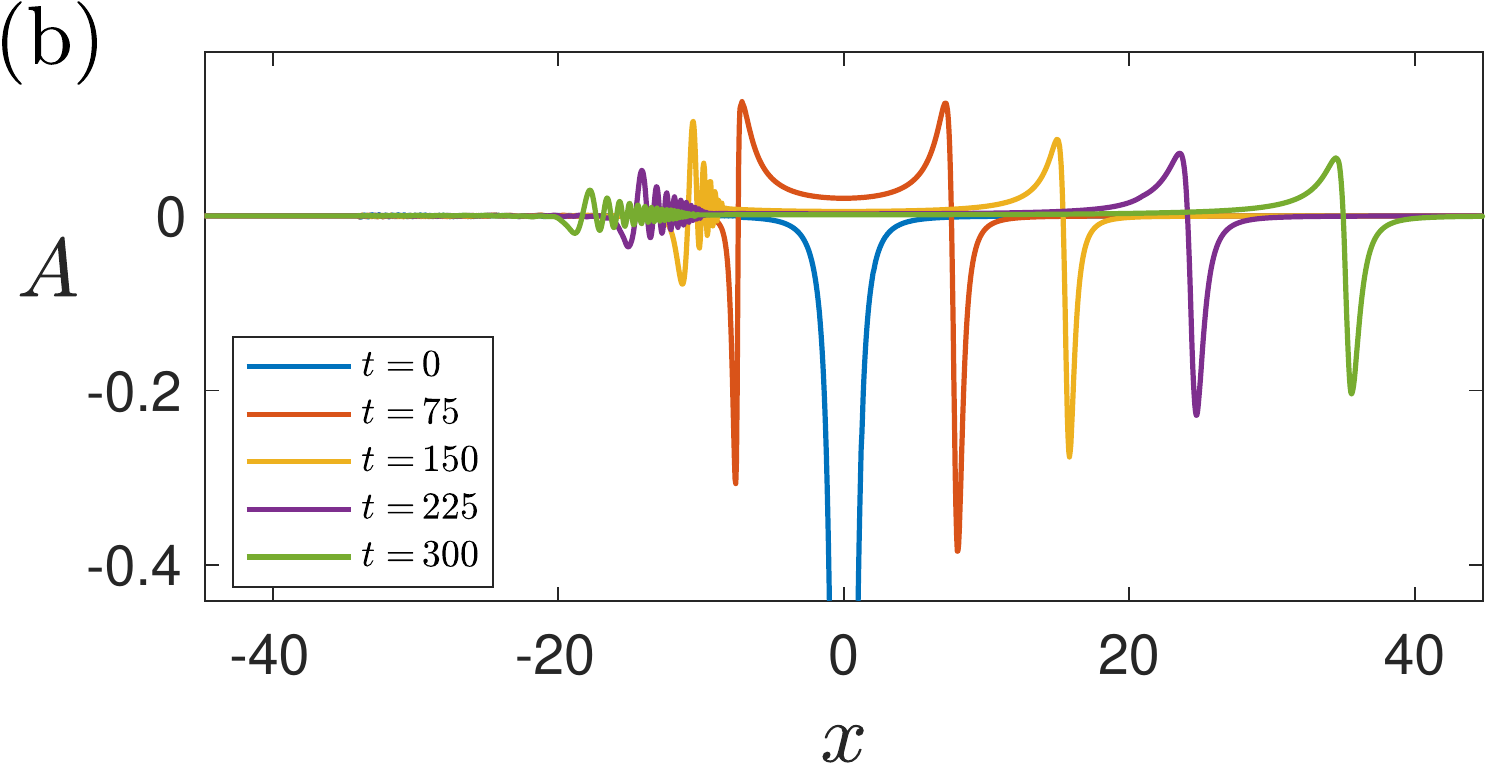}\\[0.2cm]
\includegraphics[width=0.95\linewidth]{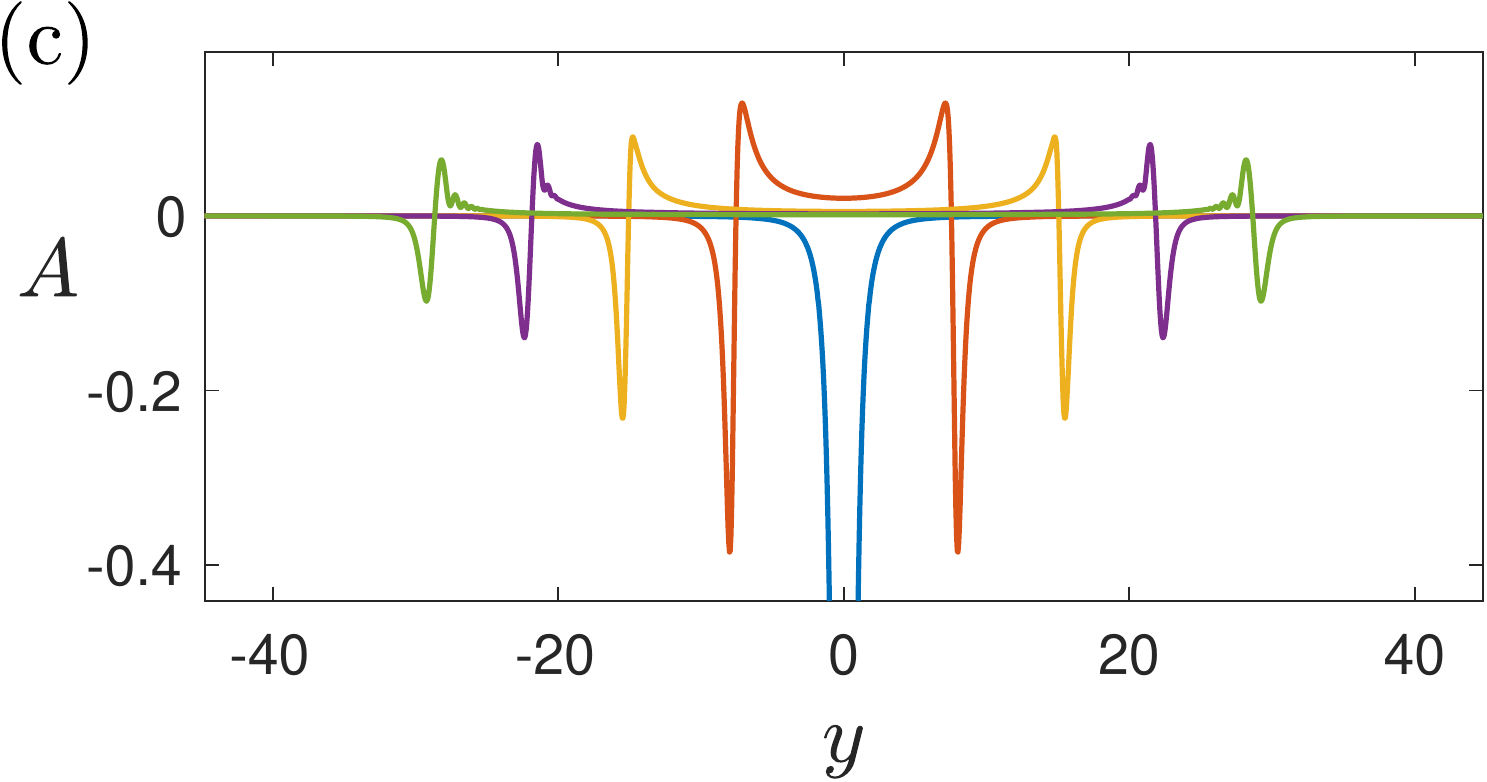}
\end{center}
\end{minipage}
\vspace{-0.7cm}
\caption{Mode-2 nonlinear ring waves for $\gamma = 0.1$, $Q = -5$, when $\rho_1=1, \rho_2=1.1, \rho_3 = 1.2$ and $d_{1} = 0.3, d_{2} = 0.7$, $d_{1}= 0.3$. Panel (a) shows the wavefronts, with the colour scheme corresponding to the wave amplitude, while panels (b) and (c) show the wave profiles in the directions $y=0$ (i.e. the $x$ axis) and $x = 0$ (i.e. the $y$ axis), respectively, at $t=0$, $t=75$, $t=150$, $t=255$ and $t=300$.}
       \label{fig:PicS56_6_31b}
\end{figure}

\begin{figure}
\begin{center}
\includegraphics[width=0.45\linewidth]{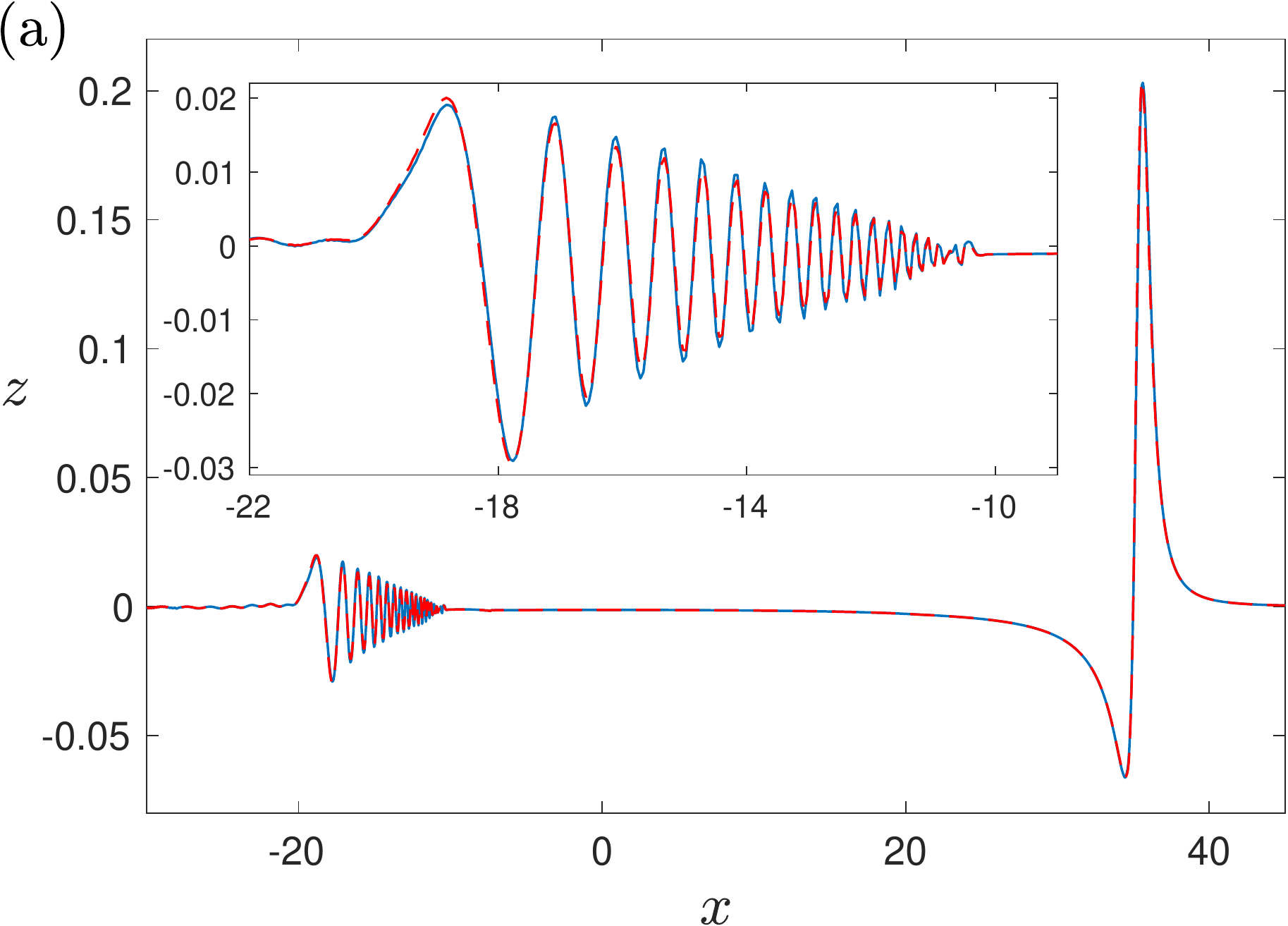}\qquad
\includegraphics[width=0.45\linewidth]{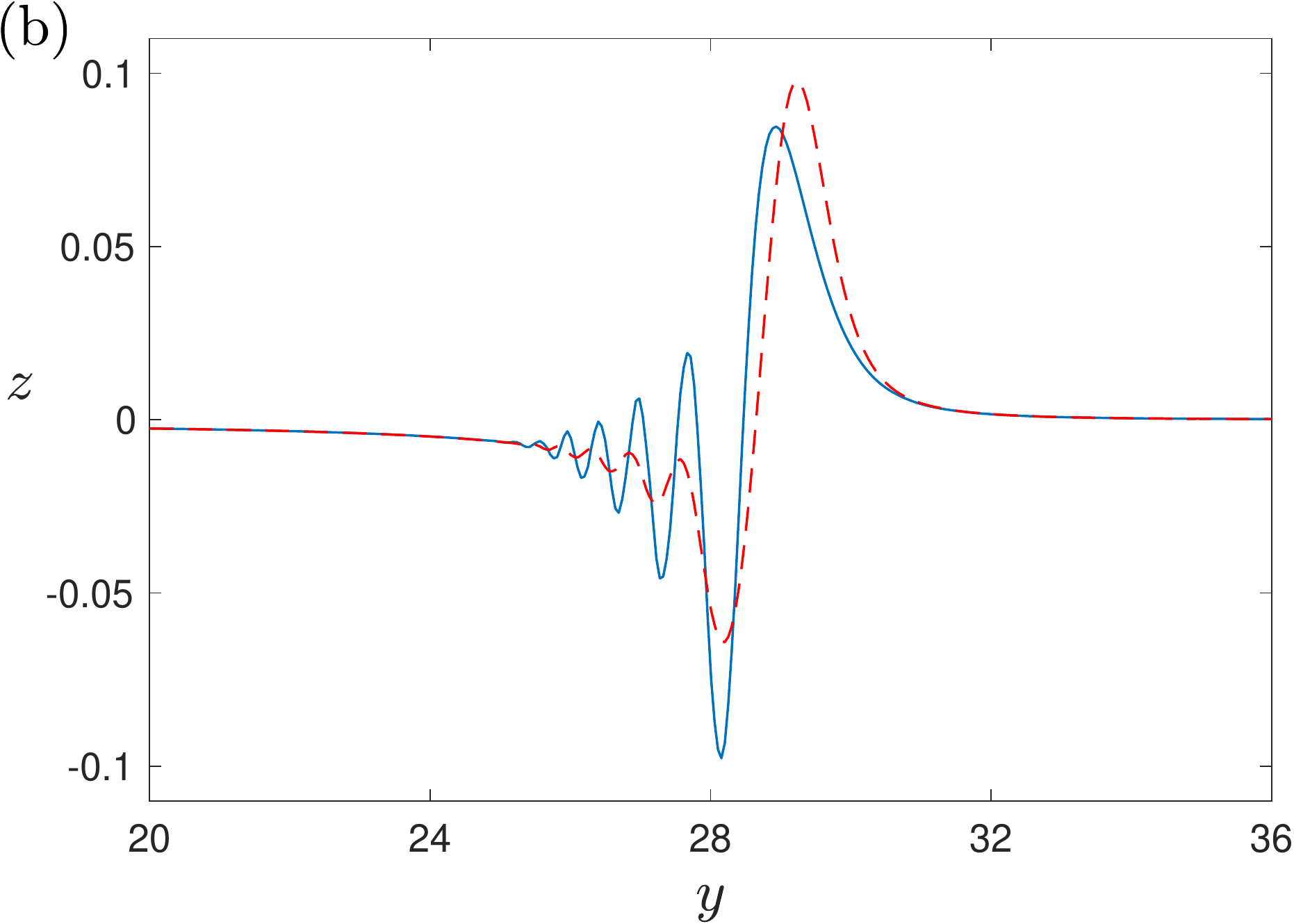}
\end{center}
\vspace{-0.3cm}
\caption{Comparison of nonlinear mode-2 wave profiles with $\gamma = 0.1$ at $t=300$ for different initial conditions ($Q=\pm 5$) and $\rho_1=1, \rho_2=1.1, \rho_3 = 1.2$.  The undisturbed interfaces are at $d_{1}= 0.3$ and $d_{2}= 0.7$.  The blue solid lines show $A$ for $Q=5$ and the red dashed lines show $-A$ for $Q=-5$.  (a) and (b)  show the profiles in the directions $y=0$ (i.e. the $x$ axis) and $x = 0$ (i.e. the $y$ axis), respectively.}
       \label{fig:comparison_profiles}
\end{figure}

Our numerical results indicate that, for weak currents, the interfacial ring waves remain nearly concentric with some elongation in the downstream direction for mode-1 waves, and squeezing for mode-2. Much more interesting features can be observed for stronger currents. Figures \ref{fig:PicS56_6_30}, \ref{fig:PicS56_6_31} present the numerical results for $\gamma = 0.1$ and an initial condition of elevation with $Q = 5$. For the same current strength, we show in 
Figures~\ref{fig:PicS56_6_30b}, \ref{fig:PicS56_6_31b} the numerical results obtained for an initial condition of depression with $Q = -5$.
As expected, in all these cases the leading waves propagate faster downstream than upstream. For the first mode, it can be observed that the wave fronts are  elongated in the flow direction, whereas for the second mode the wave fronts are squeezed in the flow direction. The deformations of the wavefronts observed in numerical simulations agree with those predicted using the analytical singular solutions for $k(\theta)$ presented in Figures \ref{fig:combined_6_10} and \ref{fig:combined_6_11}.

In previous studies involving baroclinic mode ring waves in two-layer configurations with a free surface \cite{KZ,K, HKG}, it was revealed that such waves are squeezed by the effect of current. Here, we find that one of the baroclinic modes is also squeezed (mode-2), while the other (mode-1) is elongated. To the best of our knowledge, this is the first example where such a feature is revealed. 
 
We  also observe that, in both modes, the balance of the weak nonlinearity, dispersion and cylindrical divergence generate well-developed oscillatory dispersive wave trains behind the lead wave (for both $Q=\pm5$), most significantly in the upstream direction. 
It is also rather noticeable that the second mode is suffering from stronger dispersion in the upstream direction, and also all the way up to the direction orthogonal to the current, while the lead wave is able to propagate to considerable distances downstream, resulting in a wave pattern where a part of the ring wave propagating upstream is effectively eroded by this radiation. 
 
%Changing the undisturbed positions of the interfaces results only in quantitative differences, while all key features of the solutions remain the same.

%take small positive values in the downstream direction, decreasing to its negative minimum value around $\theta = 4 \pi/3$ and then increasing to a larger negative value in the upstream direction. 
To explain in more detail the differences observed for mode-2 waves in the cases of initial conditions of elevation or depression (see Figures~\ref{fig:PicS56_6_31} and \ref{fig:PicS56_6_31b}), we go back to the coefficient behaviour presented in Figure~\ref{fig:PicS56_005_mod2}.
In particular, there is little difference in the downstream direction, as can be seen  in Figure \ref{fig:comparison_profiles}(a) showing numerical solutions at $t=300$ in the direction $y=0$ (i.e.\ the $x$ axis), and there is virtually no dispersive radiation. This is in agreement with the plots of the coefficients in this direction, shown in Figure~\ref{fig:PicS56_005_mod2}, since all three coefficients are close to zero.  The waves propagating upstream are different: the effects of the dispersion and cylindrical divergence are very strong in that direction, resulting in the emergence of similar small amplitude dispersive wave trains, which can be seen in the plots of the corresponding numerical solutions in  Figure~\ref{fig:comparison_profiles}(a). On the contrary, waves propagating in the orthogonal direction to the current (along the $y$ axis) show a much stronger dispersive radiation for $Q=5$, when compared to $Q=-5$. 
%significant differences between $Q=5$ and $Q=-5$ (see Figure~ \ref{fig:comparison_profiles}(b)): we see much stronger dispersive radiation in the first case, for the initial condition of elevation. 
This is again understandable in the view of the plots in Figure~\ref{fig:PicS56_005_mod2}: the reflection $A \to -A$ in equation (\ref{cKdV1}) maps the initial-value problem in the case of $Q=5$ into nearly the same as in the case of $Q=-5$, with the only difference being in the sign of the nonlinearity coefficient, which changes sign, resulting in a different character of the solution. 

%
%\begin{figure}
%\centering
%\includegraphics[width = 0.7\textwidth] {FIG/plots_Q_a2}
%\caption{Plots of $H-\hat{a}^2$ for mode~2 as functions of $\hat{a}$ for various values of $\gamma$, as is indicated in the legend, showing the transition from the ``elliptic'' to the ``hyperbolic'' regime for $\rho_1=1$, $\rho_2=1.1$, $\rho_3=1.2$ and $d_{1} = 0.3$, $d_{2} = 0.7$.}
%\label{fig:Q_a2}
%\end{figure}
%
%\begin{figure}
%\centering
%\includegraphics[height=6.5cm]{FIG/wavefronts_elliptic_hyperbolic}
%\caption{%\footnotesize 
%Wavefronts of the second interfacial ring mode for $\rho_1=1$, $\rho_2=1.1$, $\rho_3=1.2$ and $d_{1} = 0.3$, $d_{2} = 0.7$ described by $k(\theta)r = 50$ when $\gamma=0$ (thick black solid line), $\gamma =0.125$ (red long-dashed line), $\gamma =0.25$ (blue dash-dotted line), $\gamma =0.375$ (green short-dashed line) and $\gamma =0.524063$ (thin brown solid line), respectively.}
%\label{fig:wavefronts_elliptic_hyperbolic}
%\end{figure}
%
%\begin{figure}
%\centering
%\includegraphics[height=6.5cm]{FIG/critical_levels}
%\caption{%\footnotesize 
%The location of the critical layer $z=z_c(\theta)$ as a function of $\theta$ for $\rho_1=1$, $\rho_2=1.1$, $\rho_3=1.2$ and $d_{1} = 0.3$, $d_{2} = 0.7$ for several values of $\gamma$ as indicated in the legend.}
%\label{fig:critical_layer_d1_0_3_d2_0_7}
%\end{figure}
\subsection{Asymmetric configuration with $d_1=0.2$, $d_2=0.9$}

\begin{figure}%[!p!t]
      \centering
      \includegraphics[width=0.3\linewidth]{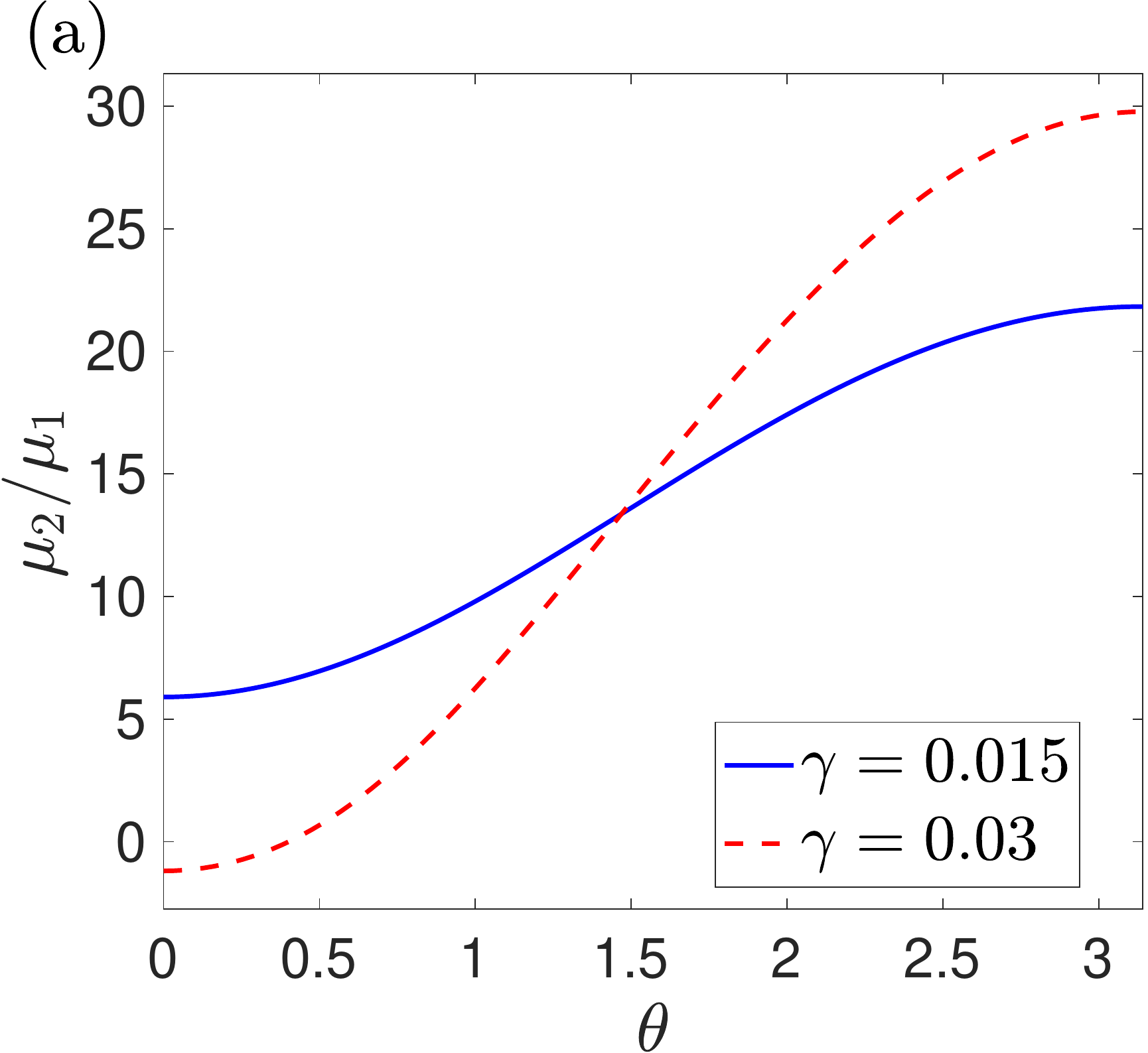} \hfill
      \includegraphics[width=0.3\linewidth]{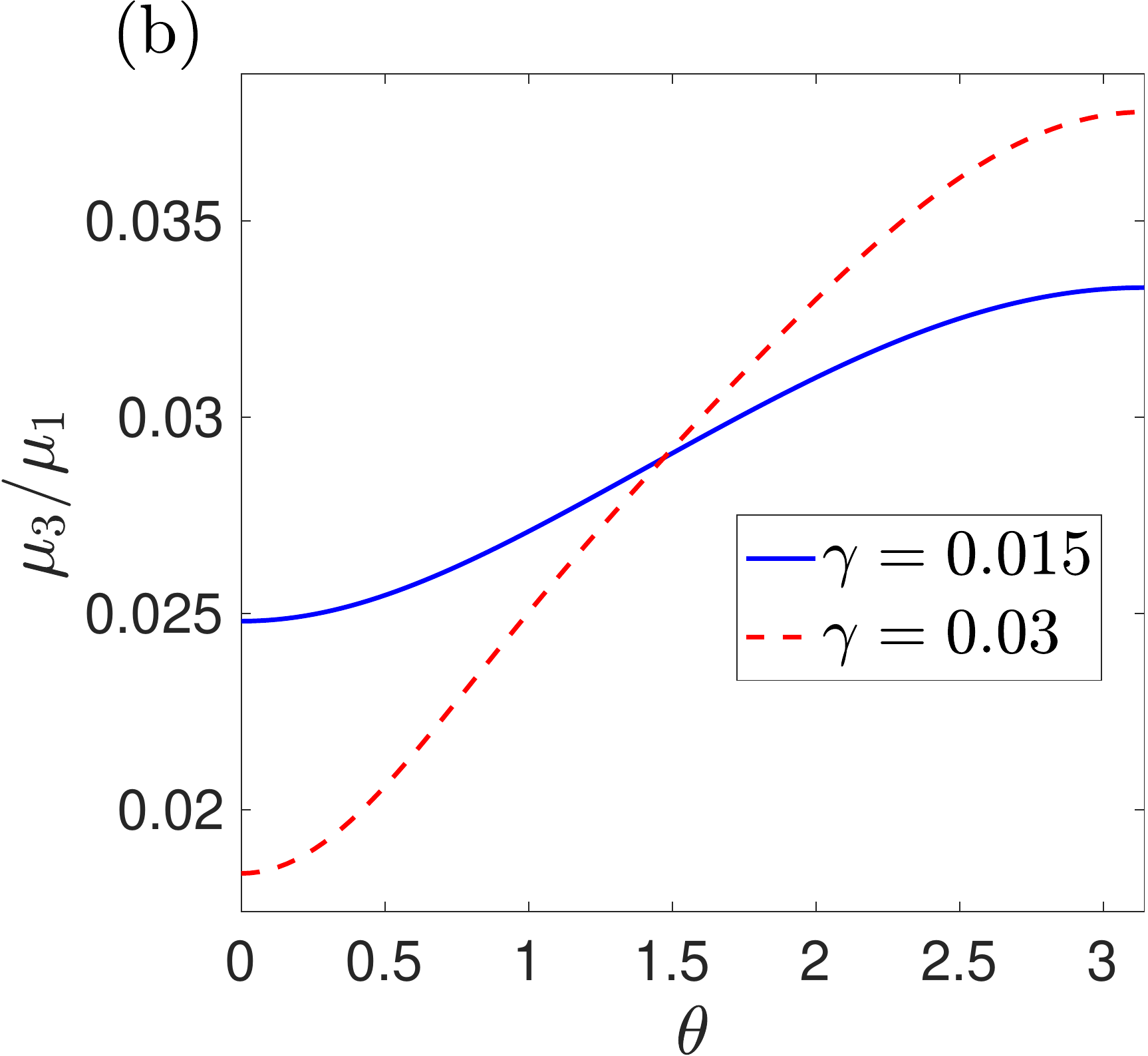} \hfill
      \includegraphics[width=0.3\linewidth]{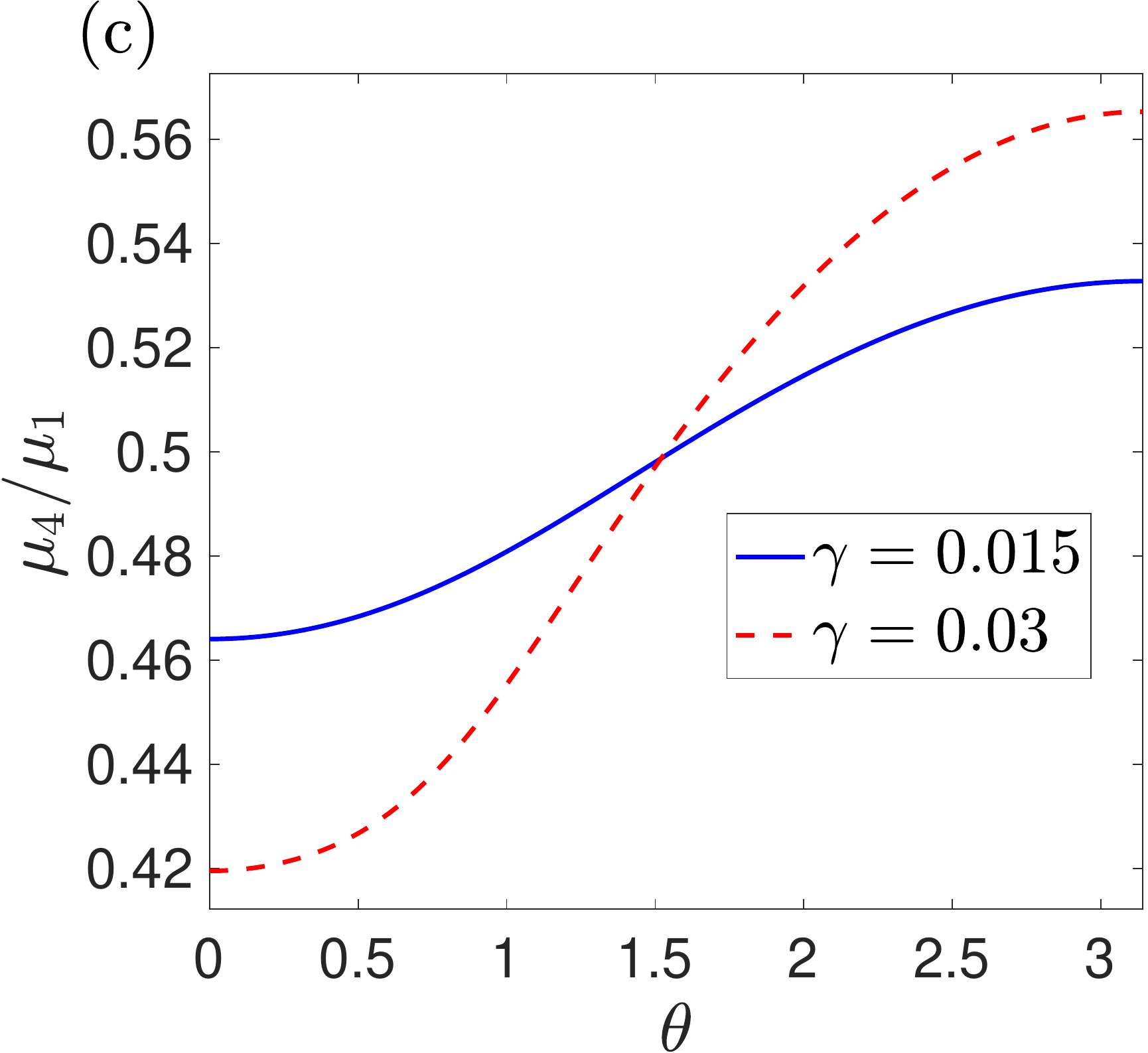} 
\vspace{-0.2cm}
       \caption{Coefficients $\mu_i(\theta)/ \mu_1(\theta), i = 2, 3, 4,$  for the mode-1 ring waves for $\rho_1=1, \rho_2=1.1, \rho_3 = 1.2$ and $d_1=0.2$, $d_2=0.9$, when $\gamma=0.015$ (blue solid lines), $\gamma=0.03$ (red dashed lines).}
       \label{fig:mu_nonlin_mod1}
\vspace{0.5cm}       
      \includegraphics[width=0.3\linewidth]{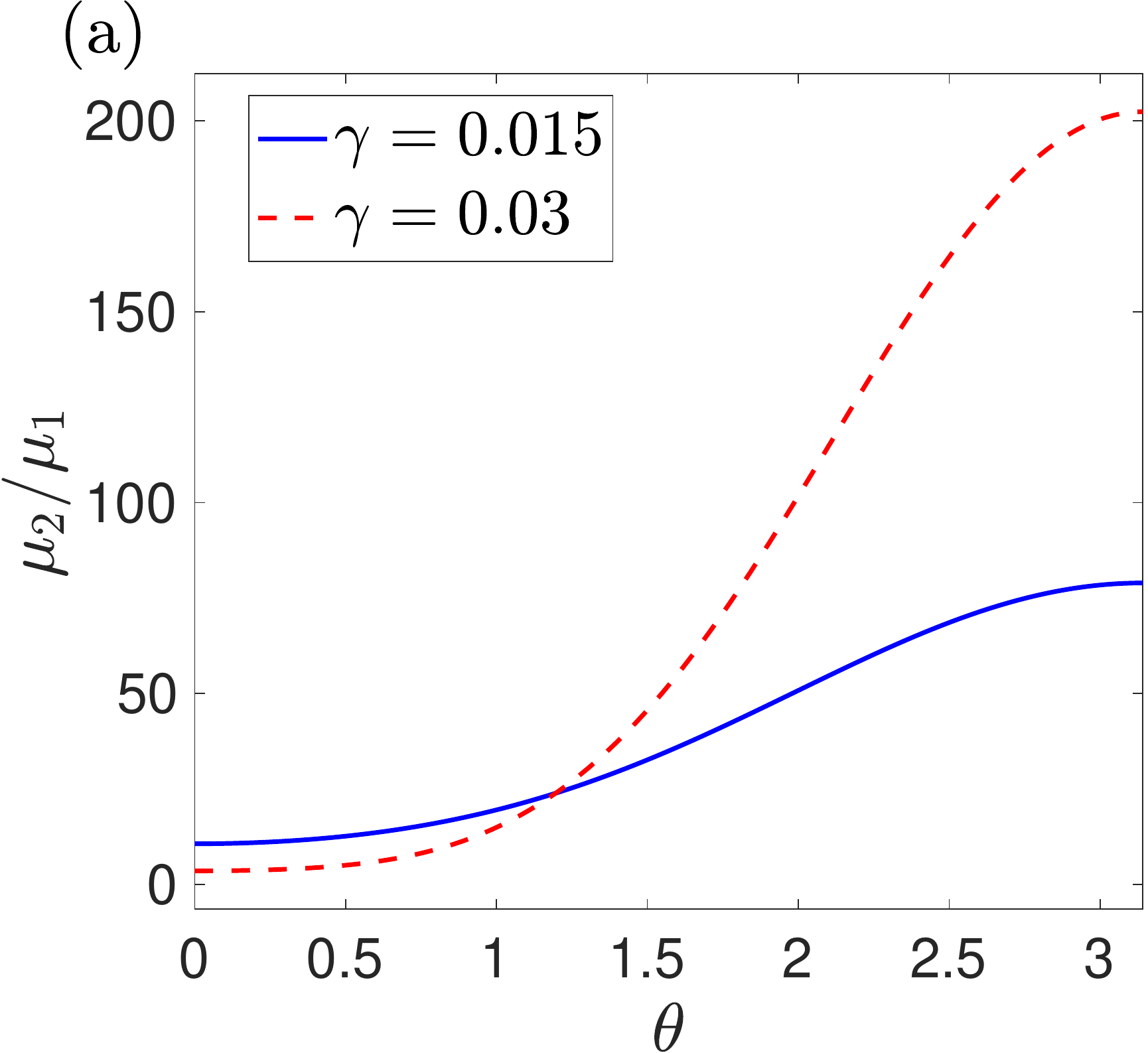} \hfill
      \includegraphics[width=0.3\linewidth]{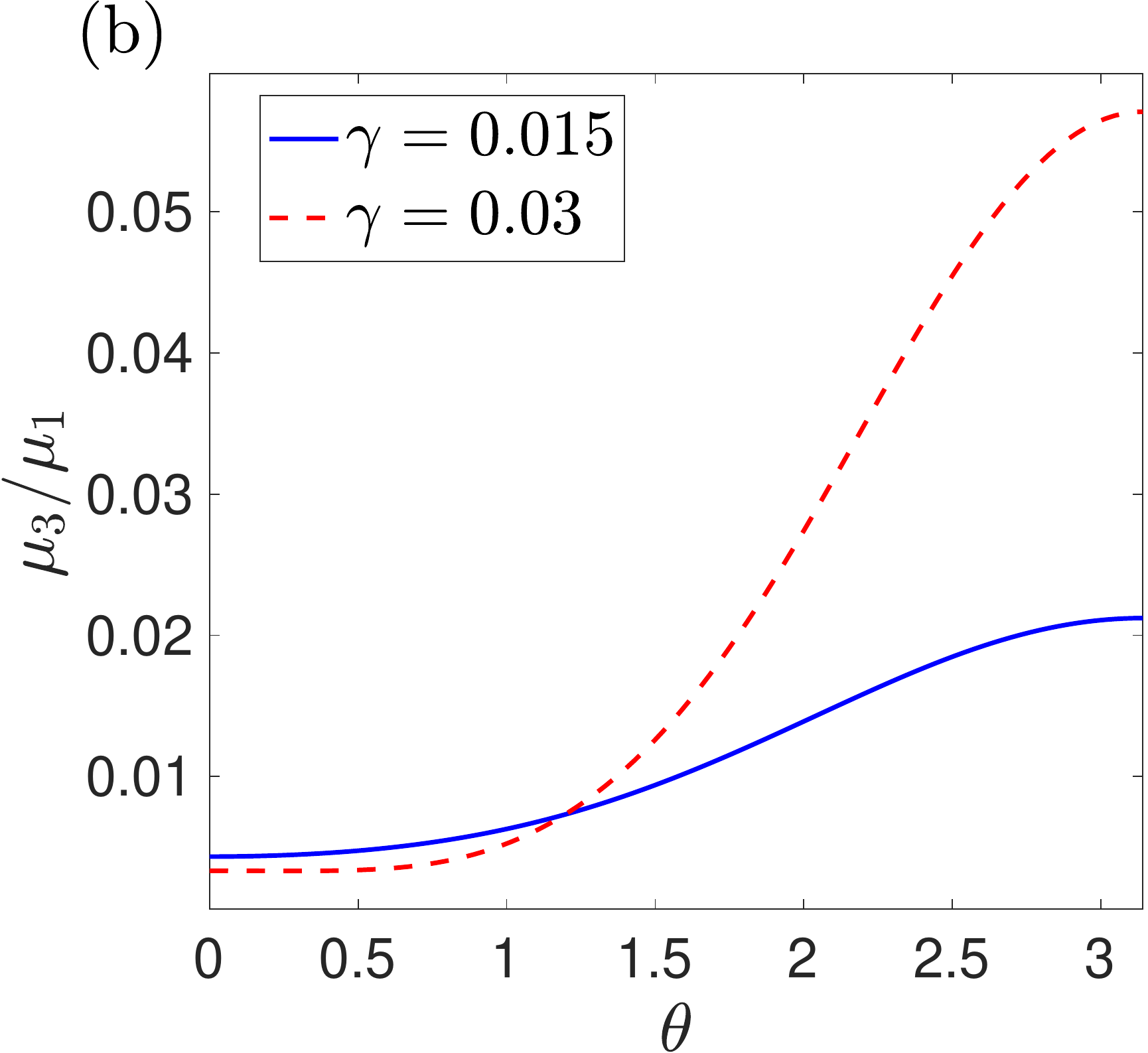} \hfill
      \includegraphics[width=0.3\linewidth]{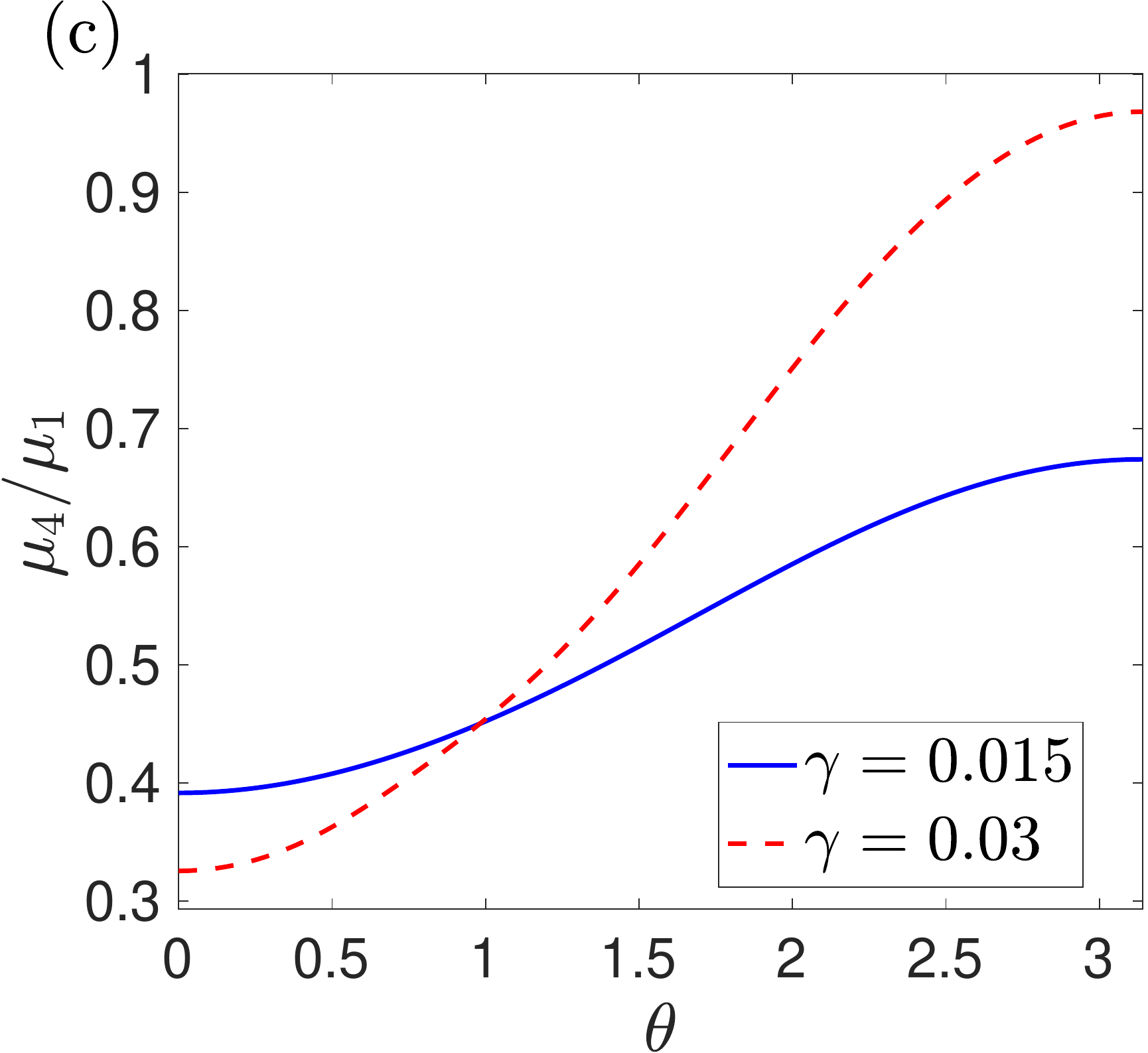} 
\vspace{-0.2cm}
       \caption{Coefficients $\mu_i(\theta)/ \mu_1(\theta), i = 2, 3, 4,$  for the mode-2 ring waves for $\rho_1=1, \rho_2=1.1, \rho_3 = 1.2$ and $d_1=0.2$, $d_2=0.9$, when $\gamma=0.015$ (blue solid lines), $\gamma=0.03$ (red dashed lines).}
       \label{fig:mu_nonlin_mod2}
\vspace{1cm}
\centering
\includegraphics[height=6.5cm]{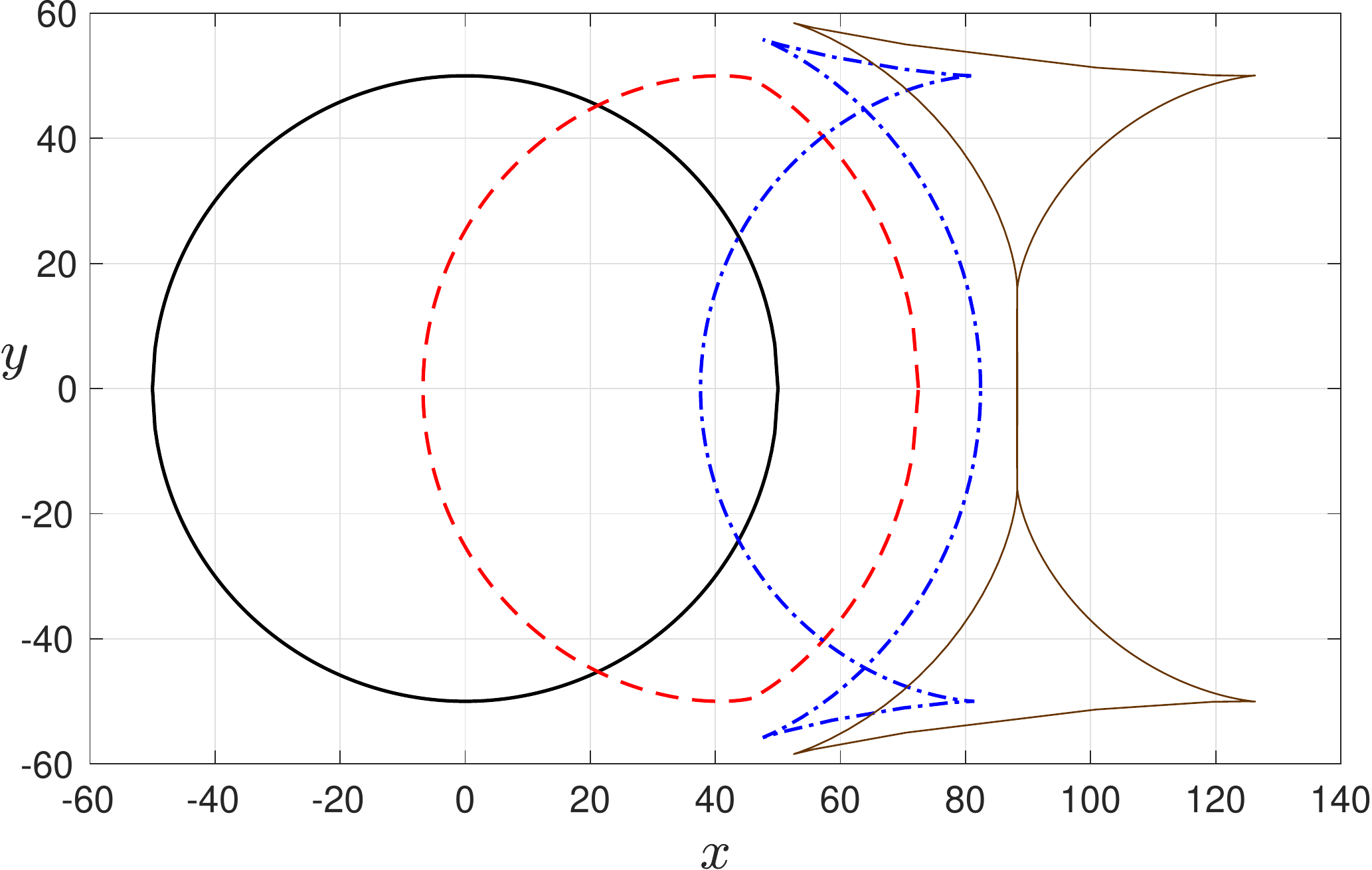}
\caption{%\footnotesize 
Wavefronts of mode-2 ring waves for $\rho_1=1, \rho_2=1.1, \rho_3 = 1.2$ and $d_{1} = 0.2$, $d_{2} = 0.9$, described by $k(\theta)r = 50$, showing the transition from the elliptic to hyperbolic regime and further up to the onset of the long-wave instability, with $\gamma=0$ (thick black solid line), $\gamma =0.09$ (red long-dashed line), $\gamma =0.18$ (blue dash-dotted line), and $\gamma =0.274145$ (thin brown solid line), respectively.}
\label{fig:wavefronts_elliptic_hyperbolic2}
\end{figure}

Here, we consider an asymmetric case by setting $d_1=0.2$, $d_2=0.9$. For weak currents, it is well known for planar waves that the symmetric configuration examined above is close to the so-called {\it criticality} condition for mode-1 waves. This corresponds to the vanishing of the nonlinearity coefficient of the KdV equation. By breaking symmetry, we expect an enhanced nonlinear behaviour of the solutions of the first mode. Incidentally, we find here that nonlinear effects on mode-2 solutions are also amplified. This is confirmed in Figures~\ref{fig:mu_nonlin_mod1} and \ref{fig:mu_nonlin_mod2}, where $\mu_i/\mu_1$ ($i=2,3,4)$ are plotted for mode-1 and mode-2, respectively.  In contrast with Figures~\ref{fig:PicS56_005_mod1},~\ref{fig:PicS56_005_mod2}, here the range of values for the nonlinearity coefficient $\mu_2/\mu_1$ is an order of magnitude larger, even though the current is weaker.

For this new set of parameters, the transition from the elliptic to the hyperbolic regime for mode-2 waves occurs at $\gamma =\gamma_{p2}\approx0.1038$. Interestingly, the swallowtail singularity appears for mode-2 waves already in the elliptic regime, at $\gamma\approx0.0877$. Furthermore, we find that no critical surfaces appear in this elliptic regime. The transition from the elliptic to the hyperbolic regime is depicted in Figure~\ref{fig:wavefronts_elliptic_hyperbolic2} where the wavefronts are shown for different values of $\gamma$ up to $\gamma=\gamma^-\approx 0.274145$, at which long-wave instability arises and mode-2 ring waves cease to exist. 
We notice in Figure~\ref{fig:wavefronts_elliptic_hyperbolic2} that for $\gamma=0.09$ (red dashed line) a swallowtail singularity is already present, although it is not visible on the scale of the figure.

\begin{figure}%[t!]
\begin{minipage}{.55\textwidth}
\begin{center}
\vspace{5mm}
\includegraphics[width=\linewidth]{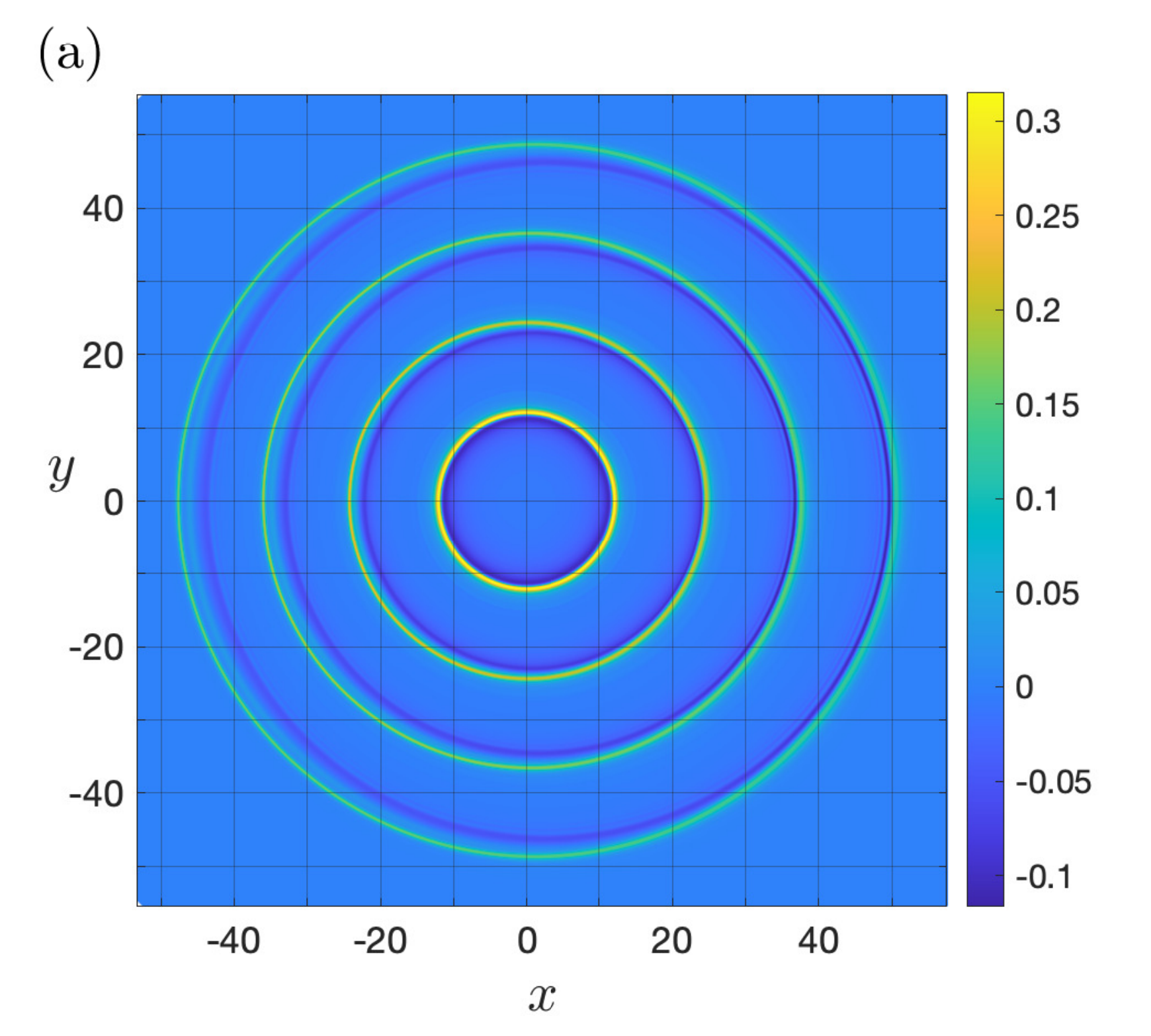}
\vspace{0.3cm}
\end{center}
\end{minipage}%\hspace{-0.04\textwidth}
\begin{minipage}{.45\textwidth}
\begin{center}
\includegraphics[width=0.95\linewidth]{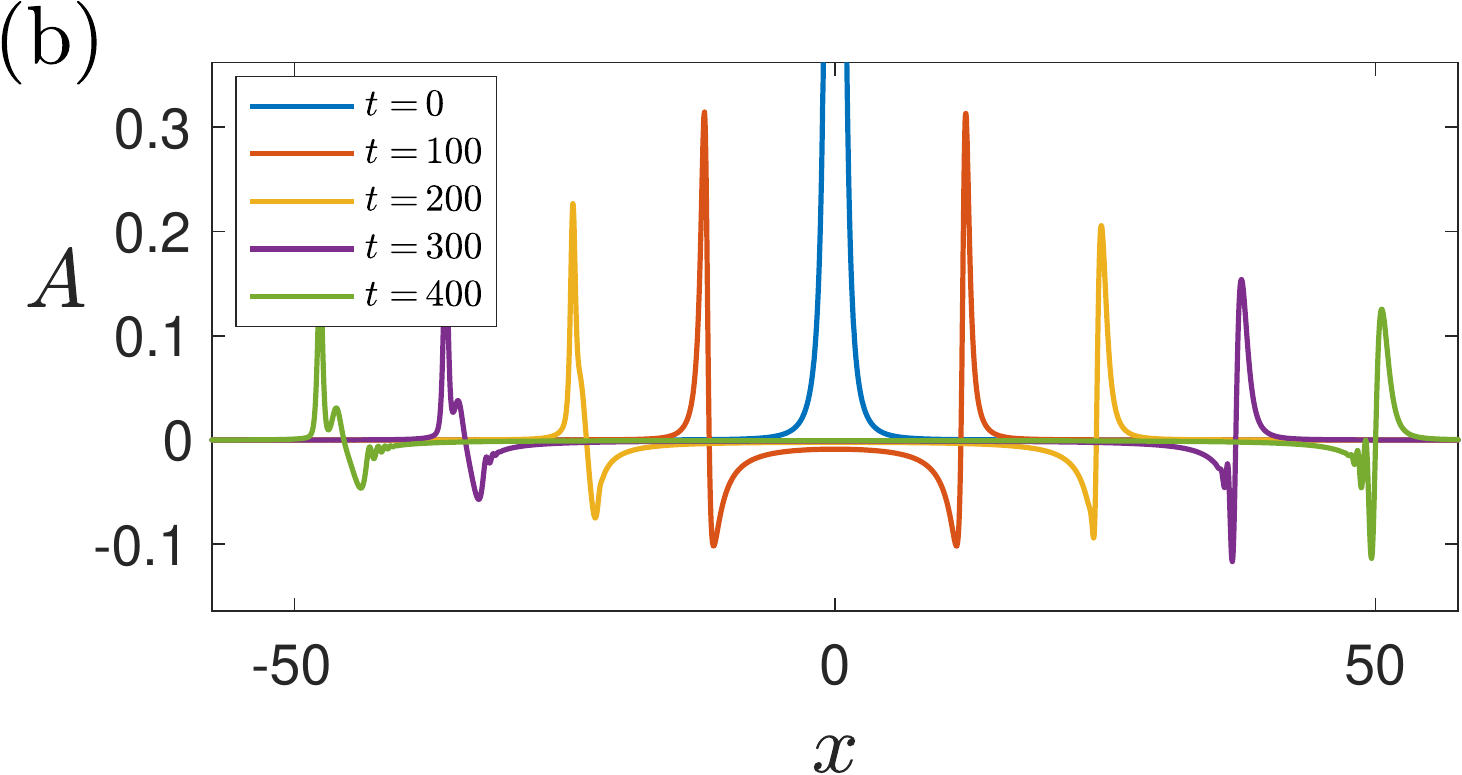}\\[0.2cm]
\includegraphics[width=0.95\linewidth]{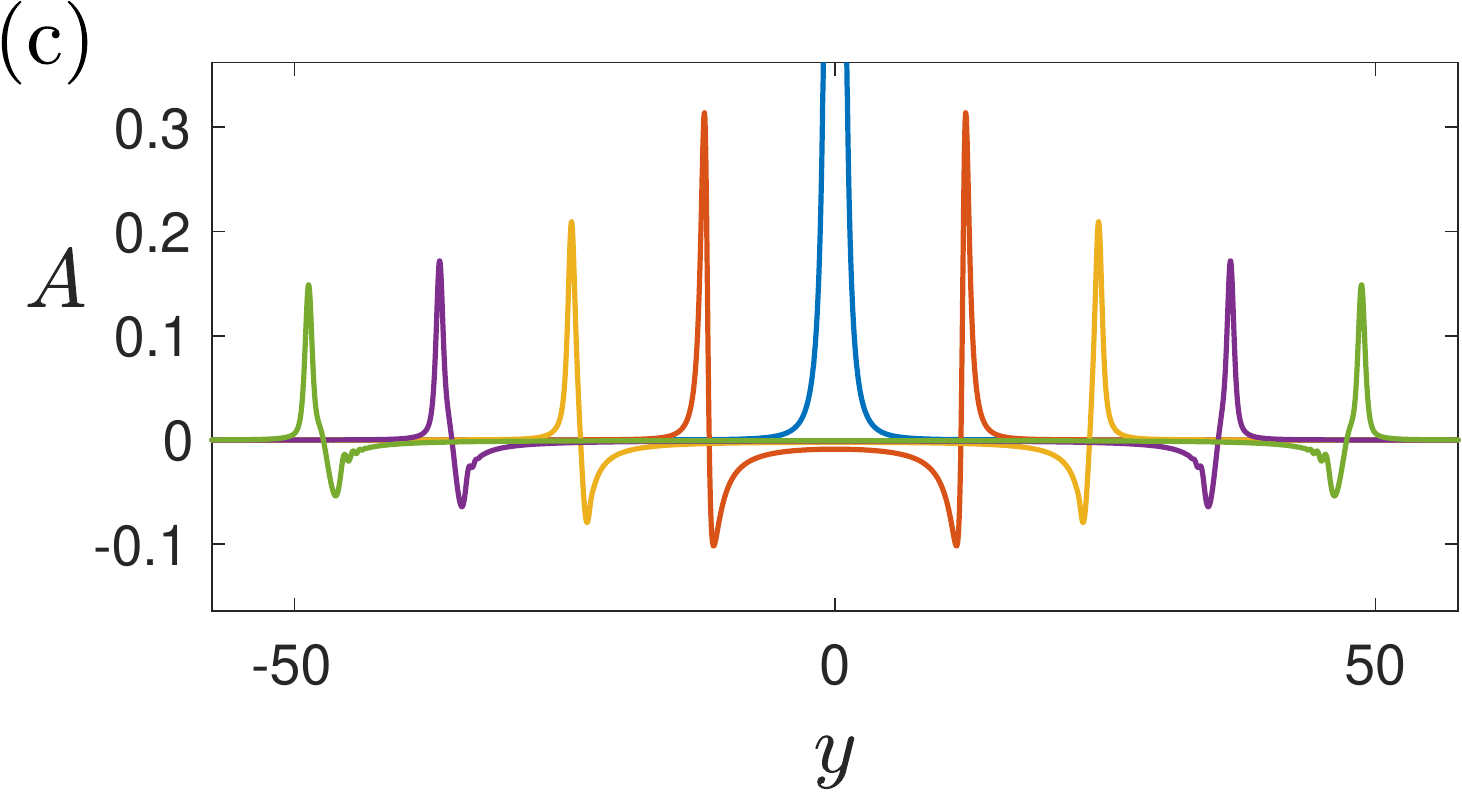}
\end{center}
\end{minipage}
\vspace{-1cm}
\caption{
Mode-1 nonlinear ring waves for $\gamma = 0.03$, $Q = 5$, when $\rho_1=1, \rho_2=1.1, \rho_3 = 1.2$ and $d_{1}= 0.2$, $d_{2}= 0.9$. Panel (a) shows the wavefronts, with the colour scheme corresponding to the wave amplitude, while panels (b) and (c) show the wave profiles in the directions $y=0$ (i.e. the $x$ axis) and $x = 0$ (i.e. the $y$ axis), respectively, at $t=0$, $t=100$, $t=200$, $t=300$ and $t=400$.}
       \label{fig:nonlin_solution_d1_0_2_d2_0_9_mod1_gam_0_03}
%\begin{figure}%[t!]
\begin{minipage}{.55\textwidth}
\begin{center}
\vspace{5mm}
\includegraphics[width=\linewidth]{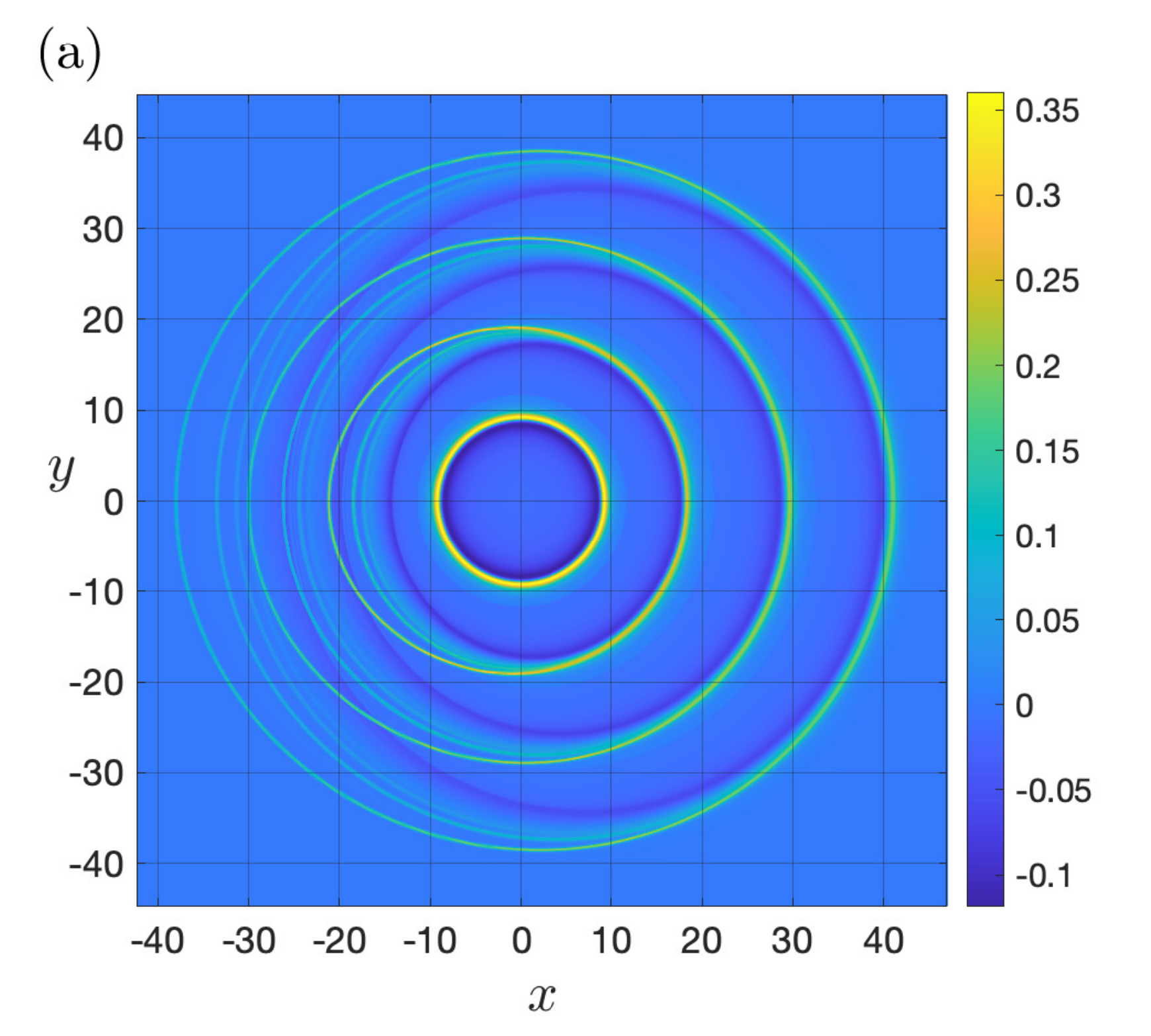}
\vspace{0.3cm}
\end{center}
\end{minipage}%\hspace{-0.04\textwidth}
\begin{minipage}{.45\textwidth}
\begin{center}
\includegraphics[width=0.95\linewidth]{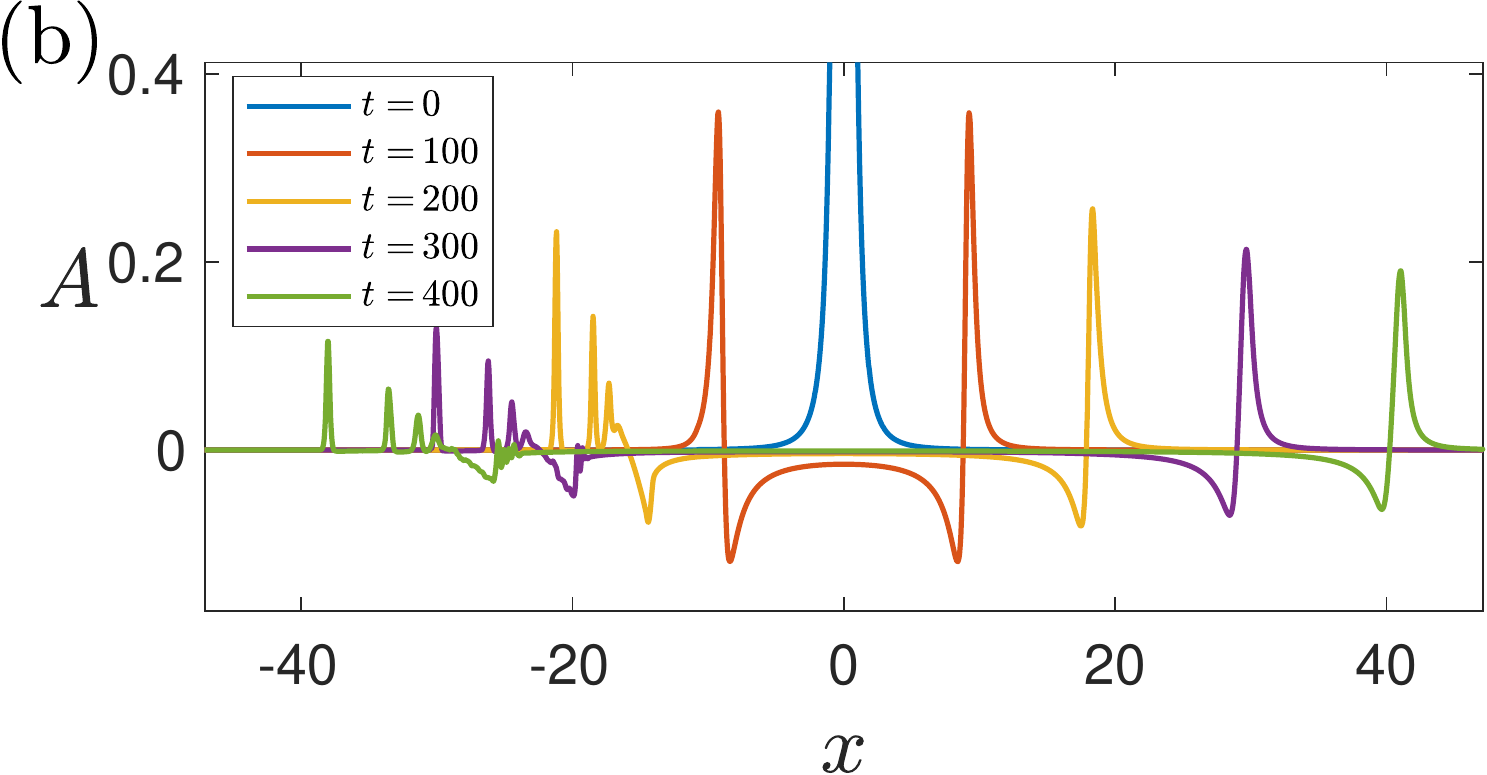}\\[0.2cm]
\includegraphics[width=0.95\linewidth]{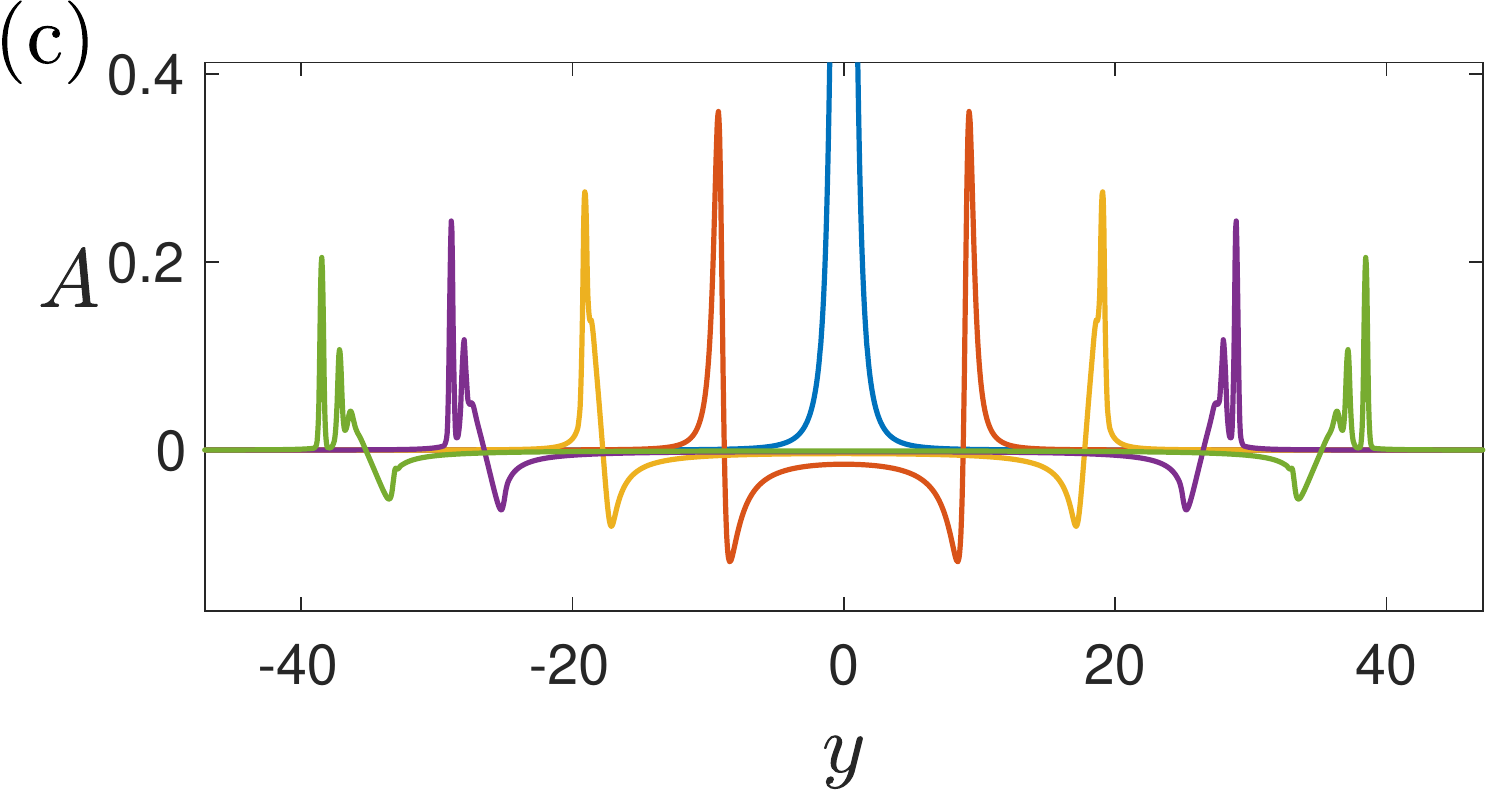}
\end{center}
\end{minipage}
\vspace{-1cm}
\caption{Mode-2 nonlinear ring waves for $\gamma = 0.03$, $Q = 5$, when $\rho_1=1, \rho_2=1.1, \rho_3 = 1.2$ and $d_{1}= 0.2$, $d_{2}= 0.9$. Panel (a) shows the wavefronts, with the colour scheme corresponding to the wave amplitude, while panels (b) and (c) show the wave profiles in the directions $y=0$ (i.e. the $x$ axis) and $x = 0$ (i.e. the $y$ axis), respectively, at $t=0$, $t=100$, $t=200$, $t=300$ and $t=400$.}
       \label{fig:nonlin_solution_d1_0_2_d2_0_9_mod2_gam_0_03}
\end{figure}

Next, we look at the nonlinear dynamics and present the solutions of the cKdV equation (\ref{cKdV1}) in Figures \ref{fig:nonlin_solution_d1_0_2_d2_0_9_mod1_gam_0_03}, \ref{fig:nonlin_solution_d1_0_2_d2_0_9_mod2_gam_0_03}, for $\gamma = 0.03$ and an initial condition of elevation with $Q = 5$. Both modes are considered and, similarly to the symmetric case described above, the wavefronts are  elongated in the flow direction for mode-1, whereas for mode-2 the wave fronts are squeezed in the flow direction. However, nonlinearity effects become more noticeable here. In the upstream direction we start observing fission of waves, especially for the second baroclinic mode. Wave fission is initiated at $\theta=\pi$ and can be observed in an increasingly wider sector, as time evolves. Hence, in this regime the wavefronts of the nonlinear ring waves at the outer front of the wave structure eventually become significantly different from the wavefronts of the linear waves in its  inner tail. We note that fission of ring waves has also manifested itself, in a different setting, in numerical experiments in \cite{HH}.

%\newpage
\section{Concluding remarks}

In this paper, we have studied the propagation of two interfacial ring modes generated by a 3D localised initial condition in a three-layer fluid with a linear shear current. The problem was studied in the rigid-lid approximation, and the emphasis was on the effects of the current. 
%(with the strongest current being on the surface of the fluid). 
Our study was based on the weakly-nonlinear theory developed in \cite{KZ}, and it included a combination of analytical and numerical results for the relevant modal and amplitude equations. 

It transpired that the current has a very different effect on these modes: the wavefronts of the faster (first baroclinic) mode become elongated in the direction of the current, while the wavefronts of the slower (second baroclinic) mode become squeezed in the same direction. To the best of our knowledge this is the first such result for the second baroclinic ring mode in any configuration.
%Previously, elongation and squeezing feature for the faster and slower mode, respectively, are aligned with those in \cite{KZ, KZ1} for a two-layer flow with a free surface. 
These effects were described analytically using the constructed singular solutions of the highly nonlinear first-order angular adjustment equation (regarder as a 2D long-wave dispersion relation) 
%constituting a further generalisation of the Burns' \cite{B} and generalised Burns' \cite{J1} conditions) 
and also observed in numerical simulations of the problem where an initially concentric ring wave enters a region with the current. %Previously, such effect was reported for the interfacial ring wave in a two-layer fluid with the piecewise-constant current  \cite{KZ}, and some power-law currents  \cite{HKG}. To the best of our knowledge, this is the first such result for a three-layer stratification.

In addition, we identified different regimes for each mode according to the vorticity strength. In particular, when the vorticity is weak, part of the wavefront is able to propagate upstream (the so-called elliptic regime). However, when the vorticity is strong enough, the whole wavefront propagates downstream (the so-called hyperbolic regime). The transition between the elliptic and hyperbolic regimes occurs when the wavefront has one fixed point at the origin -- a structurally unstable case which is referred to as the parabolic regime. We found that a richer behaviour can be observed for the slower mode which is being squeezed in the presence of a current. Namely, as the vorticity strength increases, singularities of the swallowtail-type may arise and, eventually, solutions with compact wavefronts crossing the downstream axis cease to exist. We showed that the latter is related to the long-wave instability of the base flow. 

We also analysed the vertical structure of both interfacial ring modes and found that it is strongly three-dimensional. Namely, while the structure is qualitatively similar to that without any current when the current is weak, for stronger currents the maximum of the faster mode shifts from the top interface in the downstream direction to the bottom interface in the upstream direction. Similarly, the maximum of the slower mode shifts from the bottom interface to the top interface, respectively. 

The numerical modelling with the cKdV-type amplitude equation in the weakly nonlinear regime revealed very strong dispersive effects in the upstream direction, which leads to the effective erosion of the wave fronts of both interfacial ring modes in the upstream direction when nonlinearity is weak. This feature, and the difference between the behaviour of the solutions for the localised sources of elevation and depression, can be efficiently interpreted and understood using the graphs of the analytical coefficients of the amplitude equation, dependent on the solution of the modal and directional adjustment equations. In addition, we found that when nonlinearity is enhanced, fission of waves can occur in the upstream part of the wave structure. It is initiated in the upstream direction and can be observed in an increasingly wider sector as time evolves. We found that  fission of waves is more prominent for the second (slower) baroclinic mode.

 \section*{Acknowledgments}
Karima Khusnutdinova is grateful to Professor Sir Michael Berry for useful discussions and the references \cite{Arnold, Berry1, Berry2}.  Noura Alharthi gratefully acknowledges King Abdulaziz University (KAU), the Kingdom of Saudi Arabia, Rabiqh for financial support of her research.

%%
%We would like to thank Ricardo Barros for useful discussions. 

\section*{Appendix A}
In this section, we list the coefficients of the 2+1 dimensional amplitude equation (\ref{cKdV}) for the ring waves propagating over the shear flow $u_0(z) = \gamma z$, computed using the general formulae (\ref{mu})  derived in \cite{KZ}:
\vspace{0.3cm}
{\small
\begin{eqnarray*}
&&\textstyle\hspace{-2cm} \mu_{1} =  \frac{F(0) (k^2+k'^2)}{\gamma K} \Big[\frac{B^2}{\rho_{3}} \Big(\frac{1}{F^2(d_1)} - \frac{1}{F^2(0)}\Big) +\frac{C^2}{\rho_{2}} \Big(\frac{1}{F^2(d_2)} - \frac{1}{F^2(d_1)}\Big) 
+ \frac{A^2}{\rho_{1}} \Big(\frac{1}{F^2(1)} - \frac{1}{F^2(d_2)}\Big)\Big], \\[3ex]
&&\textstyle\hspace{-2cm} \mu_{2} = \frac{(k^2+k'^2)^3}{\gamma K} \Big[\frac{B^3}{\rho_{3}^2} \Big(\frac{1}{F^3(d_1)} - \frac{1}{F^3(0)}\Big) +\frac{C^3}{\rho_{2}^2} \Big(\frac{1}{F^3(d_2)} - \frac{1}{F^3(d_1)}\Big) + \frac{A^3}{\rho_{1}^2} \Big(\frac{1}{F^3(1)} - \frac{1}{F^3(d_2)}\Big)\Big], \\[3ex]
&&\textstyle\hspace{-2cm} \mu_{3} = - (k^2+k'^2) \times \Big[\frac{B^2(k^2+k'^2)^2 d_{1}^3}{3 \rho_{3} F^2(0)}  + \frac{A^2(k^2+k'^2)^2(1-d_{2})^3}{3 \rho_{1} F^2(1)}\\
&&\textstyle+ \frac{C^2(k^2+k'^2)^2 (d_2 - d_1)^3}{3 \rho_2 F^2(d_2)} - \frac{C D (k^2 + k'^2) (d_1 - d_2)^2 (2 F(d_1) + F(d_2))}{3 F(d_2)} - \frac{\rho_2 D^2 (F^3(d_1) - F^3(d_2))}{3 \gamma K}  \Big], \\[3ex]
&&\textstyle\hspace{-2cm} \mu_4 = - k (k'' + k) \Big[\frac{B^2}{\rho_3} (L(d_1) -L(0)) + \frac{C^2}{\rho_2} (L(d_2) -L(d_1)) + \frac{A^2}{\rho_1} (L(1) -L(d_2)) \Big],  \\
&&\textstyle\hspace{-2cm} \mbox{where} \quad L(z) = \int_0^z \Big[(k^2 + k'^2) \frac{1}{F^2} + 4 \gamma k' (k^2+k'^2) \sin \theta \frac{z}{F^3} + 3 \gamma^2 (k^2 + k'^2)^2 \sin^2 \theta \frac{z^2}{F^4}\Big]\  dz, \\[3ex]
&&\textstyle\hspace{-2cm} \mu_5 = - 2 k (k^2 + k'^2) \Big[\frac{B^2}{\rho_3} (M(d_1) -M(0)) + \frac{C^2}{\rho_2} (M(d_2) -M(d_1)) + \frac{A^2}{\rho_1} (M(1) -M(d_2)) \Big], \\
&&\textstyle\hspace{-2cm} \mbox{where} \quad M(z) = \int_0^z \Big [\frac{k'}{F^2} + \gamma (k^2 + k'^2) \sin \theta \frac{z}{F^3} \Big ]\ dz.
\end{eqnarray*}
}

%\newpage
Here, one should use the following formulae (only the dependence on $z$ is indicated explicitly, while it is implicitly assumed that $k = k(\theta)$):
\vspace{0.3cm}
\begin{eqnarray*}
&& K  =  (k \cos \theta - k'  \sin \theta),  \quad  F = F(z) = -s + \gamma z (k \cos \theta - k'  \sin \theta),  \\
&& \int_0^z \frac{dz}{F^2} = \frac{z}{F(0) F} , \quad \int_0^z \frac{z dz}{F^3} =  \frac{z^2}{2 F(0) F^2}, \quad
 \int_0^z \frac{z^2 dz}{F^4} = \frac{z^3}{3 F(0) F^3}.
\end{eqnarray*}

\section*{Appendix B}

Considering the range $\theta\in(0,\pi)$, it can be shown that in the elliptic regime
\begin{equation}
\theta( a) =\left\{
\begin{array}{ll}
\arctan \left (-\frac{2\sqrt{H- a^2}}{H_{ a}-2 a}\right )&\quad \mbox{if} \quad H_{ a}-2 a<0,\\[0.3cm]
\arctan \left (-\frac{2\sqrt{H- a^2}}{H_{ a}-2 a}\right )+\pi&\quad \mbox{if} \quad H_{ a}-2 a>0,
\end{array}\right.
\label{t1}
\end{equation}
and in the hyperbolic regime
\begin{equation}
\theta( a) =\left\{
\begin{array}{ll}
\arctan \left (-\frac{2\sqrt{H- a^2}}{H_{ a}-2 a}\right )&\quad \mbox{if} \quad H_{ a}-2 a<0,\\[0.3cm]
\arctan \left (\frac{2\sqrt{H- a^2}}{H_{ a}-2 a}\right )&\quad \mbox{if} \quad H_{ a}-2 a>0.
\end{array}\right.
\label{t1}
\end{equation}
In the parabolic regime, it turns out that $H_{ a}-2 a<0$ when $H- a^2>0$, and then
\begin{equation}
\theta( a)=\arctan \left (-\frac{2\sqrt{H- a^2}}{H_{ a}-2 a}\right ).
\end{equation}

\section*{Appendix C}

%{\color{red} Add Numerical scheme (briefly) and typical time and space steps.
%\bigskip

%\vspace{0.6cm}
%\noindent In this section, we present proposed one of finite difference method, that is known implicit finite-difference method (IFDM) to numerically solve the derived 2 + 1-dimensional cKdV-type equation; which can be written in the form
\noindent In this section, we present an implicit finite-difference method to numerically solve the derived $1+1$-dimensional cKdV-type equation~(\ref{cKdV1}), see \cite{KZ1,FM}. The equation can be written in the form
\begin{eqnarray}
&&\mu_{1} A_{R} + \frac{\mu_{2}}{2} (A^2)_{\xi} +  \mu_{3} A_{\xi \xi \xi} + \mu_{4}  \frac{A}{R}  = 0, \label{ckdveq} % 27
\end{eqnarray}
\noindent where $\mu_{i} = \mu_{i}(\theta)$, $i = 1,\ldots,4$ and $A = A(R,\xi,\theta)$ with the dependence on $\theta$ resulting through the dependence of the coefficients on $\theta$. We solve the equation for each $\theta_m=m\,\Delta\theta$, $m = 0, 1, \ldots, M$, where $\Delta\theta=\pi/M$. In our numerical simulations, we typically take $M=100$.

We assume that $R\in[R_0,\,R_{\max}]$ typically choosing $R_0=0.3$ and $R_{\max}=1.4$ in our numerical simulations. In the physical space, we obtain solutions for $t\in[t_0,\,t_{\max}]$ and we typically take $t_0=0$ and $t_{\max}=300$. Since
$t=({R/\varepsilon-\xi})/s$, this implies that the domain for $\xi$ is $[\xi_{\min},\,\xi_{\max}]$, where 
\begin{equation*}
\xi_{\min} = R_0/\varepsilon-s\,t_{\max}, \qquad \xi_{\max}=R_{\max}/\varepsilon-s\,t_0.
\end{equation*}
In our numerical simulations, we take $\varepsilon=0.02$.

We discretise the $\xi$ domain into the grid $\xi_l=l\,\Delta\xi$, $l = 0,\, 1,\, \ldots,\, L$, where $\Delta \xi=(\xi_{\max}-\xi_{\min})/L$, and we typically choose $L=1000$. The numerical solution is obtained at discrete $R$ values $R_n=R_0+n\,\Delta R$, $n = 1,\, \ldots,\, N$, where $\Delta R=(R_{\max}-R_0)/N$, and we typically choose $N=1600$.

For each $\theta=\theta_m$, $m = 0, 1, \ldots, M$, we use the following second-order central-difference approximation of the partial derivative of $A_{\xi\xi\xi}$ at $R=R_n$, $\xi=\xi_l$:
\begin{eqnarray*}
%&&A_{\xi}\mid_{l,m}^{n} = \frac{A_{l+1,m}^{n}-A_{l-1,m}^{n}}{2 \Delta \xi} + \it O(\Delta \xi)^2,  \label{30}\\
%&&(A_{\xi})_{l}^{n} = \frac{A_{l+1}^{n}-A_{l-1}^{n}}{2 \Delta \xi},  \label{Axi1}\\
&& (A_{\xi\xi\xi})_{l}^{n} = \frac{A_{l+2}^{n}-2A_{l+1}^{n}+2A_{l-1}^{n}-A_{l-2}^{n}}{2 \Delta \xi^{3}},  
\end{eqnarray*}
where the subscripts and superscripts are used to indicate the index of the $\xi$ and $R$ value, respectively, at which the corresponding quantity is considered.

We implement an implicit Crank-Nicolson-type method, approximating the equation at the grid points $R=R_{n+\frac{1}{2}}\equiv R_n+\Delta R/2$, $\xi=\xi_l$, where $n = 1,\, \ldots,\, N$ and  $l = 0,\, 1,\, \ldots,\, L$, using the second-order central-difference formula for $A_R$:
\begin{equation*}
(A_R)_{l}^{n+\frac{1}{2}} = \frac{A_{l}^{n+1}-A_{l}^{n}}{\Delta R}.
\end{equation*}
The third-order $\xi$ derivative for $R=R_{n+\frac{1}{2}}$, $\xi=\xi_l$ is approximated using averages of the second-order central-difference formulas at $R=R_{n}$ and at $R=R_{n+1}$, so that
\begin{eqnarray*}
%&&(A_{\xi})_{l}^{n+\frac{1}{2}} = \frac{1}{2}\Bigl((A_{\xi})_{l}^{n}+(A_{\xi})_{l}^{n+1}\Bigr),\\
&& (A_{\xi\xi\xi})_{l}^{n+\frac{1}{2}} = \frac{1}{2}\Bigl((A_{\xi\xi\xi})_{l}^{n}+(A_{\xi\xi\xi})_{l}^{n+1}\Bigr).
\end{eqnarray*}
In addition, we have 
\begin{eqnarray*}
&&\left(\frac{A}{R}\right)_{l}^{n+\frac{1}{2}} = \frac{1}{2}\left( \frac{A_l^{n+1}}{R_{n+1}} + \frac{A_l^{n}}{R_{n}}\right),
\end{eqnarray*}
which is also second-order accurate in $\Delta R$.

Treating the nonlinear term in a similar way would results in a fully implicit scheme, where a system of nonlinear equations would need to be solved at each time step to obtain $\boldsymbol{A}^{n+1}=(A_0^{n+1},\,A_1^{n+1},\,\ldots,\,A_L^{n+1})$ from $\boldsymbol{A}^{n}=(A_0^{n},\,A_1^{n},\,\ldots,\,A_L^{n}).$ To simplify this, we linearise the nonlinear term. Firstly, we have the following second-order accurate (in $\Delta R$) approximation:
$$f_l^{n+\frac{1}{2}}=\frac{1}{2}(f_l^{n+1}+f_l^{n}).$$
Using the Taylor series expansion of $f$ with respect to $R$, we obtain
\begin{eqnarray*}
&& f_{l}^{n+1} = f_{l}^{n} + \Delta R \Big(\frac{\partial f}{\partial R} \Big)_ {l}^{n}  +  O(\Delta R^2) = f_{l}^{n}+ D_{l}^{n} \; \Delta A_{l}^{n+1} +  O(\Delta R^2), 
\end{eqnarray*}
where 
\begin{eqnarray*}
D_{l}^{n} = \Big(\frac{\partial f}{\partial A} \Big)_{l}^{n} = 2A_{l}^{n}\quad     \mbox{and} 	\quad  \Delta A_{l}^{n+1} = A_{l}^{n+1}-A_{l}^{n}. %\\[3ex]
\end{eqnarray*}
This then implies that 
$$f_l^{n+\frac{1}{2}}=f_l^{n}+A_{l}^{n}(A_{l}^{n+1}-A_{l}^{n})= A_{l}^{n}A_{l}^{n+1},$$
which is accurate to the second order in $\Delta R$. The $\xi$ derivative of $f$ at $R=R_{n+\frac{1}{2}}$, $\xi=\xi_l$ is then approximated using the second-order central-difference formula, giving
\begin{equation*}
(f_\xi)_l^{n+\frac{1}{2}}=\frac{A_{l+1}^{n}A_{l+1}^{n+1}-A_{l-1}^{n}A_{l-1}^{n+1}}{2\Delta\xi},
\end{equation*}
with the truncation error $O(\Delta R^2)+O(\Delta \xi^2)$.

The resulting discretised equation at $R=R_{n+\frac{1}{2}}$, $\xi=\xi_l$ for $\theta=\theta_m$ takes the form
\begin{eqnarray}
\mu_{1}(\theta_m) \left(\frac{A_{l}^{n+1}-A_{l}^{n}}{\Delta R}\right) 
+ \frac{\mu_{2}(\theta_m)}{2} \left(\frac{A_{l+1}^{n}A_{l+1}^{n+1}-A_{l-1}^{n}A_{l-1}^{n+1}}{2\Delta\xi}\right)\nonumber\\ 
\qquad+  \frac{\mu_{3}(\theta_m)}{2}\Bigl((A_{\xi\xi\xi})_{l}^{n}+(A_{\xi\xi\xi})_{l}^{n+1}\Bigr)  + \frac{\mu_{4}(\theta_m)}{2}  \left( \frac{A_l^{n+1}}{R_{n+1}} + \frac{A_l^{n}}{R_{n}}\right)  = 0, \label{discretised_eq}
\end{eqnarray}
with the truncation error  $O(\Delta R^2)+O(\Delta \xi^2)$.

For each $m=0,\,\ldots,\,M$, $n=0,\,\ldots,\,N-1$, considering equation (\ref{discretised_eq}) for  $l=0,\,\ldots,\,L$, we obtain a system of linear equations for $A_0^{n+1},\,A_1^{n+1},\,\ldots,\,A_L^{n+1}$, assuming that $A_0^{n},\,A_1^{n},\,\ldots,\,A_L^{n}$ have been determined from the previous $R$ step (or from the initial condition for $n=0$). We note that we choose the domain in $\xi$ in such a way that $A$ becomes close to $0$  as $\xi$ tends to $\xi_{\min}$ or $\xi_{\max}$, so that we may assume $A_{l}^{n}=0$ for $l<0$ and for $l>L$. The resulting system of linear equations can be appropriately rearranged and solved using Gaussian elimination for each $n=0,\,\ldots,\,N-1$.

% \section*{New References}
% \bibliography{%
% $HOME/Home/Bibliography/uwelitall,%$
% $HOME/Home/Bibliography/books} %$

\end{document}